\documentclass{jfm}
\usepackage{graphicx}
\usepackage{epstopdf, epsfig,color}
\usepackage{amssymb}
\usepackage{wrapfig,txfonts}
\usepackage{natbib}
\usepackage{url}

  \newcommand{\thalf}{\textstyle{\frac{1}{2}} \displaystyle }

  \newcommand{\bfomega}{\mbox{\boldmath $\omega$}} 
 
   \newcommand{\be}{\begin{equation}}
      \newcommand{\ee}{\end{equation}}
         \newcommand{\tn}{\textnormal}
  \newcommand{\bea}{\begin{eqnarray}}
   \newcommand{\eea}{\end{eqnarray}}

\title[ Finite-time singularity]{ Towards a finite-time singularity of the \\Navier-Stokes equations}

\author[ H.K.Moffatt  and Y.Kimura ]{ H. K. Moffatt$^1$ and Yoshifumi Kimura$^2$ }

\affiliation{$^1$Department of Applied Mathematics and Theoretical Physics, \\
Wilberforce Road, Cambridge CB3 0WA, UK\\
[\affilskip]$^2$Graduate School of Mathematics, Nagoya University,\\
Furo-cho, Chikusa-ku, Nagoya 464-8602 Japan}

\pubyear{??} \volume{??} \pagerange{??--??}
\date{}
\begin{document}

\maketitle

\begin{abstract}
The evolution towards a finite-time singularity of the Navier-Stokes equations for flow of an incompressible fluid of kinematic viscosity $\nu$ is studied, starting from a finite-energy configuration of two vortex rings of circulation $\pm \Gamma$ and radius $R$, symmetrically placed on two planes at angles $\pm\alpha$ to a plane of symmetry $x=0$.  The minimum separation of the vortices $2s$ and the scale of the core cross-section $\delta$ are supposed to satisfy the initial inequalities $\delta\ll s\ll R$, and the vortex Reynolds number $R_{\Gamma}=\Gamma/\nu$ is supposed very large. It is argued that in the subsequent evolution, the behaviour near the points of closest approach of the vortices (the `tipping points') is determined solely by the curvature $\kappa(\tau)$ at the tipping points and by $s(\tau)$ and  $\delta(\tau)$, where $\tau=(\Gamma/R^2)\,t$ is a dimensionless time variable.  The Biot-Savart law is used to obtain analytical expressions for the rate-of-change of these three variables, and a nonlinear dynamical system relating them is thereby obtained.  The solution shows a finite-time singularity, but the Biot-Savart law breaks down just before this singularity is realised, when $\kappa s$ and $\delta/\!s$ become of order unity.  The dynamical system admits  `partial Leray scaling' of just $s$ and $\kappa$, and ultimately full Leray scaling of $s,\kappa$ and $\delta$, conditions for which are obtained. The tipping point trajectories are determined; these meet at the singularity point at a finite angle.

An alternative model is briefly considered, in which the initial vortices are ovoidal in shape, approximately hyperbolic near the tipping points, for which there is no restriction on the initial value of the parameter $\kappa$; however it is still the circles of curvature at the tipping points that determine the local evolution, so the same dynamical system is obtained, with breakdown again of the Biot-Savart approach just before the incipient singularity is realised.  

 The Euler flow situation ($\nu=0$) is considered, and it is conjectured on the basis of the above dynamical system that  a finite-time singularity can indeed occur in this case.   
  
 \end{abstract}


\section{Introduction}\label{Sec_introduction}
In this paper, we address the Clay Mathematics Prize question articulated by \cite{Fefferman2006}, one variation of which may be paraphrased as follows:  given at some initial instant a smooth velocity field ${\bf u}_{0}({\bf{x})}$ of finite kinetic energy in an incompressible fluid filling all space, can a singularity of the field appear within a finite time under evolution governed by the Navier-Stokes equations?  We shall show by explicit example that, with high probability, a point singularity for the Euler equations can indeed appear within a finite time;  and we shall show that consideration of vortex reconnection is required to determine whether a Navier-Stokes singularity does or does not occur.

The singularity question was originally posed by \cite{Leray1934}, and has since then provoked intense investigation, both theoretical and numerical, but no definitive answer has hitherto emerged. The  complex situation with regard to the Navier-Stokes equations has been reviewed by \cite{Doering2009} and in the recent book, both entertaining and instructive, of \cite {Lemarie-Rieusset2016}.  Much parallel work has been devoted to the Euler equations for ideal flow of an incompressible fluid of zero viscosity; but even here the `regularity' question remains open.  It is widely believed that, even if the Euler equations can admit the appearance of a finite-time singularity, the diffusive effect of viscosity, no matter how weak, must always smooth out an incipient singularity before it forms.  We shall show that the nonlinear process leading to a singularity can be stronger than the viscous smoothing effect throughout nearly all of the evolution that we consider, until the very last moment at which viscous reconnection of vortex lines must be taken into account

The theorem of \cite{Beale1984} states in effect that if a finite-time singularity occurs at time $t=t_c$, then the vorticity $\bfomega=\nabla\times\bf u$ must be unbounded as $t\!\uparrow \!t_c$;   more precisely, that
\be
\int_{0}^{t_c}\tn{sup}\,|\bfomega({\bf x},t)|_{{\bf x}\in \mathbb{R}^3}=\infty\,.
\ee
This theorem, proved in the Euler context, but equally applicable (as the authors state) to the Navier-Stokes equations, encourages us to specify the initial conditions in terms of the vorticity field and to focus on the development of this vorticity field  $\bfomega({\bf x}, t)$ for $t>0$.  

Many similar and tighter results, have since been obtained by powerful methods of functional analysis.  For example, \cite{Seregin2002} have proved that if a finite-time singularity of the Navier-Stokes equations occurs at time $t=t_c$, then the pressure field $p({\bf x}, t)$ is unbounded below, a result that makes good physical sense, given that the pressure is minimal in the core of a stretched vortex, and if $|\bfomega({\bf x},t)|\rightarrow\infty$ at some point in the core due to extreme stretching, then $p$ may be expected to tend to $-\infty$ at the same point.  \cite{Escauriaza2003} have similarly  proved that  the $L_{3}$-norm of $\bf u$ must blow up as $t\uparrow t_{c}$:
\be
\tn{lim\,sup}_{t\uparrow t_c}\,\int_{\mathbb{R}^3}|{\bf {u}}({\bf {x}},t)|^{3}\,\tn{d}V=\infty\,.
\ee

\cite{Caffarelli1982} had earlier proved  that, if a singularity occurs, then the space-time Hausdorff dimension of the singularity is such that it can be at most a point singularity. For this reason, if a singularity is to form, then the length-scale of the vorticity field must collapse to zero from all directions towards the point in question.  This is difficult to reconcile with the fact that the rate-of-strain tensor which is responsible for vorticity intensification must always, in an incompressible fluid,  have at least one positive eigenvalue suggesting \emph {increase} rather than decrease of scale in the direction of the corresponding eigenvector.  This  might lead one to believe that a point singularity cannot form; but we shall show that this belief is misplaced, although for rather subtle reasons.

A  clue is provided by the theorem of \cite{Constantin1996}, who proved, again in the context of the Euler equations, that any singularity must involve a singularity of the direction as well as the magnitude of the vorticity, so that the direction of the vorticity vector is indeterminate at the singularity.  This suggests that one should look to the interaction of non-parallel vortex tubes  in seeking a route to a possible singularity -- rather like looking for the proverbial needle in a haystack!

\begin{figure}
\begin{center}
\begin{minipage}{0.99\textwidth}
\includegraphics[width=0.30\textwidth, trim=0mm 0mm 0mm 0mm]{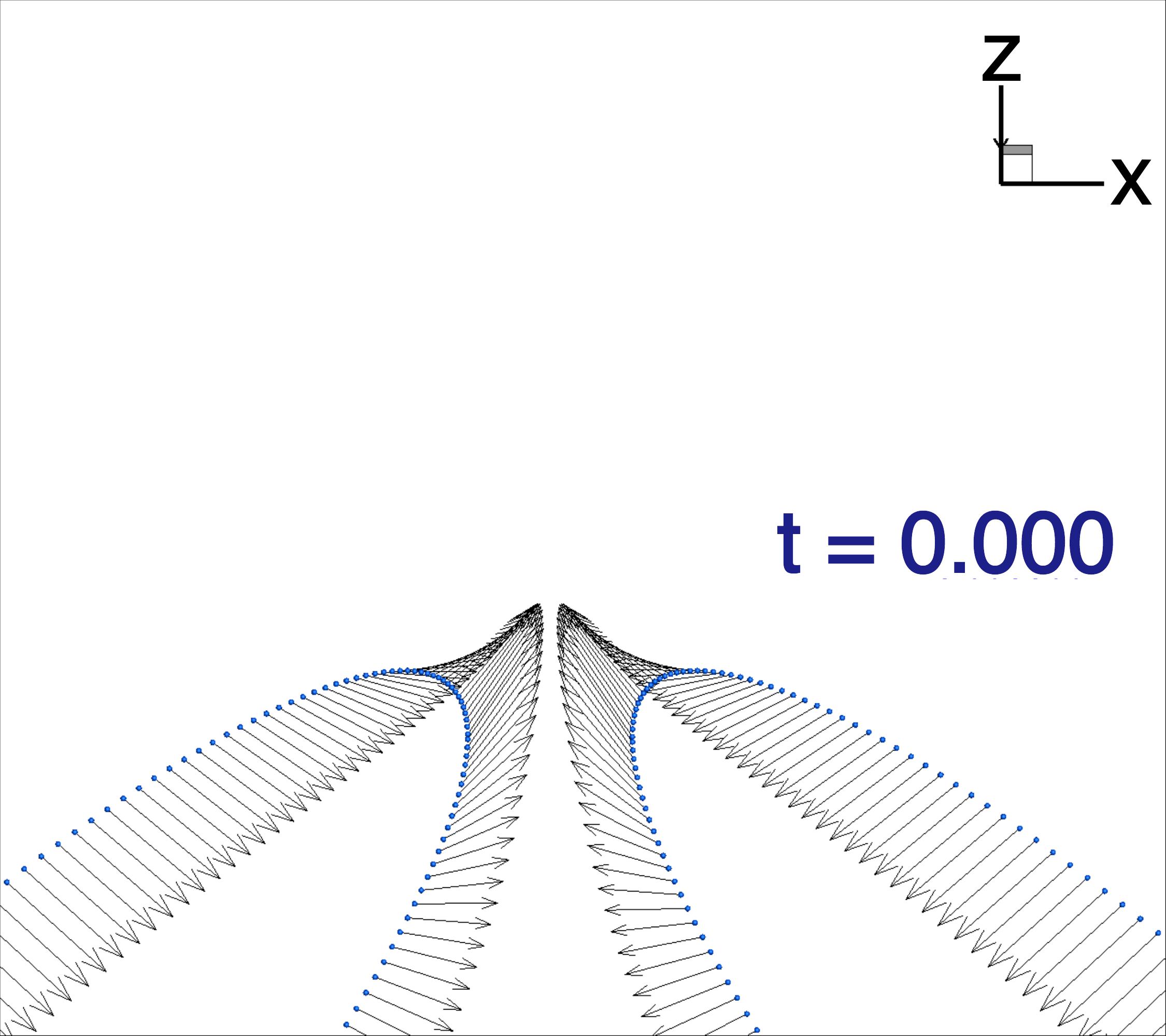}
\hspace*{12pt}
\includegraphics[width=0.30\textwidth,  trim=0mm 0mm 0mm 0mm]{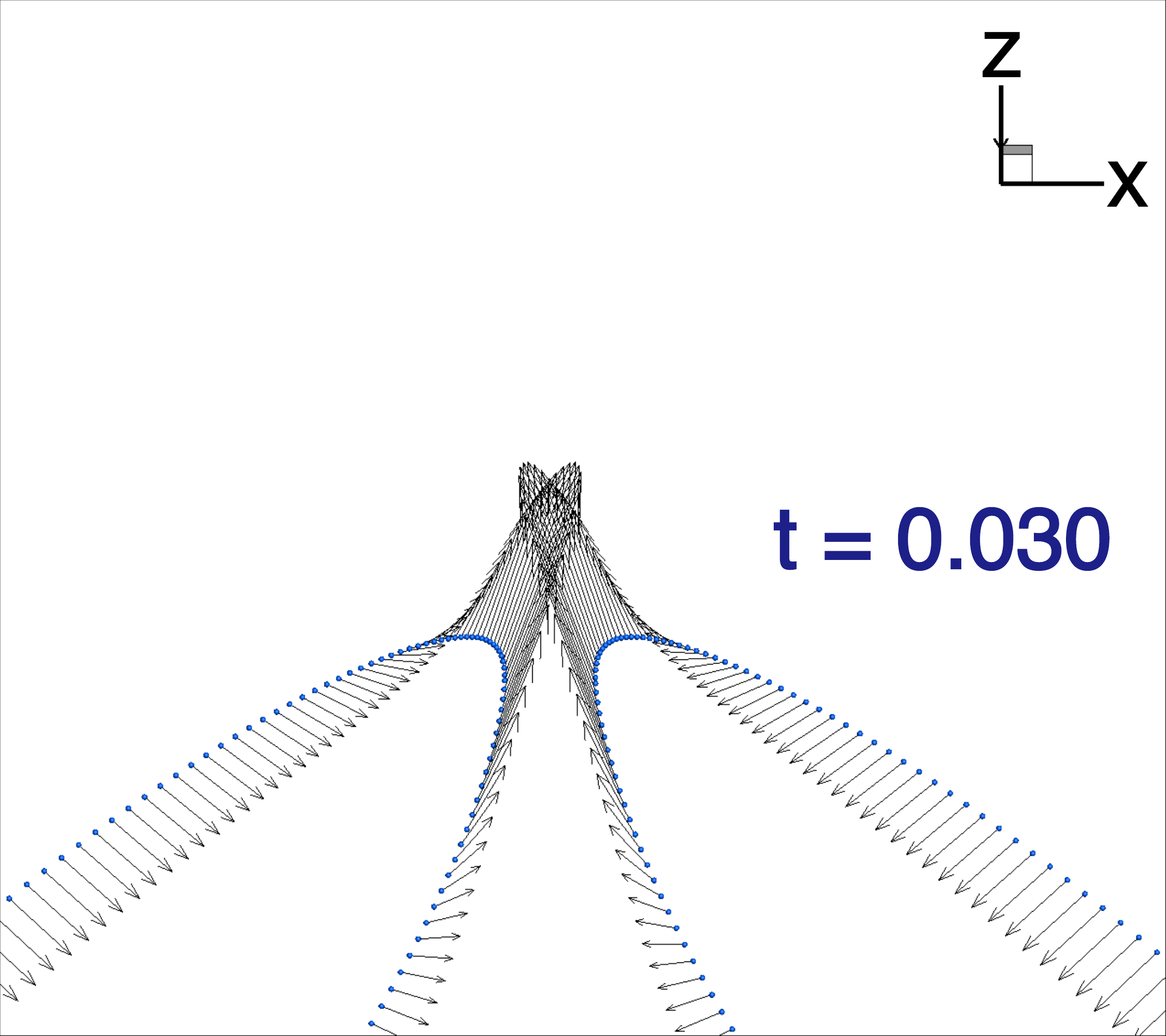}
\hspace*{12pt}
\includegraphics[width=0.30\textwidth,  trim=0mm 0mm 0mm 0mm]{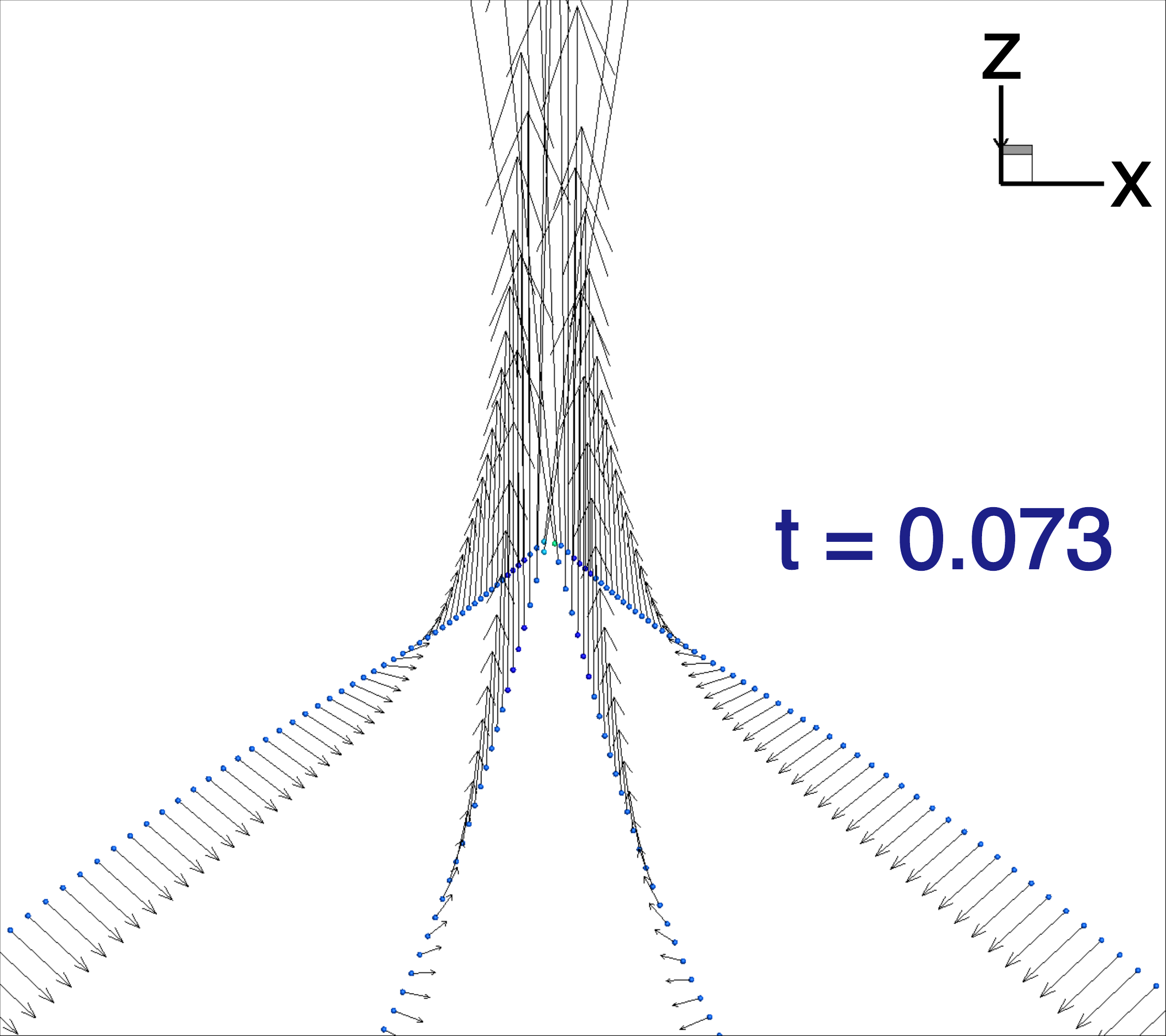}\\
\vskip 2mm
\hskip 18mm (a) \hskip 42mm (b) \hskip 42mm (c) 
\vskip 2mm
\includegraphics[width=0.30\textwidth, trim=0mm 0mm 0mm 0mm]{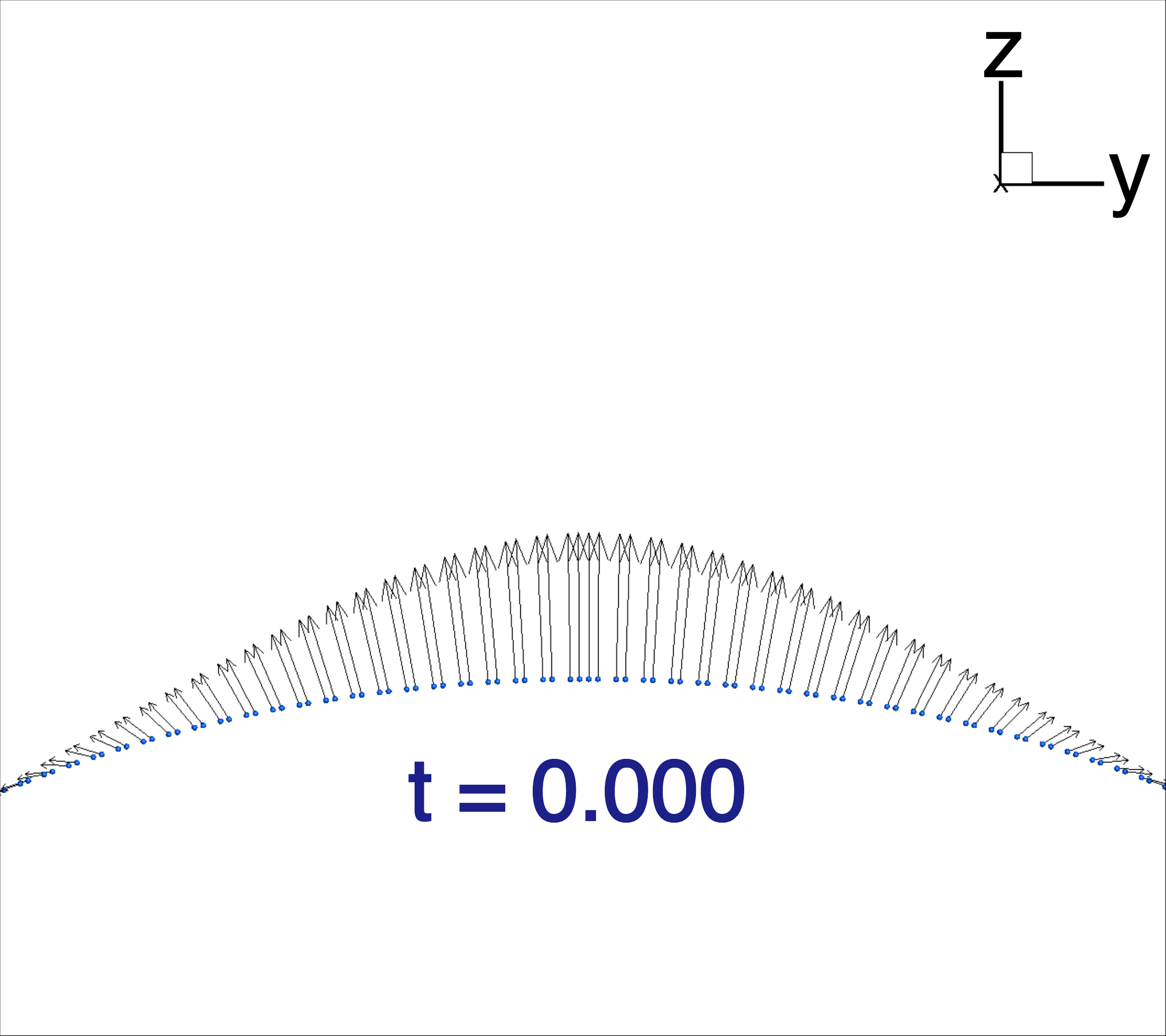}
\hspace*{12pt}
\includegraphics[width=0.30\textwidth,  trim=0mm 0mm 0mm 0mm]{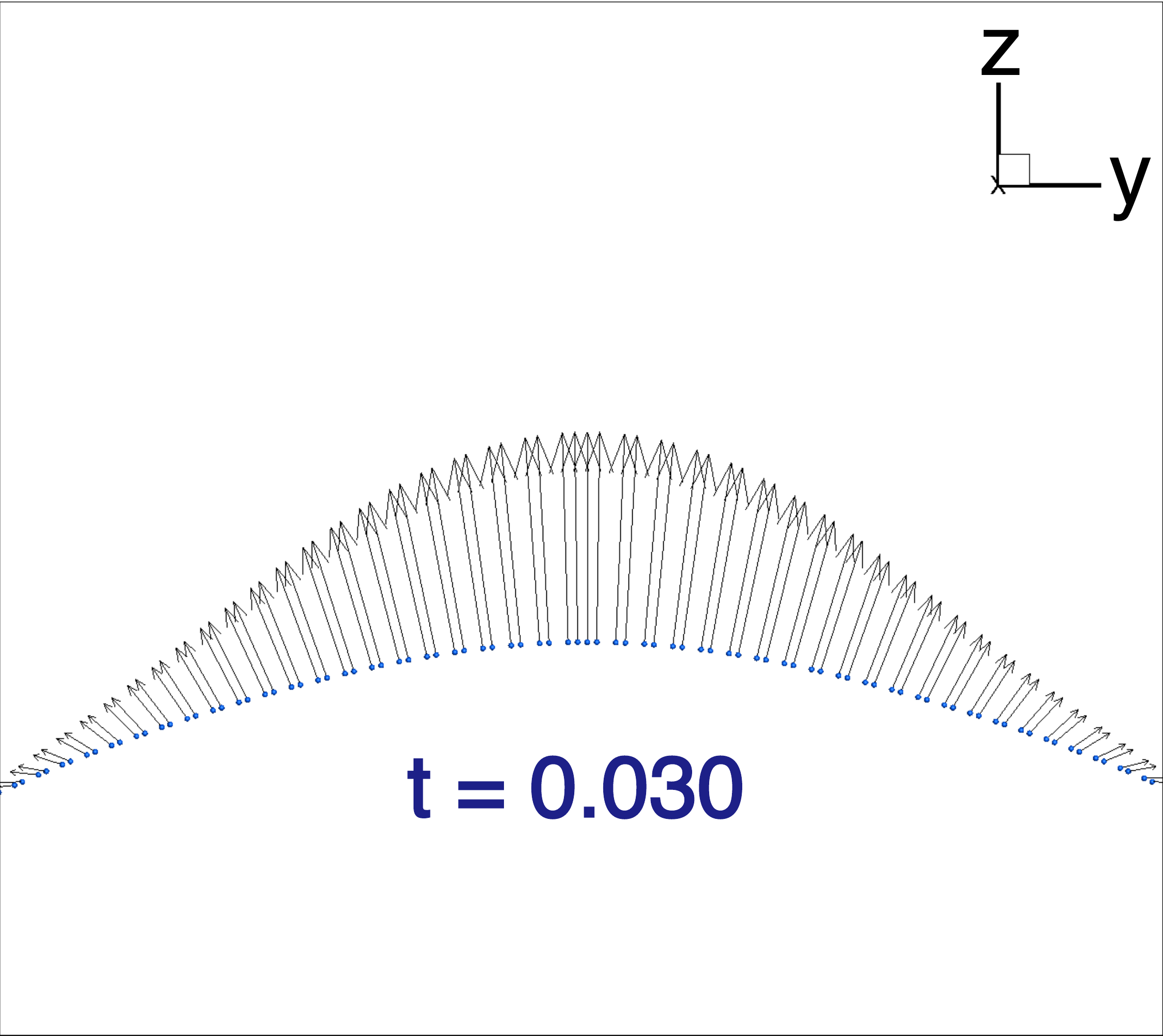}
\hspace*{12pt}
\includegraphics[width=0.30\textwidth,  trim=0mm 0mm 0mm 0mm]{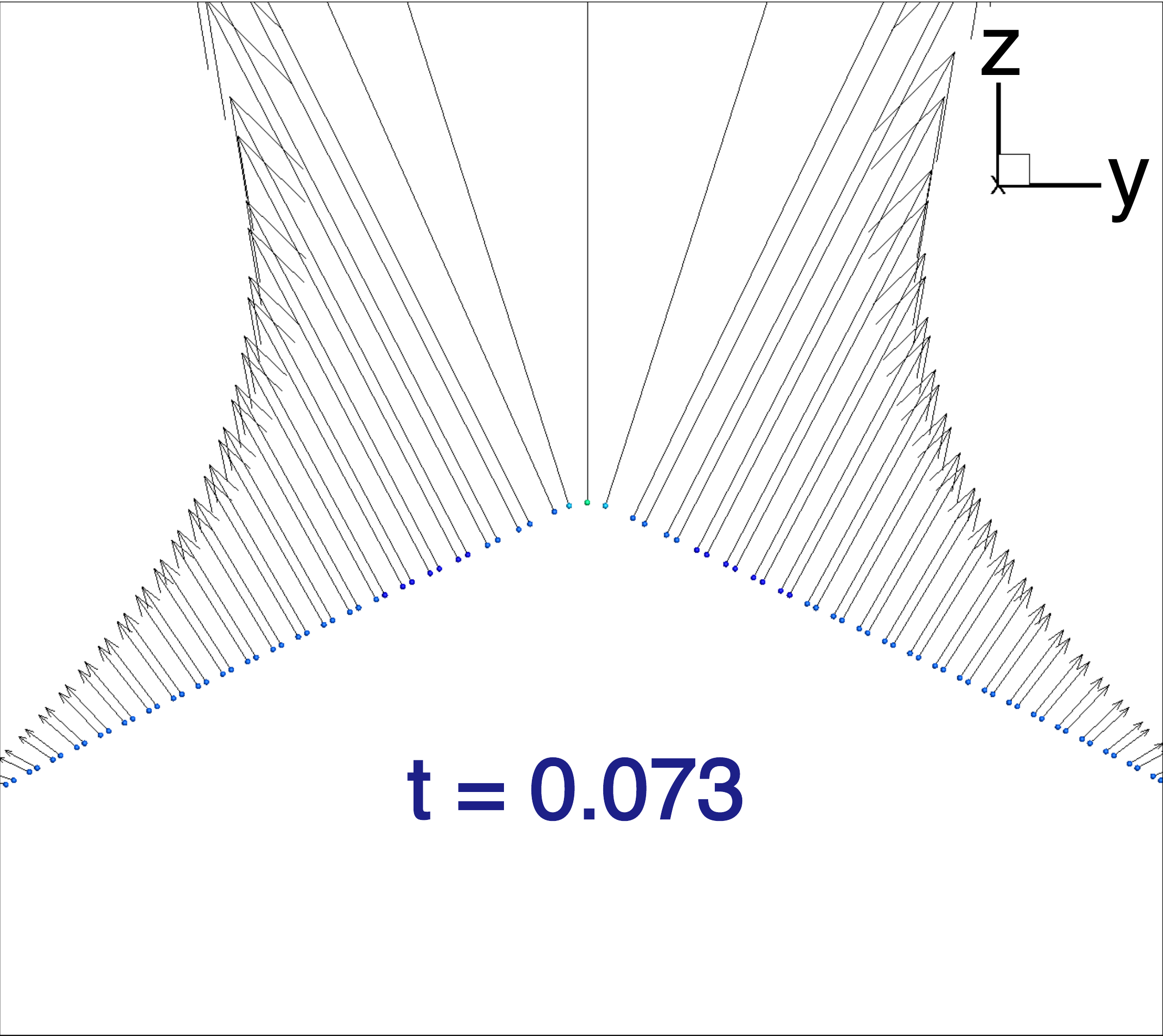}
\vskip 2mm
\hskip 18mm (d) \hskip 42mm (e) \hskip 42mm (f) 
\vskip 2mm
\end{minipage}
\end{center}
\caption{Biot-Savart approach  to a singularity (using `cut-off' regularisation) starting with two circular vortices (as in figure \ref{Fig_sketch}), radius $R$ and circulations $\pm\Gamma$, symmetrically placed on  planes inclined at angles $\pm\pi/4$ to the vertical, with initial separation parameter $s(0)=0.03$ (as defined in \S\ref{Sec_Initial-condition}); time is non-dimensionalised relative to $R^2/\Gamma$; the arrows represent the velocity distribution on the vortices; (a) $t=0$: the initial state showing only the portion of the circles near the points of closest approach; (b) $t=0.03$: the tip separation has decreased and the local velocity has increased; (c) $t=0.073$: very near to the singularity time, when the vortices approach `collision' at the origin; (d-f) $xz$-projections of the same evolution, showing the increase of curvature at the tip, and approach to locally hyperbolic form.  A movie showing this process may be found in the supplementary material.}
\label{Fig_rings_3D}
\end{figure}
Such a search was initiated by the numerical investigations of  \cite{Schwarz1985} and  \cite{Siggia1985} on the behaviour of interacting vortex filaments in Euler flow, an approach that has particular relevance for the evolution and reconnection of quantised vortices in superfluid 
$^{4}$He (see for example \citealt{Bewley2008}, who observed the motion of  particles trapped on the quantised vortices).  \cite{Siggia1985a} set out a procedure for computing the Biot-Savart evolution of a vortex filament, taking into account the decrease of the cross-section of the filament associated with vortex stretching; they provided evidence that the cross-sectional area may decrease to zero in a finite time, or at least for so long as the filament remains genuinely `tube-like' without strong deformation in the plane of cross-section.  Similarly, \cite{Pumir1987} showed that a vortex filament can pair with itself with length-scale decreasing like $(t_{c}-t)^{1/2}$ and velocity increasing like   $(t_{c}-t)^{-1/2}$ (`Leray scaling' that implies singularity at time $t=t_{c}$), and concluded that  a finite-time Navier-Stokes singularity ``cannot be easily dismissed".  In the present paper we adopt an approach similar to that of these pioneering papers.

The work of Pumir \& Siggia led to the important paper of \cite{Waele1994} on the evolution of two vortex filaments treated as line-singularities, in a variety of initial configurations. These authors again focussed on Euler flow, using a regularised form of the Biot-Savart law to determine the velocity field induced by the vortex filaments; they found  that a `pyramid'  or `tent-like' structure tended to emerge in every case. We have shown in the same way (\citealt{Kimura2018a, Kimura2018})  how a singularity may be approached by counter-rotating vortices centred on the two branches of a `tilted hyperbola'.  A  similar Biot-Savart computation for two initially circular vortices on planes inclined at angles $\pm \pi/4$ shows the approach to a singularity displayed in figure \ref{Fig_rings_3D}(a--c);  the $xz$-projections (figure \ref{Fig_rings_3D}(d--f)) show a strong increase of curvature at the tip and approach to a locally hyperbolic form.  Discretisation of the vortices and the `cut-off' regularisation adopted clearly limits the validity of this type of computation as the singularity is approached. The challenge that we now confront is to analyse the details of this phenomenon under Navier-Stokes evolution, taking account of the finite vortex-core structure and the associated effects of viscous diffusion.

There have been many direct numerical simulations (DNS) of the processes of vortex interaction and reconnection in viscous fluids, some showing trends towards a singularity (e.g. \citealt{Kerr1989, Kerr2005}), and some showing near approach but ultimate evasion of a singularity (e.g. \citealt{Deng2005, Hou2006, Hou2008}) --- we might describe this as the \emph{adventus interruptus} of singularity formation! But whatever the outcome of such investigations, it must be admitted  that, even if a genuine singularity does exist for suitable initial conditions, any brute-force computational approach must fail in the final stage of approach to it, simply because the length-scale of the phenomenon must always decrease to less than the computational resolution available. The best that can be hoped is detection of self-similarity during the initial stages of the collapse process, which could be then coupled with analysis incorporating appropriate scaling assumptions in the manner well described by \cite{Eggers2015}.  This desirable outcome has not yet been achieved, the indications being rather that there are significant departures from self-similarity as a putative singularity is approached (\citealt{Kerr2005a, Bustamante2008}).

The computational difficulties are evident in the recent work of \cite{Hormoz2012} and 
\cite{Brenner2016}, who find that the Biot-Savart description of approaching vortices  has a tendency to break down some time before a singularity is reached because of the flattening of the vortex cross-sections that is observed  to occur. They propose a model in which counter-rotating vortex pairs are flattened to sheets that become unstable through some kind of Kelvin-Helmholtz or roll-up mechanism, forming a new sequence of vortex pairs on a smaller scale on which the whole process can then iterate. DNS has already achieved  evidence of the earliest stages of such an iterative process (\citealt{McKeown2018}).

We contend however that the vortex core flattening encountered by \cite{Kerr2005a}, \cite{Brenner2016} and others does not in fact occur under Navier-Stokes evolution  if the separation of the vortices is small compared with their radius of curvature at closest approach \emph{and} if the vortex Reynolds number R$_{\Gamma}=\Gamma/\nu$ is large enough;  this assertion is based on the asymptotic analysis of \cite{Moffatt1994a}, by which it was proved that a vortex subjected to non-axisymmetric strain is \emph {not} disrupted but retains its compact cross-sectional structure when R$_{\Gamma}\!\gg\!1$ (for details, see \S \ref{non-axisymmetric strain} below).  We shall find that the above dual requirement requires an exceptionally large R$_{\Gamma}$, far beyond current DNS possibilities.

On the experimental side, the pioneering work of \cite{Kleckner2013} on the visualisation of knotted and linked vortices and the reconnection processes that they undergo has had an electrifying effect on research in this area,  although, as in computational approaches, the vortex Reynolds number  (of order $10^{4}$ -- $10^{5}$ in these experiments) is limited for reasons of practicality.   In recent work, \cite{Scheeler2017} have shown experimentally that the helicity of a vortex tube, consisting of the sum of its twist and writhe ingredients, is conserved to good approximation on time-scales O$(R^{2}\!/\Gamma)$, where $R$ is the \emph{mean} radius of curvature of the vortex, but that on longer time-scales,  the  twist of vortex lines within the core decays to zero, while the writhe helicity remains approximately constant.  We shall provide an explanation of this observation, which has obvious relevance for vortex interactions, in 
\S\ref{Preferential_twist_decay} below.

As regards the remaining structure of the paper, and the conclusions that may be drawn from the detailed analysis, we refer the reader immediately to the concluding \S \ref{Conclusions}.

\section{ Some relevant background}\label{Sec_Background}
\subsection{Vortex subjected to time-dependent strain}\label{Sec_time-dependent strain}
We first recall  the simple idealised problem (\citealt{Moffatt2000b}) of a Burgers-type vortex 
\begin{equation}\label{Burgers}
\bfomega=(0,0,\omega(r,t))
\end{equation}
in cylindrical polar coordinates $(r,\,\theta,\,z)$,
subjected to an axisymmetric but time-dependent strain field
\begin{equation}\label{axisymmetric_strain}
{\bf U} = (-\thalf \lambda(t)\,r, 0, \lambda(t)\,z),
\end{equation}
with $\lambda(t)>0$ for $t\ge 0$, to ensure positive vortex stretching.
The vorticity equation in this situation is linear
\begin{equation}\label{Burgers_vorticity}
\frac{\partial \omega}{\partial t}=\frac{\lambda(t)}{2r}\frac{\partial}{\partial r}(r^{2} \omega)
+ \frac{\nu}{r}\frac{\partial}{\partial r}\left(r\frac{\partial\omega}{\partial r}\right)\,,
\end{equation}
and with the  initial condition
\begin{equation}\label{initial_condn}
\omega(r,\,0)=\omega_0\exp\left({-r^2/\delta_{0}^2}\right),
\end{equation}
a similarity solution of the form
\begin{equation}\label{similarity_soln}
\omega(r,\,t)=\frac{\Gamma (\mu-1)}{4\pi\nu(t_{c}-t)}\exp\left(\frac{-(\mu-1)r^2}{4\nu(t_{c}-t)}\right)
\end{equation}
exists provided
\begin{equation}\label{gamma(t)}
\lambda(t)=\frac{\mu}{t_{c}-t}\,,\quad 0<t<t_{c}\,,
\end{equation}
where $\mu$ is a  constant satisfying
\begin{equation}\label{c}
\mu=1+\lambda_{s} t_{c},\quad \lambda_s=4\nu/\delta_{0}^2.
\end{equation}
Here, $\lambda_s $ is the strain rate that would be required to maintain a steady Burgers vortex against viscous erosion. Provided $\lambda(0)\,(=\!\mu/t_{c})\! >\!\lambda_s$,  the solution (\ref{similarity_soln})  blows up at the finite time  $t\!=\!t_{c}\!=\!(\lambda(0)\!-\!\lambda_s)^{-1}$, because the effect of the increasing strain-rate dominates over viscous diffusion for all $t\in (0,t_{c})$.  

For this very special exact solution of the Navier-Stokes equation, the cross-sectional scale of the vortex is  $\delta (t)=[4\nu (t_{c}-t)/(\mu-1)]^{1/2}$ scaling  like 
$(t_{c}-t)^{1/2}$,  and the vorticity $\omega\sim \delta^{-2}$ scales like $(t_{c}-t)^{-1}$.  This particular scaling, first identified by \cite{Leray1934},  is  appropriately  described as  `Leray scaling'. 
 
If the straining flow  ${\bf U}$ is induced by other vortices, then we can imagine a situation in which these other vortices approach the strained vortex in such a way that $\lambda $ increases without limit within a finite time. The problem is first to define an appropriate vortex configuration such that the length-scale of the interaction region of the vortices decreases in all directions towards a point singularity, despite the conflicting tendency associated with the direction of positive strain.  We shall show how this difficulty is overcome in the model developed in the following sections.  

We shall need a slight generalisation of the result (\ref{similarity_soln}) -- (\ref{c}) above.  For general $\lambda(t)>0$, guided by (\ref{similarity_soln}), we may look for a similarity solution of (\ref{Burgers_vorticity}) of the form
\begin{equation}\label{new_similarity}
\omega(r,\,t)=\frac{1}{4\pi  \delta^{2}(t)}\exp\left(-\eta^{2}\right)\quad \textnormal{where}\,\,\eta=\frac{r}{2\,\delta(t)}\,.
\end{equation}
Substituting in  (\ref{Burgers_vorticity}), we find that this equation is satisfied provided 
\begin{equation}\label{gamma_delta}
\frac{d}{dt}\delta^2=\nu- \lambda(t)\,\delta^2\,.
\end{equation}
If $\lambda(t)=\mu/(t_{c}-t) $ where $\mu>0$, then, with $\tau=t/t_{c}$,
the solution of this equation satisfying the initial condition $\delta(0)=\delta_{0}$ is 
\be\label{delta_squared(t)}
\frac{{\delta^{2}(\tau)}}{\nu\, t_{c}}=\frac{1-\tau }{\mu-1}+\left(\frac{{\delta^{2}_{0}}}{\nu \, t_{c}}-\frac{1}{\mu-1}\right)(1-\tau)\,^{\mu}\,,\quad\tn{if}\,\,\,\,\,\mu\ne 1,
\ee
\be\label{delta_squared(t)_1}
\frac{{\delta^{2}(\tau)}}{\nu\, t_{c}}=\frac{\delta_{0}^{2}}{\nu \,t_{c}}(1-\tau)-(1-\tau)\log{(1-\tau)} \,,\quad\tn{if}\,\,\mu = 1.
\ee
The solution (\ref{delta_squared(t)}) has a different character according as $\mu <$ or $>1$.\,
If  $0\!<\!\mu\!<\!1$, then the second term proportional to $(1-\tau)\,^{\mu}$ dominates as $\tau\rightarrow 1$, and the solution has infinite negative gradient in this limit.  If $\mu>1$, then the first term, proportional to $1-\tau$, dominates in the limit, and the solution has limiting gradient 
$-(\mu-1)^{-1}$ for any value of  $\delta_{0}^{2}/\nu \,t_{c}$. If $\mu>1$ and $\delta_{0}^{2}/\nu t_{c}=  1/(\mu-1)$, then we have the special solution considered above with the Leray scaling
\be\label{Leray_case}
\delta^{2}(\tau)=\delta_{0}^{2}(1-\tau).
\ee
\begin{figure}
\begin{center}
\begin{minipage}{0.99\textwidth}
\hspace*{20pt}
\includegraphics[width=0.45\textwidth, trim=0mm 0mm 0mm 0mm]{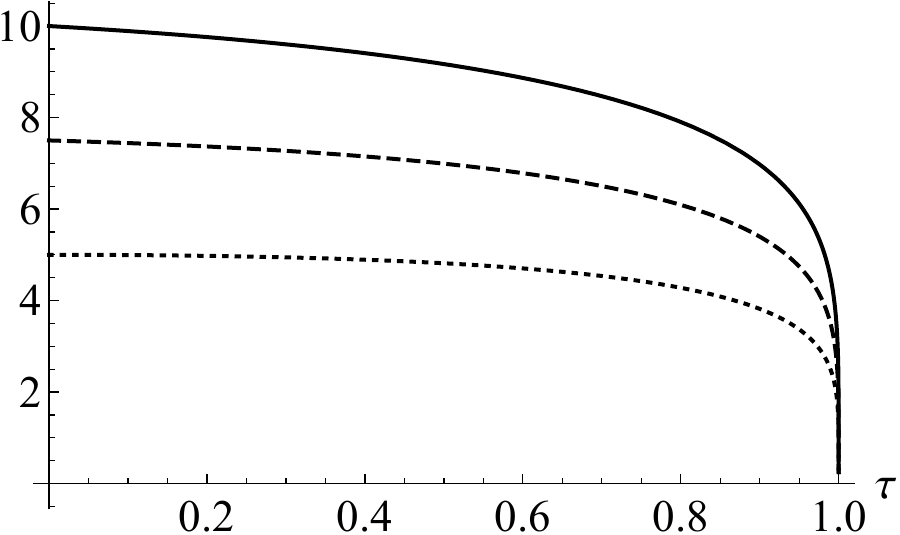}
\hspace*{10pt}
\includegraphics[width=0.45\textwidth,  trim=0mm 0mm 0mm 0mm]{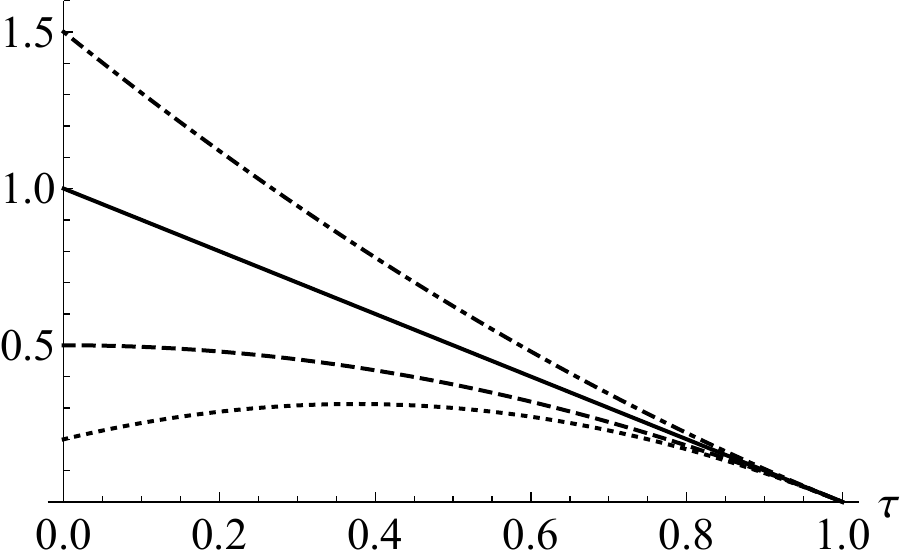}
\end{minipage}
\end{center}
\vskip 2mm
\hskip 37mm (a) \hskip 61mm (b)  
\vskip 2mm
\caption{The solution $\delta^{2}/\nu\,t_{c}$ of (\ref{delta_squared(t)}) as a function of $\tau=t/t_{c}$: (a) $\mu=0.2$; $\delta_{0}^2/\nu\,t_{c}=5$ (dotted), $7.5$ (dashed) and $10$ (solid); (b) $\mu=2$; $\delta_{0}^2/\nu\,t_{c}=0.2$ (dotted), $0.5$ (dashed),  and $1.5$ (dash-dotted); the case of Leray scaling, $\delta_{0}^2/\nu\,t_{c}=1$ is shown by the straight solid line.}
\label{Fig_delta_squared_2_9}
\end{figure}
The situation is illustrated in Figure \ref{Fig_delta_squared_2_9}.  In (a), $\mu<1$, and the solution `plunges to zero' at $\tau=1$ with gradient $-\infty$; (the limiting solution (\ref{delta_squared(t)_1}) also has this property).  In (b), $\mu >1$, and the solution `glides to zero' with finite gradient $-(\mu-1)^{-1}$ at $\tau=1$; here, the  solution with the Leray scaling (\ref{Leray_case}) is the straight line (solid). In all cases, the solution exhibits a finite-time singularity ($\delta=0$, so $\omega=\infty$) at $\tau=1$.

\subsection{Preferential decay of twist}\label{Preferential_twist_decay}
Suppose we add a velocity $(0,0,w(r,0))$ parallel to the vorticity (\ref{initial_condn}) in the initial condition for the stretched Burgers vortex, and with the same Gaussian structure
\begin{equation}\label{initial_condn_axial_velocity}
w(r,0)=w_0\exp\left({-r^2/\delta_{0}^2}\right)\,.
\end{equation}
The associated flux in the $z$-direction is
\begin{equation}\label{axial_flux}
Q_{0}=\int_{0}^\infty w(r,0) \,2\pi\, r\tn\,{d}r = \pi\,w_{0}\,\delta_{0}^2\,.
\end{equation}
The vortex lines are now helical within the vortex, and the `twist helicity' per unit axial length is
\begin{equation}\label{axial_flux}
{\mathcal H}_{0}=\int_{0}^\infty w(r,0)\,\omega(r,0) \,2\pi r\tn{d}r= \frac{\Gamma Q_{0}}{2\pi \delta_{0}^2}.
\end{equation}
Under the time-dependent straining (\ref{gamma(t)}), and with no additional applied pressure gradient, the $z$-component of the Navier-Stokes equation takes the form
\begin{equation}\label{Burgers_vorticity}
\frac{\partial w}{\partial t}+2\lambda(t)\,w=\frac{\lambda(t)}{2r}\frac{\partial}{\partial r}(r^{2} w) + \frac{\nu}{r}\frac{\partial}{\partial r}\left(r\frac{\partial w}{\partial r}\right)\,.
\end{equation}
From this, with $\lambda(t)=\mu/(t_{c}-t)$, it is evident that $(t_c-t)^{2\mu}w(r,t)$ satisfies the same equation as $\omega(r,t)$, so that the solution for $w(r,t)$ analogous to (\ref{similarity_soln}) is 
\begin{equation}\label{similarity_soln_axial_vel}
w(r,t)=\frac{Q(t)(\mu-1)}{4\pi\nu(t_{c}-t)}\exp\left(\frac{-(\mu-1)r^2}{4\nu(t_{c}-t)}\right)\,,
\end{equation}
\vskip -3mm
\noindent where
\vskip -3mm
\begin{equation}\label{similarity_soln_axial_vel}
Q(t)=(t_{c}-t)^{2\mu}\,Q_{0}.
\end{equation}
Thus $w(r,t)$ and the associated flux $Q(t)$ decay to zero as $t\!\rightarrow \!t_c$, while becoming ever more concentrated towards the axis $r=0$.
The helicity per unit axial length ${\mathcal H}(t)=\Gamma Q(t)/2\pi\delta^{2}(t)$  decays similarly:
\be
{\mathcal H}(t)=\frac{\Gamma Q_{0}\lambda_{s}t_{c}}{4\nu}(t_{c}-t)^{1+2\lambda_{s}t_{c}}\,.
\ee

In general, the helicity of a vortex that is deformed out of a plane through interaction with other vortices is given by
\be
{\mathcal H} =\Gamma^2(Wr +Tw),
\ee
where $Wr$ is the `writhe' of the vortex axis, and $Tw$ is the twist, which consists in part of the total torsion $\mathcal {T}$ of the axis, and in part of the internal twist of vortex lines within the core (\citealt{Moffatt1992a}). ${\mathcal H}$ is a pseudo-scalar invariant of the Euler equations  representing the net linkage of vortex lines of the flow (\citealt{Moffatt1969}), but it is not an invariant of the Navier-Stokes equations.  Viscosity can dissipate the internal core contribution to twist helicity by the  above mechanism, while torsion and writhe remain relatively unaffected.  The preferential decay of twist helicity observed by \cite{Scheeler2017} may be at least in part attributable to this mechanism.

We note that the same effect does not occur for a Lamb vortex, i.e. a line vortex decaying in a viscous fluid with no imposed strain.  When an axial flux $Q$ is included, both $\Gamma$ and 
$Q$ are conserved, while the radial length scale increases as $\sqrt{\nu t}$.  The kinetic energy per unit axial length for such a vortex diverges logarithmically as $r\rightarrow\infty$. However enstrophy $\Omega(t)$ and helicity ${\mathcal H}(t)$ per unit  length are  finite and decay like $t^{-1}$:
\be
\Omega(t)\sim\Gamma^2/8\nu t\,,\quad \mathcal {H}(t)\sim \Gamma Q/8\nu t\,, \quad \tn{as}\,\, t\rightarrow\infty\,.
\ee

\subsection{Vortex subjected to non-axisymmetric strain}\label{non-axisymmetric strain}
We shall be concerned with  straining of vortex tubes of Gaussian vorticity profile, as considered above, at high but finite Reynolds number  $\textnormal {R$_{\Gamma}$}=\Gamma/\nu$, under locally non-axisymmetric irrotational strain  
\begin{equation}\label{non-axisymm}
{\bf U}=(\lambda_{1}x, \lambda_{2}y,\lambda_{3}z)\,,
\end{equation}
where $\lambda_{1}<\lambda_{2}<\lambda_{3}$.  The incompressibility condition implies that $\lambda_{1}+\lambda_{2}+\lambda_{3}=0$, so that $\lambda_{1}<0$ and $\lambda_{3}>0$.
We suppose now that the vortex tube is aligned with the $y$-axis, i.e. $\bfomega=(0,\,\omega(x,z),\,0)$, and that $\lambda_{2}>0$ so that the tube is subject to positive stretching.  

The strain rates in the plane transverse to the vortex axis are  $\lambda_{1}\!<\!0$ and $\lambda_{3}\!>\!0$, so that there is a tendency to flatten the cross-section to oval shape, as found by \cite{Brenner2016}.  However, as  recognised by \cite{Lin1984} and 
\cite{Neu1984} for the particular case of plane strain, and as analysed more generally by \cite{Moffatt1994a}, the cross-section of the vortex remains circular at leading order when R$_{\Gamma}\gg1$, essentially because the spin is then so rapid that the vortex cross-section experiences just the average strain rate 
$\thalf(\lambda_{1}+\lambda_{3})=-\thalf \lambda_{2}<0$, thus maintaining its compact structure  as for a Burgers vortex subjected to axisymmetric strain.  The stretched vortex can therefore survive for a long time even when $\lambda_{3}>\lambda_{2}>0$, provided R$_{\Gamma}$ is large enough.
\begin{figure}
\begin{center}
\begin{minipage}{0.99\textwidth}
\includegraphics[width=0.30\textwidth, trim=0mm 0mm 0mm 0mm]{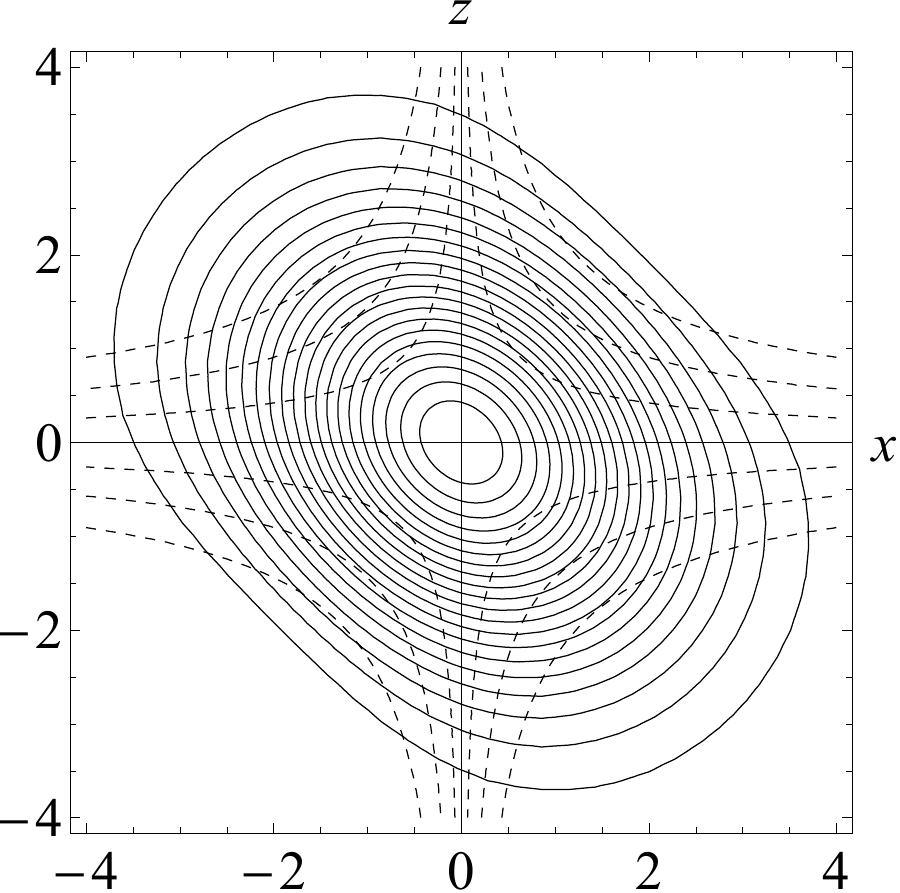}
\hspace*{10pt}
\includegraphics[width=0.30\textwidth,  trim=0mm 0mm 0mm 0mm]{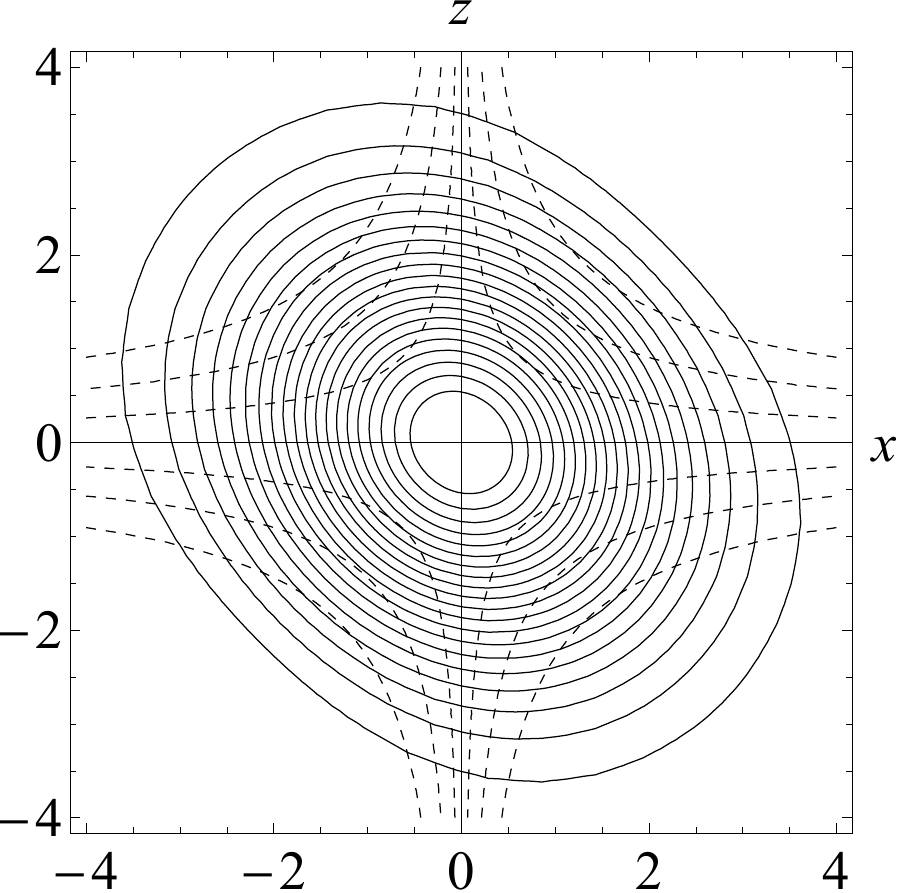}
\hspace*{10pt}
\includegraphics[width=0.30\textwidth,  trim=0mm 0mm 0mm 0mm]{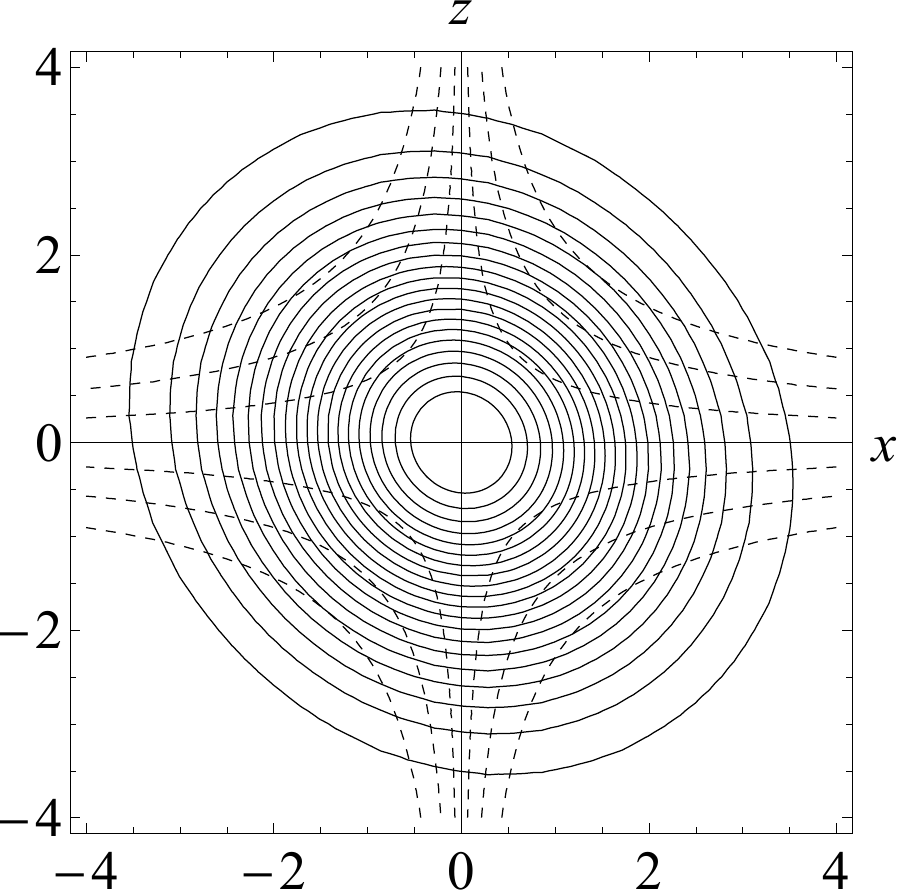}
\\[-1.8pt]
\end{minipage}
\end{center}
\vskip 2mm
\hskip 18mm (a) \hskip 38mm (b) \hskip 37mm (c) 
\vskip 2mm
\caption{Isovorticity contours in the core of a vortex subjected to the non-axisymmetric irrotational strain field (\ref{non-axisymm}); $\epsilon_{1}$ defined by (\ref{strain-field_parameter}): (a) $\epsilon_{1}=0.008$; (b) $0.005$; (c) $0.0025$: the flattening decreases as 
$\epsilon_{1}\rightarrow 0$. The streamlines of the strain field in the plane $y=$ const. are shown by the dashed curves, for the choice $\{\lambda_1:\lambda_2:\lambda_3\}=\{-3:1:2\}$. [After \citealt{Moffatt1994a}.]}
\label{Fig_isovorticity}
\end{figure}
The asymptotic analysis  was found to require that \footnote{In \cite{Moffatt1994a}, the vortex was in the $z$-direction, and subject to a uniform strain $(\alpha x,\beta y, \gamma z)$; the correspondence with our present notation is therefore $(\alpha,\beta,\gamma)\rightarrow (\lambda_1, \lambda_3,\lambda_2)$}
\begin{equation}\label{strain-field_parameter}
\epsilon_{1} \equiv c \, \epsilon\ll 1,\quad \tn{where}\,\,\epsilon= \nu/\Gamma=1/{\textnormal{R}_{\Gamma}}\quad \tn{and}\,\, c = 1+2\lambda_3/\lambda_2\,.
\end{equation}
Figure \ref{Fig_isovorticity} shows isovorticity contours $\omega(x,z)=$ const.~in the vortex core for three values of $\epsilon_{1}$, showing how the flattening effect decreases as $\epsilon_{1}\rightarrow 0$.

At the leading order O($\epsilon_{1}^0$),  it was found that the axial component of vorticity in the vortex core has just the usual Gaussian profile of the traditional Burgers vortex,
\begin{equation}\label{strain-field_parameter2}
\omega(r) = \frac{\lambda_{2}\Gamma}{4\pi\nu}\exp\left[-\frac{\lambda_{2} r^2}{4\nu}\right]\,,
\end{equation}
where $\lambda_2\,(\,>\!0)$ is the axial strain rate, and $r=(x^2+z^2)^{1/2}$ is the radial coordinate in the plane of cross-section of the vortex.  This result, obtained through the need to satisfy a solvability condition at O$(\epsilon_{1})$,  was found to be independent of the value of $c $, and is still valid, with one qualification, when the rate-of-strain tensor has two positive eigenvalues ($c >1$); this qualification (as discussed in \S 5 of \citealt{Moffatt1994a}) is that the solution fails at a large distance $r={\textnormal O}(\epsilon_{1}^{-1/2}\delta)$ from the vortex axis, where however the vorticity is already transcendentally small, of order 
$\exp{(-1/\epsilon_{1})}$.  In this `remote' region, since $\lambda_3>0$, vorticity is stripped away in the $z$-direction, but the resulting decrease of circulation $\Gamma$ of the vortex is extremely slow.  

A similar analysis applies when the axial strain rate $\lambda_2$ is time-dependent in the manner considered in \S \ref{Sec_time-dependent strain} above, as proved in \S 3 of \cite{Moffatt2000b} again by an asymptotic treatment  in the limit $\Gamma/\nu\rightarrow\infty$.

 \section{Initial condition for two interacting vortices}\label{Sec_Initial-condition}
\subsection{Initial configuration}\label{Sec_Initial_configuration}
\begin{figure}
\begin{center}
\includegraphics[width=6cm]{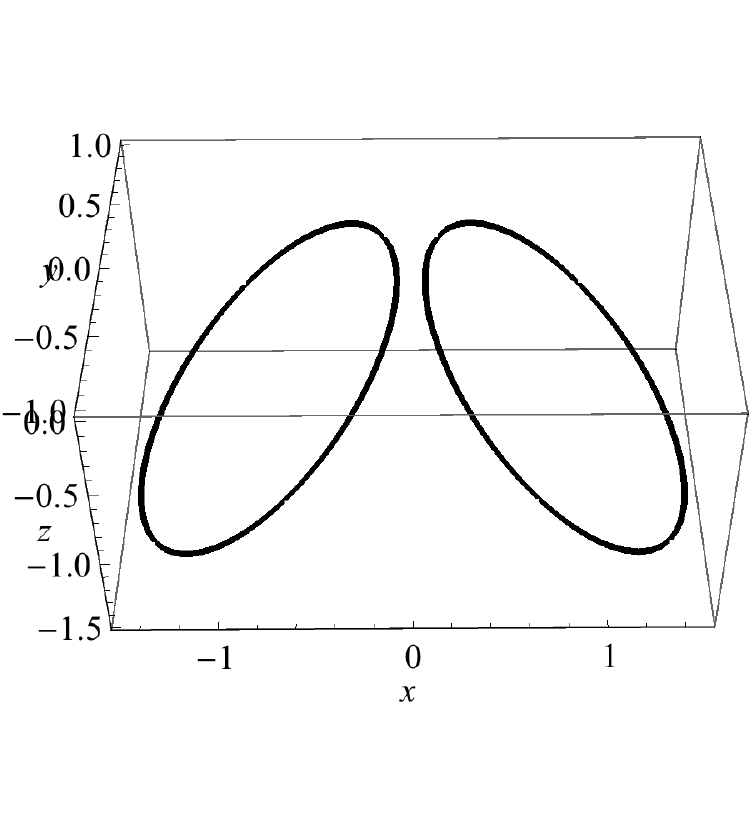}
\end{center}
\vskip  -10mm
\caption{Two circular vortices  of equal radii $R$ and circulations $\mp\Gamma$ are located on inclined planes $x=\pm z\tan\alpha$; the centres are at $(\mp (R\!+\!d)\sin\alpha,\,0,\, -(R\!+\!d)\cos\alpha)$  and the `tipping points', i.e.~the points of nearest approach, are at $(\mp d\sin\alpha,\,0,\,-d\cos\alpha)$; the separation of the tipping points is $2s$ where $s=d\sin\alpha$; the vortices have equal core cross-sections of scale $\delta$, and it is supposed that $\delta \ll s\ll R$; the configuration is symmetric about the planes $x=0$ and  $y=0$, and, neglecting any possible instabilities,  these symmetries persist for $t>0$ under Navier-Stokes evolution; the circulations are such that the vortices propagate towards each other, while being distorted upwards near the tipping points.}
\label{Fig_sketch}
\end{figure}

Consider now the configuration sketched in figure \ref{Fig_sketch}. Suppose that at time $t=0$, with $\Gamma>0$, two vortex tubes ${\mathcal V}_{1}$ and ${\mathcal V}_{2}$ of small but finite Gaussian core cross-sections and  of circulations 
$\Gamma_1=-\Gamma$ and  $\Gamma_2=+\Gamma$  
 are centred on circles $C_1$ and $C_2$ of radius $R$ (curvature $\kappa_{0}\equiv R^{-1}$), and that they are imbedded in an incompressible fluid of kinematic viscosity $\nu$ and located on inclined surfaces $x=\pm z \tan\alpha$.With this notation, we distinguish between a vortex tube (or briefly, a vortex) $\mathcal V$ and the curve $C$ on which its axis is located. We suppose that the fluid is of infinite extent in all directions and that the Reynolds number $\textnormal {R$_{\Gamma}$}=\Gamma/\nu\equiv \epsilon^{-1}$ is large but finite:
\begin{equation}
1\ll \textnormal {R$_{\Gamma}$} <\infty, \quad \tn{or equivalently}\quad 0<\epsilon\ll 1.
\end{equation}
We can be more specific later (see \S\ref{Navier-Stokes_sing} and figure \ref{Fig_max_vort}) about just how large R$_{\Gamma}$ must be if a singularity or near singularity is to form.  We note that a similar starting point was adopted by \cite{Boue2013}, but with hyperbolic, rather than circular, vortices.  The circular configuration as considered here was studied by DNS at R$_{\Gamma}=577$ by \cite{Kida1991} who focussed on the detailed character of the reconnection process. 

Let  ${\bf e}_1= (-\sin\alpha,\,0,\,-\cos\alpha)$ and ${\bf e}_2= (\sin\alpha,\,0,\,-\cos\alpha)$ be unit vectors down the lines of steepest slope on the two planes from the origin O$(0,\,0,\,0)$.  We suppose that the centres of the circles $C_1$ and $C_2$ are at $(R+d)\,{\bf e}_1$ and $(R+d)\,{\bf e}_2$, and we shall require that $d \ll R$, so that the interaction between the vortices is mainly localised in an O$(d)$ neighbourhood of the origin.  We suppose further that each vortex has Gaussian core cross-section ($\sim \exp[-r^{2}/4\delta^{2}]$) of radial scale $\delta > 0$, and we require that $\delta\ll d$, so that, at least initially, the effect of  each vortex on the other can be represented in terms of the Biot-Savart law. The circulations $\pm \Gamma$ are oriented so that the vortices tend to propagate towards the plane of symmetry and so towards each other.  The initial velocity field corresponding to the assumed vorticity distribution is clearly a smooth $C^{\infty}$ function of position $\bf x$.

In summary, we assume
\begin{equation}\label{basic_inequalities}
0 < \delta \ll d \ll R.
\end{equation}
Note that, near their points of nearest approach $(\pm d\sin\alpha,0,-d\cos\alpha)$,  the vortices are approximately anti-parallel; their minimum separation is $2s$, where 
\begin{equation}\label{separation}
s=d\sin\alpha.
\end{equation}
We shall in places use the notation $\xi\equiv s/R=\kappa_{0} s$.

\subsection{Induced velocity field}
The vortices induce a velocity field ${{\bf u}}({\bf x},{t})$ that is finite and everywhere analytic, at least for small times (\citealt{Foias1989}; for a recent survey of formal mathematical approaches to the Navier-Stokes equations, see \citealt{Foias2017}).
The evolution of  ${\bf u}({\bf x},t)$  for $t>0$ is governed by the Navier-Stokes equation
\be
\frac{\partial {\bf u}}{\partial t}+{\bf u}\cdot\nabla{\bf u}=-\frac{1}{\rho_{0}}\nabla p +\nu\nabla^2 {\bf u}\,,\quad \nabla\cdot{\bf u}=0\,,
\ee
where $p({\bf x},\,t)$ is the pressure field, and $\rho_{0}$ is the density, assumed uniform.
There are obvious mirror symmetries in the planes $x=0$ and $y=0$; writing ${\bf u}=(u_x,\,u_y,\,u_z)$, these are 
\be\label{symmetry_x}
u_{x}(-x,y,z)=-u_{x}(x,y,z), \,\,u_{y}(-x,y,z)=u_{y}(x,y,z), \,\,u_{z}(-x,y,z)=u_{z}(x,y,z), 
\ee
\vskip -2mm
and 
\vskip -4mm
\be \label{symmetry_y}
u_{x}(x,-y,z)=u_{x}(x,y,z), \,\,u_{y}(x,-y,z)=-u_{y}(x,y,z), \,\,u_{z}(x,-y,z)=u_{z}(x,y,z). 
\ee
Ignoring any symmetry-breaking instability to which the vortices may be subject, these symmetries persist for all $t>0$ for so long as ${\bf u}({\bf x}, t)$ remains analytic.
 
The momentum associated with each vortex is of order $\rho_{0}\, \pi R^{2}\, \Gamma$ in the direction normal to its plane towards the $z$-axis.  The total momentum of the two-vortex system is therefore
\begin{equation}
{\bf P} \sim \left(0,\,0,\, -2\pi\,\rho_{0}\,\Gamma \,R^2 \,\sin\alpha \right),
\end{equation}
and this is constant for $t>0$.  The main contribution to this downwards momentum comes from the parts of the vortices at a distance much greater than $d$ from the $z$-axis.  The angular momentum is zero by symmetry, and so is the helicity as there is no linkage of vortex lines.

The initial kinetic energy of the system is  
\begin{equation}
K_{0} =\thalf\int \rho_{0}\,{\bf u}({\bf x},0)^2\,\textnormal{d}V\, \sim \thalf\rho_{0}\, \Gamma^2\, R \log (R/\delta)<\infty\,.
\end{equation}
 Energy is dissipated by viscosity, so that $ \tn{d}K(t)/\tn{d}t <0$ for all $t>0$. 
\subsection{Self-induced velocity of ${\mathcal V}_{1}$}
The  self-induced velocity ${\bf v}_1$ of the vortex ${\mathcal V}_{1}$ (with circulation $-\Gamma$ and  curvature $\kappa_{0}\equiv R^{-1}$)  is 
\begin{equation}\label{self-induced_velocity}
{\bf v}_1=(-\kappa_{0}\,\Gamma/{4\pi } ) \left[\log (1/\kappa_{0}\,\delta)+\beta\right]\,{\bf b}\end{equation}
where $\bf b$ is the unit binormal vector, and, for a Gaussian core, 
$\beta= \log{4}-0.558=0.828$  (\citealt{Saffman1970}, \citealt{Sullivan2008}\footnote{\cite{Sullivan2008} (p.324) quote Saffman's result (with $\sqrt{\nu t}=\delta$ in our notation) in the form 
$V=(\Gamma/{4\pi R } ) \left[\log \left(8R/\!\sqrt{4\nu t}\right)-0.558\right]\,{\bf b}$, }. [We shall also need the value $\beta= \log{4}-0.250=1.136$ for a uniform vorticity core.]  The additional induced velocity $\bf v_2$ due to  $C_2$ in the `tip region' of $C_1$ where $|{\bf x}|=\tn{O}(s)$ is of order $\Gamma/s$.  We  seek first to find the velocity field $\bf v_2$  in the immediate neighbourhood of ${\mathcal V}_{1}$ in order to determine how this vortex moves and how its core is deformed.  Of course the effect is mutual, and both vortices must deform in symmetric manner.  

Although we consider here only the initial configuration, we shall see that the analysis remains relevant as the vortices are progressively deformed for $t > 0$, because the interaction is predominantly local, involving only the instantaneous curvature $\kappa(t)$ at the points of nearest approach,  T$_{1}$ and T$_{2}$, and the separation 
$2s(t)$ of these points, where $s(t)=d(t)\sin\alpha(t)$; the analysis will remain valid for so long as 
\begin{equation}\label{self-induced_velocity1}
 \delta(t) \ll s(t) \ll 1/\kappa(t).
\end{equation}
We describe T$_{1}$ and T$_{2}$ as the `tipping points', because they obviously play a crucial role in the approach to a singularity. 

The parametric equations of the circles $C_1$  and $C_2$ at time $t=0$ are
\begin{eqnarray}
{\bf x}_1(\theta_1)&=&\left[-(R+R \cos\theta_1+d)\sin\alpha, -R \sin\theta_1,\, -(R+R\cos\theta_1+d)\cos\alpha\right],\label{parametricC_1}\\
{\bf x}_2(\theta_2)&=&\left[+(R+R\cos\theta_2+d)\sin\alpha, -R \sin\theta_2, \,-(R+R\cos\theta_2+d)\cos\alpha\right],\label{parametricC_2}
\end{eqnarray}
where the parameters $\theta_1$ and $\theta_2$ run from $-\pi$ to $\pi$. The tipping points 
T$_{1}$ and T$_{2}$ at time $t=0$ are then at $\theta_{1}=\pi$ and $\theta_{2}=\pi$, \,i.e.~ at
\begin{equation}
(\mp d \sin\alpha,0, -d\cos\alpha)=(\tilde {x},0,\tilde {z}),\,\,\textnormal{say,\quad on}\,\, C_{1},C_{2}\,\, \textnormal{respectively}.
\label{tipping_points}
\end{equation}
We shall use the tilde $\,{\tilde .}$\,  in this way to denote variables that are evaluated at the tipping points.

\section{Analysis of the Biot-Savart integral}\label{Sec_Biot-Savart}
\begin{figure}
\begin{center}
\includegraphics[width=10cm]{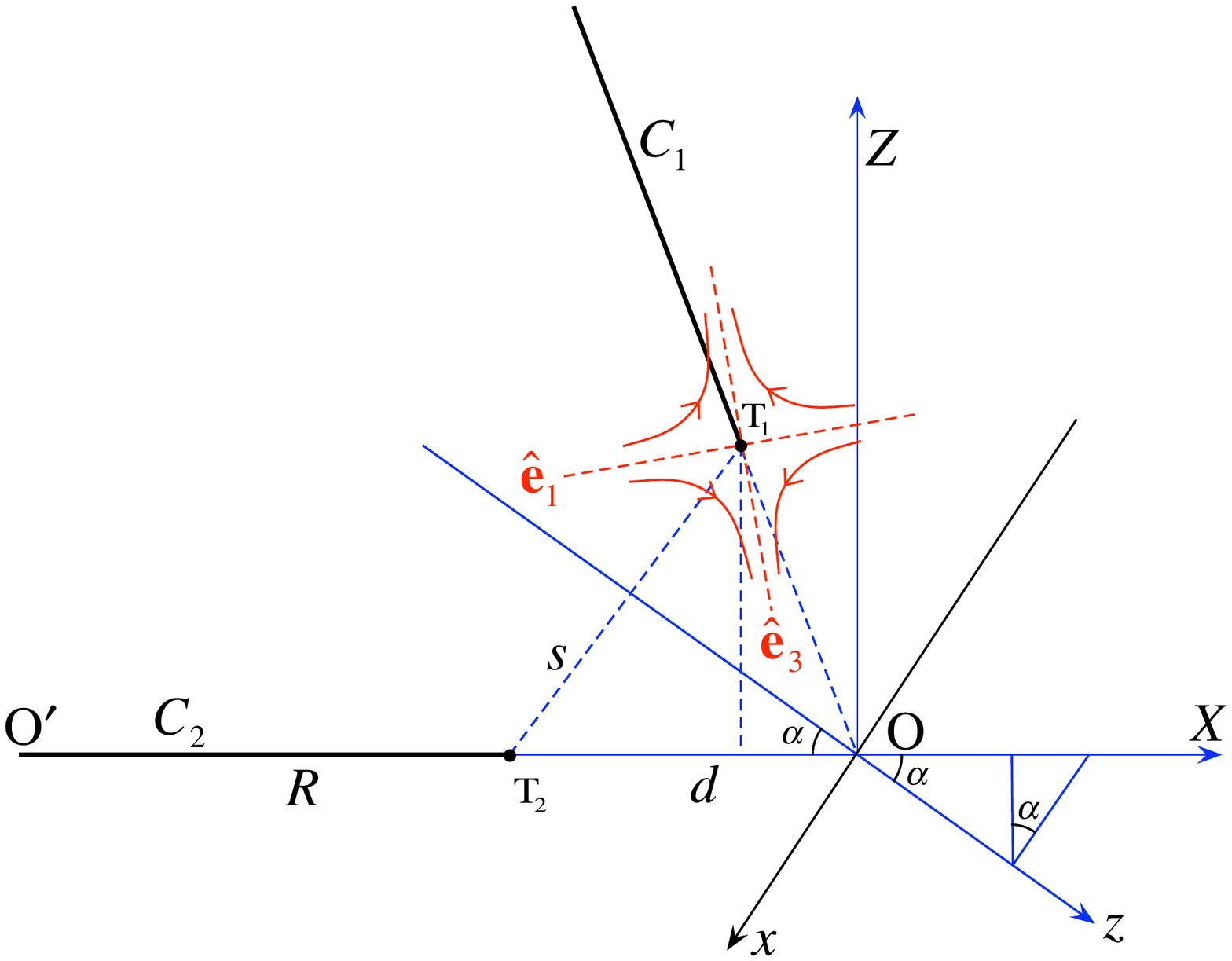}
\end{center}
\vskip -60mm
\caption{Sketch of projection on the $\{x,z\}$-plane (not to scale). The frame of reference O$XZ$ is obtained from O$xz$ by anti-clockwise rotation through an angle $\pi/2 +\alpha$ ; O$'$ is the displaced origin. The projections of $C_1$ and $C_2$ are shown in black; T$_{1}$ and T$_{2}$ are the tipping points, with separation $2 s$.  The rate-of-strain field at T$_{1}$ is as indicated; ${\hat{\bf e}}_{1}$ and ${\hat{\bf e}}_{3}$ are the eigenvectors corresponding to eigenvalues $\lambda_{1}<0$ and $\lambda_{3}>0$; the strain field tends to orient the $xz$-plane projection of the neighbourhood of the tip of $C_{1}$ towards the direction of ${\hat{\bf e}}_{3}$. Similar distortion occurs symmetrically at T$_{2}$.}
\label{Fig_sketch_2}
\end{figure}

The analysis that follows is valid for arbitrary $d>0$; it is not restricted to small $d$, although we shall later find good reasons for limiting attention to this situation. It is convenient to adopt a frame of reference O$'XYZ$ rotated anti-clockwise through  $\pi/2+\alpha$ from O$xyz$ and with displaced origin O$'$ so that O is at $X=R+d,\,Y=0,\,Z=0$ (see figure \ref{Fig_sketch_2}); thus
\begin{equation}\label{Biot_simple0}
X(x,\,z) = R+d-x\sin \alpha+ z\cos\alpha, \quad Y=y,\quad Z(x,\,z)=-x\cos\alpha -z\sin \alpha\,.
\end{equation}
In the frame O$'XYZ$, the vortex ${\mathcal V}_{2}$  lies in the  plane $Z=0$ and has parametric representation
\begin{equation}\label{Biot_simple01}
{\bf X}_{2}=(R\cos\theta,\, R\sin\theta,\,0),\quad (-\pi<\theta\le\pi),
\end{equation}
and the tipping point T$_{1}$ of $C_{1}$ is at 
\begin{equation}\label{Tipping_point_1}
{\bf X}={\tilde{\bf X}}= (R+2s \sin\alpha,\, 0,\, 2s \cos\alpha). 
\end{equation}
We  aim first to determine the induced velocity and rate-of-strain tensor  at this point.

The velocity field induced by ${\mathcal V}_{2}$ at any point ${\bf X}=(X,Y,Z)$ satisfying $|{\bf X}-{\bf X}_{2}|\gg\delta$ is given by the Biot-Savart integral
\begin{equation}\label{Biot_simple0}
{\bf v}_{2}({\bf X})=\frac{\Gamma}{4 \pi}\oint_{C_{2}}\frac{\textnormal d{\bf X}_{2}\wedge({\bf X}-{\bf X}_{2})}{|{\bf X}-{\bf X}_{2}|^{3}}\,,
\end{equation}
where, from (\ref{Biot_simple01}),
\begin{equation}\label{Biot_simple1}
{\textnormal d}{\bf X}_{2}=(-R\sin\theta,\,\,R\cos\theta,\,\,0)\,{\textnormal d}\theta \quad\textnormal {and}\quad 
{\bf X}-{\bf X}_{2} =(X-R\cos\theta, \,\,Y-R\sin\theta, \,\,Z)\,.
\end{equation}
Hence
\begin{equation}\label{Biot_simple2}
{\textnormal d}{\bf X}_{2}\wedge({\bf X}-{\bf X}_{2})=(Z\cos\theta, \,\,Z\sin\theta,\,\,R-X\cos\theta-Y\sin\theta) \,R\tn\,{d}\theta\,,
\end{equation}
and
\begin{equation}\label{Biot_simple3}
|{\bf X}-{\bf X_{2}}|^2 = (R\cos\theta-X)^2 + (R\sin\theta-Y)^2 + Z^2 =\left(R^{2}+r^2\right)[1-m \cos(\theta-\gamma)]\,,
\end{equation}
where 
\begin{equation}\label{Biot_simple4}
r^2=X^2 +Y^2 +Z^2,\quad m=\frac{2R(X^2 +Y^2)^{1/2}}{R^2 +r^2},\quad \tan\gamma=Y/X\,.
\end{equation}
Note that $0<m<1$ for points ${\bf X}\notin C_{2}$. 

With  $\phi=\theta-\gamma$, the Biot-Savart integral (\ref{Biot_simple0}) can be reduced to the form
\begin{equation}\label{Biot_simple5}
{\bf v}_{2}({\bf X})=\frac{\Gamma R}{2 \pi(R^2+r^2)^{3/2}}\left[I_1(m)\,Z \cos\gamma,\,I_1(m)\,Z\sin\gamma, \, R\,I_0(m)-I_1(m)\,(X\cos\gamma +Y\sin\gamma) \right],
\end{equation}
where
\begin{equation}\label{Biot_simple6}
I_{0}(m)=\int_{0}^{\pi}\frac{\textnormal{d}\phi}{(1-m\cos\phi)^{3/2}},\quad
I_{1}(m)=\int_{0}^{\pi}\frac{\cos\phi \,\textnormal{d}\phi}{(1-m\cos\phi)^{3/2}}\,. 
\end{equation}
These integrals may be expressed in terms of complete elliptic integrals $K(k), \,E(k)$ of the first and second kinds, defined  in the notation of \cite{Abramowitz1964} by
\be
K(k)=\int_{0}^{\pi/2}(1-k \sin^2\psi)^{-1/2}\,\textnormal{d}\psi,\quad E(k)=\int_{0}^{\pi/2}(1-k \sin^2\psi)^{1/2}\,\textnormal{d}\psi.
 \ee
The results are 
\begin{equation}\label{Biot_simple7}
I_{0}(m)=\frac{-2E(k)}{(m-1)(1+m)^{1/2}},\quad I_{1}(m)= \frac{-2[E(k)+(m-1)K(k)]}{m(m-1) (1+m)^{1/2}},\quad \tn{where}\,\, k=\frac{2m}{m+1}\,.
\end{equation}
The induced velocity field (\ref{Biot_simple5}) is thus expressible in terms of known functions; and it may be verified that this field does indeed satisfy $\nabla\!\cdot \!{\bf v}_{2}=0$, $\nabla\!\wedge\!{\bf v}_{2}=0$, identities that must therefore be satisfied at each order under expansion in the neighbourhood of any point. 

Exact expressions for the components $(\varv_{2x},\varv_{2y},\varv_{2z})$, as determined  by  (\ref{Biot_simple4}) --  (\ref{Biot_simple7}), can now be found by converting back to the frame of reference O$xyz$ through the transformation
\begin{eqnarray}\label{transformed_velocity}
\varv_{2x}(x,y,z)&=&-\varv_{2X}(X(x, z),\,y,\,Z(x, z)) \sin\alpha -\varv_{2Z}(X(x, z),\,y,\,Z(x, z)) \cos\alpha\,,\nonumber\\
\varv_{2y}(x,y,z)&=&\varv_{2Y}(X(x, z),\,y,\,Z(x, z)) \,,\nonumber\\
\varv_{2z}(x,y,z)&=&+\varv_{2X}(X(x, z),\,y,\,Z(x, z)) \cos\alpha -\varv_{2Z}(X(x, z),\,y,\,Z(x, z)) \sin\alpha\,.
\end{eqnarray}

\subsection{Velocity on the $z$-axis}
\begin{figure}
\begin{center}
\includegraphics[width=8cm]{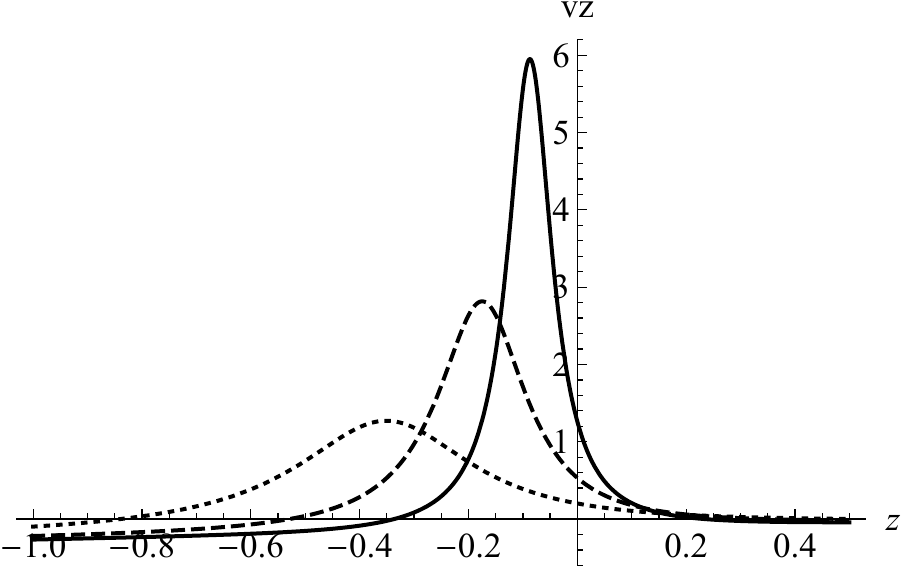}
\end{center}
\caption{Curves of  ${\varv}_{z}(0,\,0,\,z)$ for $\xi \,(=\kappa_{0} s)=0.2$ (dotted), $0.1$ (dashed), and $0.05$ (solid); dimensionless variables are used here and in other figures, based on length scale $R=\kappa_{0}^{-1}$ and velocity scale $\kappa_{0}\Gamma$; the value $\alpha=\pi/4$ is adopted throughout.}
\label{Fig_vzaxis}
\end{figure}
Based on this exact solution, we can first determine the velocity $\varv_{z}(0,0,z)$ on the $z$-axis induced by both ${\mathcal V}_{1}$ and ${\mathcal V}_{2}$; by symmetry this is simply $2\varv_{2z}(0,0,z)$. This is shown in figure \ref{Fig_vzaxis} for three values of the separation parameter $\xi=\kappa_{0} s=0.2,\,0.1$ and $0.05$.  [In the figures, we use dimensionless variables  based on length-scale $\kappa_{0}^{-1}$ and velocity-scale $\kappa_{0}\Gamma$.]  As the separation of the tipping points T$_{1}$ and T$_{2}$ decreases, it is evident that the upward velocity becomes more sharply peaked, its maximum moving towards $z=0$ and increasing without limit; actually 
${\varv}_{z}(0,0,0)\sim \kappa_{0}\Gamma /\pi\xi$ as $\xi\rightarrow 0$.

\subsection{Induced velocity at the tipping point of $C_{1}$}
\begin{figure}
\begin{center}
\includegraphics[width=6cm]{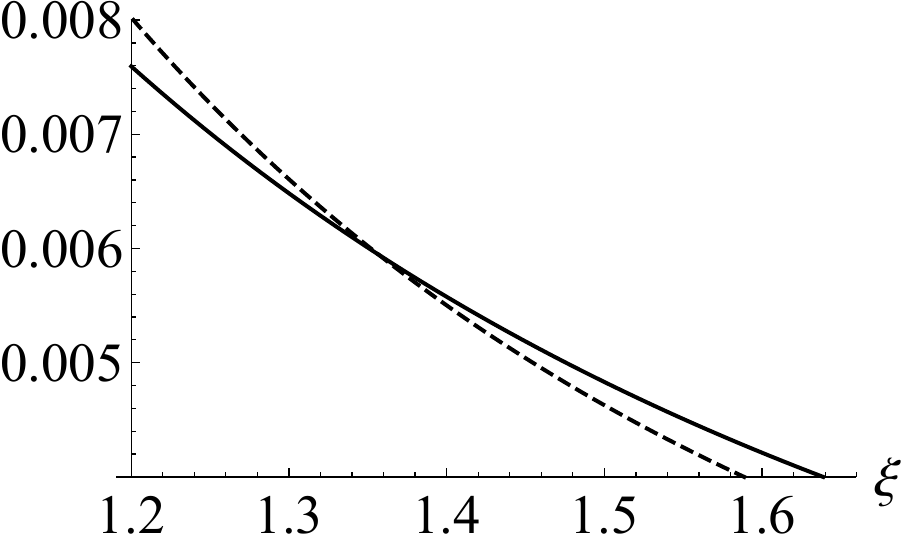}
\end{center}
\caption{Curves of $|{{\tilde{\varv}}}_{2x}(\xi,\alpha)|$ (solid) and ${{\tilde{\varv}}}_{2z}(\xi,\alpha)$ (dashed) ($\alpha=\pi/4$) showing the crossover at $\xi \approx1.357$.}
\label{Fig_crossover}
\end{figure}

Similarly, we may now obtain exact expressions for the  components of ${\bf v}_2$ 
at the tipping point T$_{1}$ of $C_{1}$, with coordinates
\be\label{Tipping_Point_X}
{\tilde{\bf X}}=\left(R(1+2\xi\sin\alpha),\,0,\,R\xi\cos\alpha\right). 
\ee
 Here,
${\tilde{\varv}}_{2y}=0$ by symmetry, and after some simplification we find
\begin{equation}\label{Tipping_point_vel2x}
{\tilde{\varv}}_{2x}(\xi,\alpha)\equiv {\varv}_{2x}({\tilde{\bf x}})= - \frac{\kappa_{0}\,\Gamma}{4\pi}\,\frac{\cos\alpha[K(k)-E(k)]}
{(1+2\xi \sin\alpha)(1+\xi^2+2\xi \sin\alpha)^{1/2}}\,,
\end{equation}
\vskip -2mm
\noindent and
\vskip -4mm
\begin{equation}\label{Tipping_point_vel2z}
\qquad\,\,\,\,{\tilde{\varv}}_{2z}(\xi,\alpha)\equiv {\varv}_{2z}({\tilde{\bf x}})= 
\frac{\kappa_{0}\,\Gamma}{4\pi}\,\frac{E(k)\,(1+2\xi^2+3\xi \sin\alpha)-\xi K(k)\,(2\xi+\sin\alpha)}{\xi(1+2\xi \sin\alpha)(1+\xi^2+2\xi \sin\alpha)^{1/2}}\,,
\end{equation}
\vskip -2mm
\noindent  where 
\vskip -4mm
\be \label{Tipping_point_k}
k=\frac{1+2\xi\sin\alpha}{1+\xi^2+2\xi\sin\alpha}\,.
\ee
\vskip 2mm
\noindent The asymptotic behaviour of these functions is, as $\xi\rightarrow 0$,
\begin{equation}\label{Tipping_point_s_small}
\frac{{\tilde{\varv}}_{2x}(\xi,\alpha)}{\kappa_{0}\Gamma}\sim \!- \frac{\cos\alpha}{4\pi}\left[\left(\log\frac{4}{\xi}\!-\!1\right) 
-\!\frac{\xi}{4\pi}\left(3\log\frac{4}{\xi}\!-\!4\right)\sin\alpha\right]\,,\quad
\frac{{\tilde{\varv}}_{2z}(\xi,\alpha)}{\kappa_{0}\Gamma}\sim\!\frac{1}{4\pi} 
\left[\frac{1}{\xi} \!-\!\left(\log{\frac{4}{\xi}}\right)\sin\alpha\right]\,,
\end{equation}
\noindent and, as $\xi\rightarrow \infty$,
\begin{equation}\label{Tipping_point_s_large}
\frac{{\tilde{\varv}}_{2x}(\xi,\alpha)}{\kappa_{0}\Gamma}\sim  -\frac{\cos\alpha}{16\, \xi^3}\,,\quad
\frac{{\tilde{\varv}}_{2z}(\xi,\alpha)}{\kappa_{0}\Gamma}\sim \frac{\sin\alpha}{32\, \xi^3}\,.
\end{equation}
In (\ref{Tipping_point_s_small}), we see the characteristic logarithmic singularity in ${{\tilde{\varv}}}_{2x}$, and the
stronger $\xi^{-1}$ singularity in ${{\tilde{\varv}}}_{2z}$; while in  (\ref{Tipping_point_s_large}), we see the  $\xi^{-3}$ behaviour in both components, characteristic of dipole behaviour at large $\xi$. [The far-field velocity induced by \emph{both} vortices has quadrupole behaviour O$(r^{-4})$ as $r\rightarrow\infty$.]
We note immediately that, since ${\tilde{\varv}}_{2z}>|{\tilde{\varv}}_{2x}|$ for small $\xi$ and 
${\tilde{\varv}}_{2z}<|{\tilde{\varv}}_{2x}|$ for large $\xi$, there must be a crossover point at some intermediate value of $\xi$.  Figure \ref{Fig_crossover} shows that this crossover in fact occurs at  $\xi \approx 1.357$, where  ${\tilde{\varv}}_{2z}=|{\tilde{\varv}}_{2x}|\approx 0.00594$.  
\begin{figure}
\begin{center}
\begin{minipage}{0.99\textwidth}
\hspace*{20pt}
\includegraphics[width=0.40\textwidth, trim=0mm 0mm 0mm 0mm]{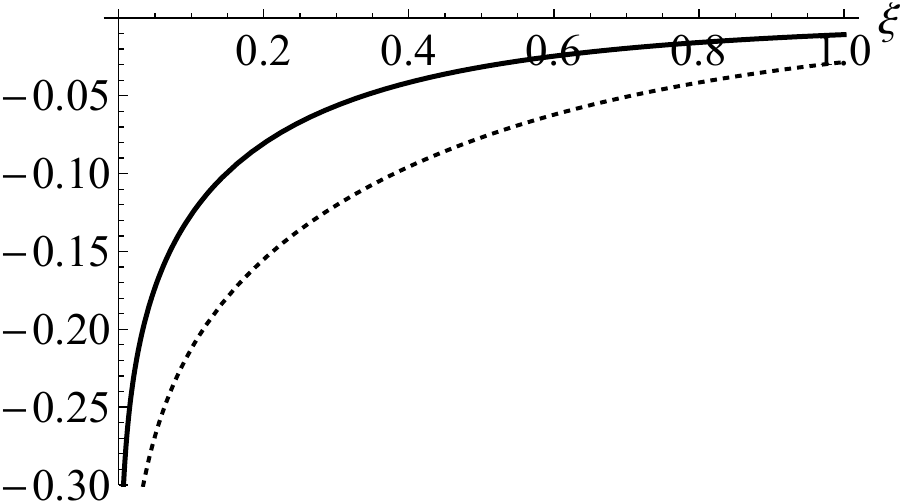}
\hspace*{30pt}
\includegraphics[width=0.40\textwidth,  trim=0mm 0mm 0mm 0mm]{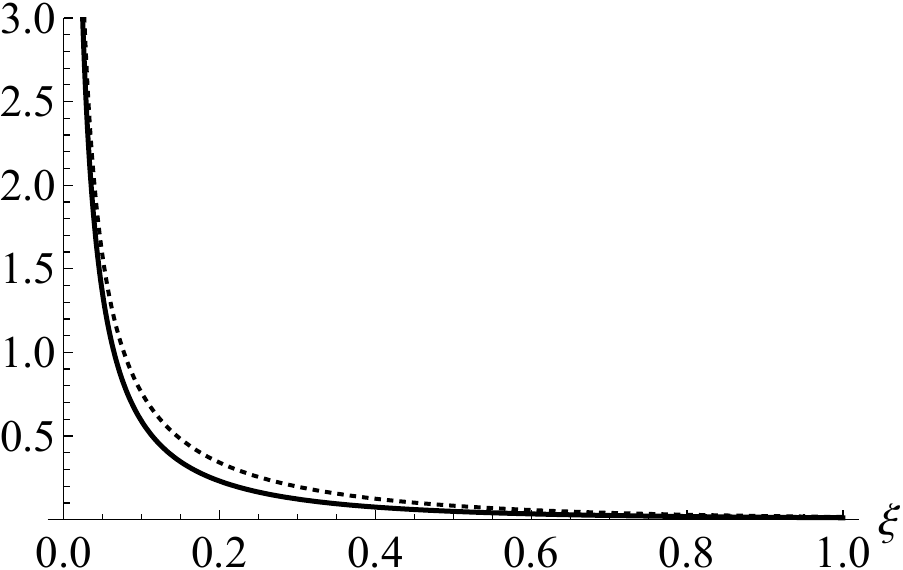}
\end{minipage}
\end{center}
\vskip 2mm
\hskip 30mm (a) \hskip 65mm (b) 
\vskip 2mm
\caption{Velocity components at T$_{1}$; (a) ${{\tilde{\varv}}}_{2x}$ and (b) ${{\tilde{\varv}}}_{2z}$,  given by (\ref{Tipping_point_vel2x}) and (\ref{Tipping_point_vel2z}), for $\alpha=\pi/4$ (solid) and $\alpha=0$ (dotted); curves for other values of $\alpha$ fit continuously between these bounds; ${{\tilde{\varv}}}_{2x}$ shows the logarithmic singularity and  ${{\tilde{\varv}}}_{2z}$  the $\xi^{-1}$ singularity  as $\xi\rightarrow 0$,  as indicated by (\ref{Tipping_point_s_small}).}
\label{Fig_tip_comp_anal}
\end{figure}

\begin{figure}
\begin{center}
\includegraphics[width=0.45\textwidth, trim=0mm 0mm 0mm 0mm]{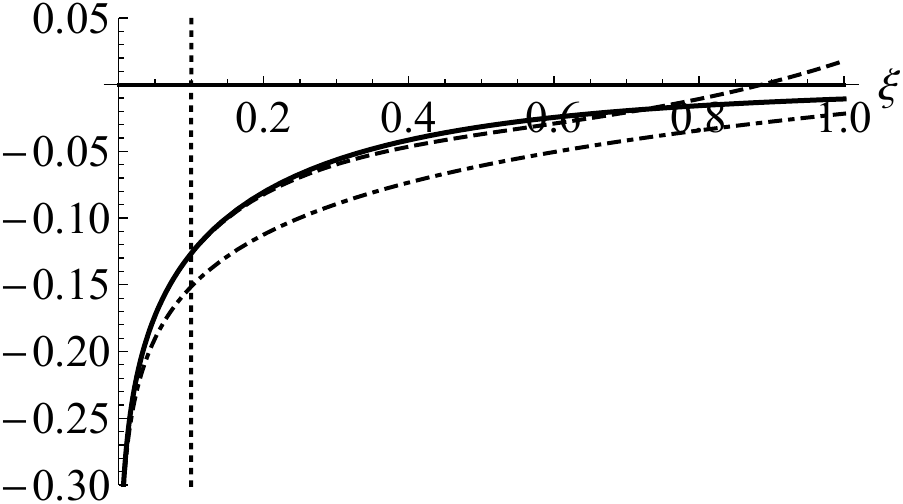}
\end{center}
\vskip 2mm
\vskip 2mm
\caption{The induced component ${{\tilde{\varv}}}_{2x}$  at the tipping point T$_{1}$ (solid)  as a function of $\xi$  (for $\alpha=\pi/4$); the two-term asymptotic result (\ref{Tipping_point_s_small})  is shown by the dashed curve which almost coincides with the exact solution for $\xi\lesssim 0.7$; the dash-dotted curve shows just the first-term of (\ref{Tipping_point_s_small}); the `small-$\xi$' range of interest lies to the left of the dotted line at  $\xi=0.1$.}
\label{Fig_vxtip_anal}
\end{figure}
\begin{figure}
\begin{center}
\begin{minipage}{0.99\textwidth}
\includegraphics[width=0.42\textwidth, trim=0mm 0mm 0mm 0mm]{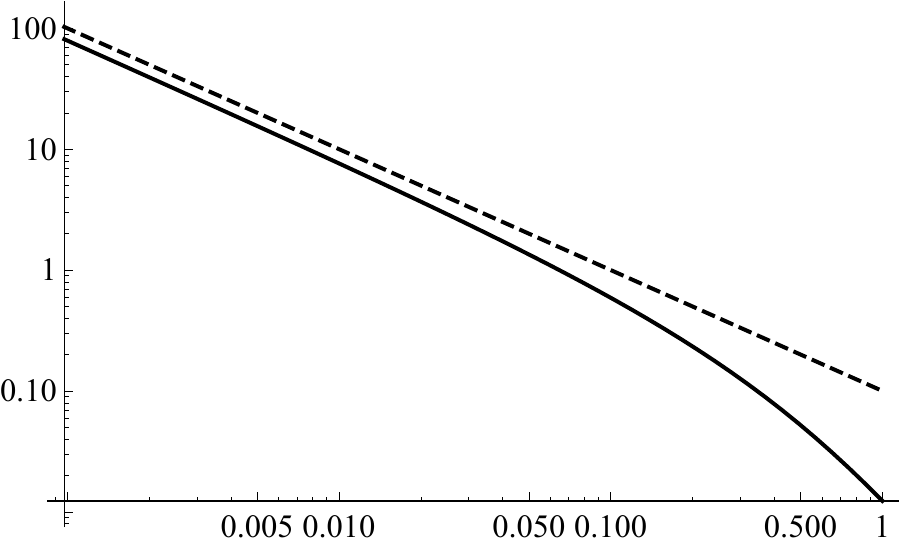}
\hspace*{30pt}
\includegraphics[width=0.37\textwidth,  trim=0mm 0mm 0mm 0mm]{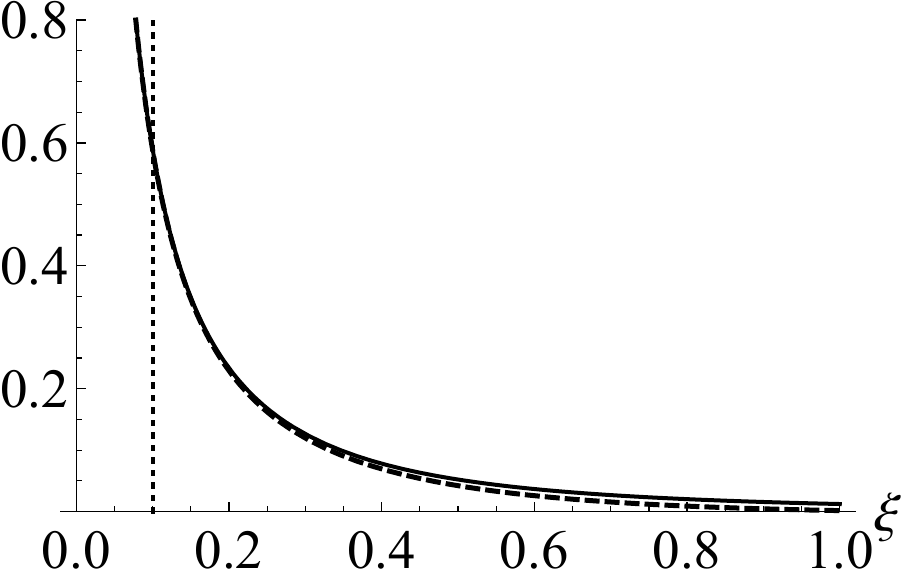}
\end{minipage}
\end{center}
\vskip 2mm
\hskip 30mm (a) \hskip 60mm (b) 
\vskip 2mm
\caption{(a) Log-log plot of the induced component ${{\tilde{\varv}}}_{2z}$ at the tipping point T$_{1}$  as a function of $\xi$ ($\alpha=\pi/4,\,0<\xi<0.6$); the slope of the dashed line is $-1$;   (b) ${{\tilde{\varv}}}_{2z}$ for the interval $0<\xi<0.12$ (black) with the asymptotic curve of  (\ref{Tipping_point_s_small}) superposed (dashed);  again the `small-$\xi$' range of interest lies to the left of the dotted line at  $\xi=0.1$.}
\label{Fig_vztip_anal}
\end{figure}
Figure \ref{Fig_tip_comp_anal} shows ${{\tilde{\varv}}}_{2x}$ and ${{\tilde{\varv}}}_{2z}$ at  T$_{1}$ as  given by (\ref{Tipping_point_vel2x}) and (\ref{Tipping_point_vel2z}) (note the different vertical scales in (a) and (b)), which make evident the logarithmic singularity in ${{\tilde{\varv}}}_{2x}$  and the much stronger $\xi^{-1}$ singularity in ${{\tilde{\varv}}}_{2z}$ as $\xi\rightarrow 0$. Figure \ref{Fig_vxtip_anal} shows ${{\tilde{\varv}}}_{2x}$ at  T$_{1}$ (solid) as a function of $\xi$ for $0<\xi<1$ and $\alpha=\pi/4$, together with the two-term asymptotic result (\ref{Tipping_point_s_small}) (dashed); this coincides well with the exact solution for $\xi\lesssim 0.7$. The first term of the asymptotic result (\ref{Tipping_point_s_small}) (dash-dotted) is reasonably accurate only for $\xi\lesssim 0.005$. The dotted vertical line indicates the upper boundary of the region of  interest. 

Turning now to the component ${{\tilde{\varv}}}_{2z}$, the situation is very different.  Figure \ref{Fig_vztip_anal}(a) shows a log-log plot (for $0<\xi<1$) of the exact analytic solution for ${{\tilde{\varv}}}_{2z}$ at  T$_{1}$, which confirms the $\xi^{-1}$ scaling of the asymptotic result  (\ref{Tipping_point_s_small}).   Figure \ref{Fig_vztip_anal}(b) shows a plot of ${{\tilde{\varv}}}_{2z}$ over the  range $0<\xi\lesssim 1$ (solid);  the two-term asymptotic result  (\ref{Tipping_point_s_small}) is superposed (dashed), showing remarkably good agreement over this whole range. Again, the dotted vertical line indicates the upper boundary of the region of  interest.

\subsection{Inclusion of ${\bf v}_1$}\label{Sec_Inclusion_v1}
As previously noted, the total initial velocity $\bf v$ on  $C_1$ is the sum of  the induced ingredient ${\bf v}_2$ as determined above and the self-induced velocity ${\bf v}_1$ given by (\ref{self-induced_velocity}), i.e. 
\begin{equation}\label{self-induced_velocity2}
{\bf v}={\bf v}_{1}+{\bf v}_{2}\quad\textnormal{with}\,\,{\bf v}_{1}= (\kappa_{0}\,\Gamma/4\pi)\left[\log \left(1/\kappa_{0}\delta\right)+\beta\right]\,(\cos\alpha,\,0,\,-\sin\alpha)\,.
\end{equation}
For the onset of `close interaction', two conditions must be satisfied. First, since 
${{\tilde{\varv}}}_{2x}<0$, the self-induced component ${{\tilde{\varv}}}_{1x}$, which is positive on $C_1$, must exceed $|{{\tilde{\varv}}}_{2x}|$, so that the tips are definitely moving towards the plane of symmetry. Since in any case we require that $\delta_{0}\!\ll\! s_0$ in order to justify use of the Biot-Savart law, this condition is automatically satisfied.

Second, although ${{\tilde{\varv}}}_{1z}<0$ on $C_1$, this should not exceed ${{\tilde{\varv}}}_{2z}$ in magnitude, because we want the tips to move upwards, i.e. we must also satisfy 
\begin{equation}\label{vz_cond}
{{\tilde{\varv}}}_{z}={{\tilde{\varv}}}_{1z}+{{\tilde{\varv}}}_{2z}=\frac{\kappa_{0}\Gamma}{4\pi}\left[\frac{1}{\kappa_{0}s}-\left[\log \left(4/\kappa_{0}^{2}s\delta\right)+\beta\right]\sin\alpha\right]>0.
\end{equation}
Again using (\ref{self-induced_velocity}) together with (\ref{Tipping_point_s_small}), this requires that
\begin{equation}\label{delta_second_constraint}
\kappa_{0}s\sin\alpha <\left[\log \left(4/\kappa_{0}^{2}s\delta\right)+\beta\right]^{-1}\,,
\end{equation}
i.e.~the tips must be close enough to ensure this upward movement. For two approaching vortices, the condition (\ref{delta_second_constraint}) in effect defines the onset of the close interaction process.
\begin{figure}
\begin{center}
\begin{minipage}{0.99\textwidth}
\includegraphics[width=0.45\textwidth, trim=0mm 0mm 0mm 0mm]{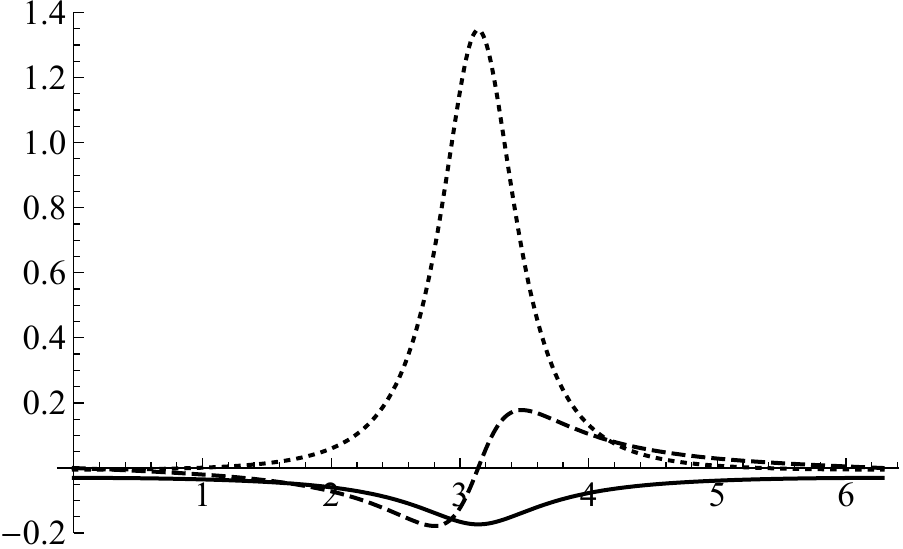}
\hspace*{20pt}
\includegraphics[width=0.45\textwidth,  trim=0mm 0mm 0mm 0mm]{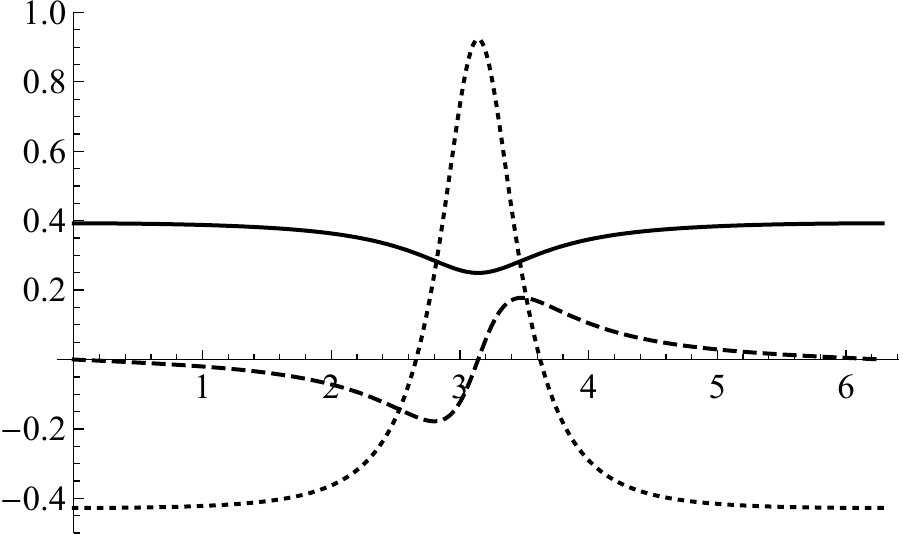}
\end{minipage}
\end{center}
\vskip 2mm
\hskip 30mm (a) \hskip 65mm (b) 
\vskip 2mm
\caption{(a) Components of ${\bf v}_{2}$ as functions of the parameter $\theta_{1}$ on $C_1$: ${\tilde{\varv}}_{2x}(\theta_{1})$ (solid),\, ${\tilde{\varv}}_{2y}(\theta_{1})$ (dashed),\,${\tilde{\varv}}_{2z}(\theta_{1})$ (dotted);   (b) the same, for components of 
${\bf v}={\bf v}_{1}+{\bf v}_{2}$. \,$(\alpha=\pi/4,\,\xi=0.05)$; note how $\varv_{z}$ is positive only in a small neighbourhood of the tip T$_{1}$ at $\theta_{1}=\pi$.}
\label{Fig_v_C1}
\end{figure}
\section{Early deformation of $C_{1}$}\label{Sec_Initial_deformation}
\begin{figure}
\begin{center}
\begin{minipage}{0.99\textwidth}
\includegraphics[width=0.30\textwidth, trim=0mm 0mm 0mm 0mm]{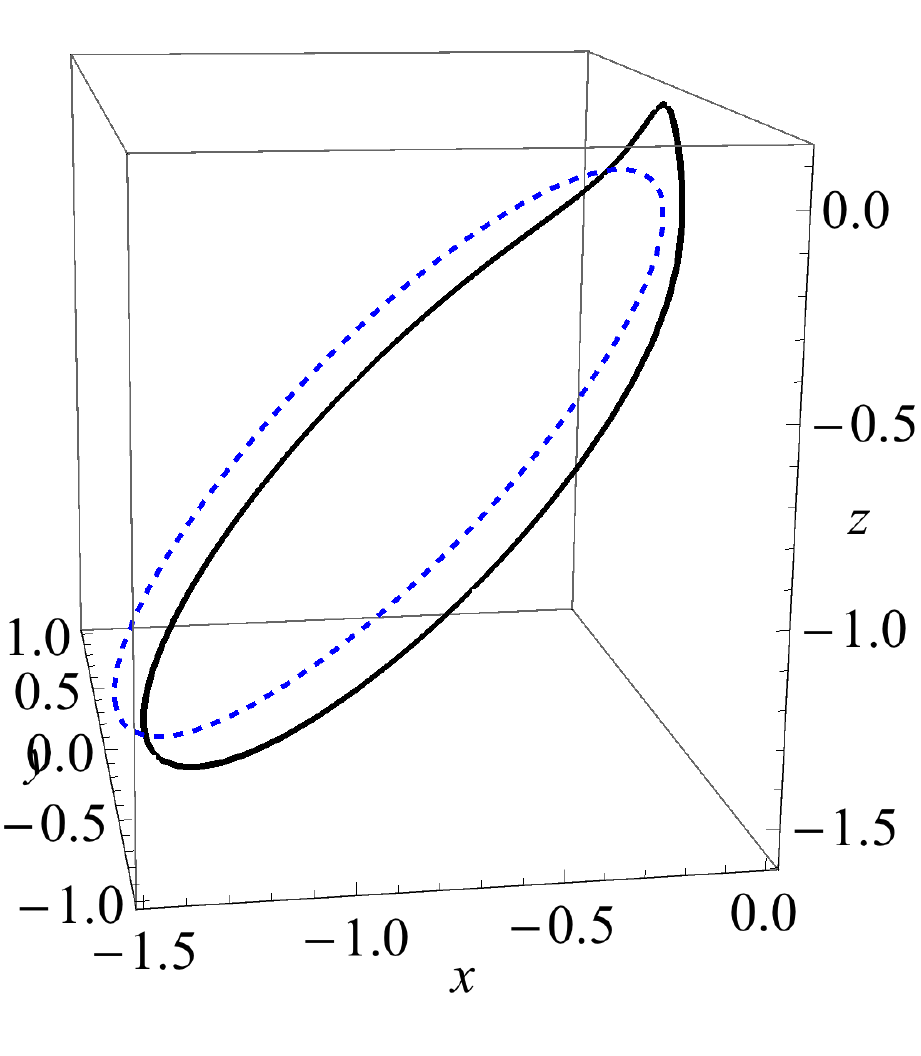}
\hspace*{12pt}
\includegraphics[width=0.30\textwidth,  trim=0mm 0mm 0mm 0mm]{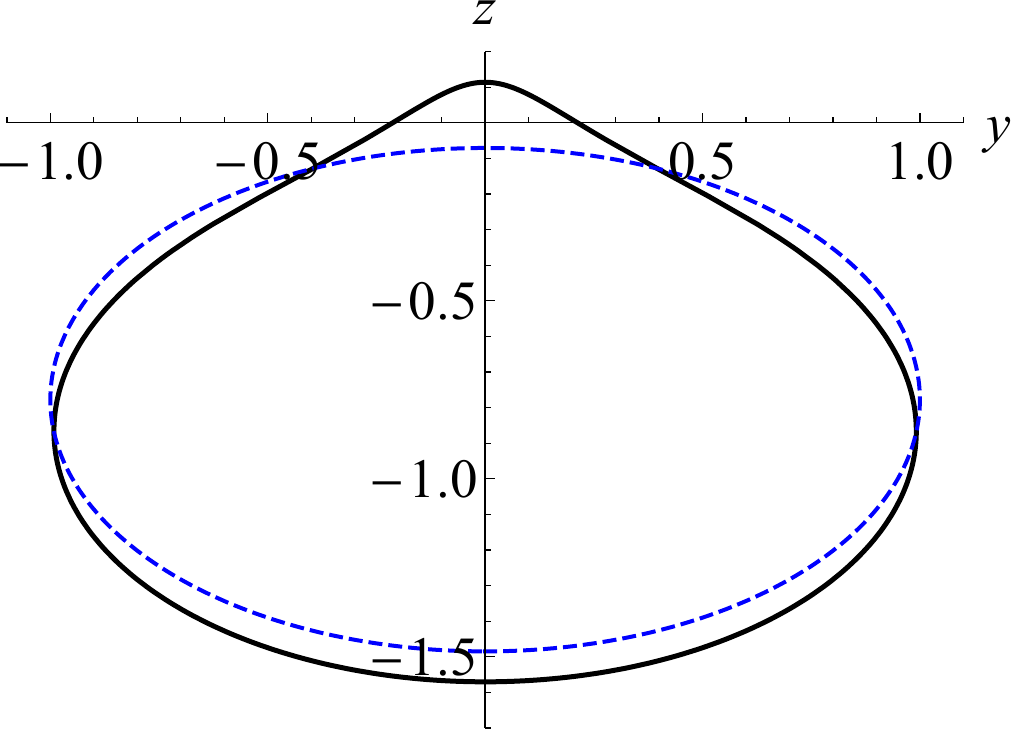}
\hspace*{12pt}
\includegraphics[width=0.30\textwidth,  trim=0mm 0mm 0mm 0mm]{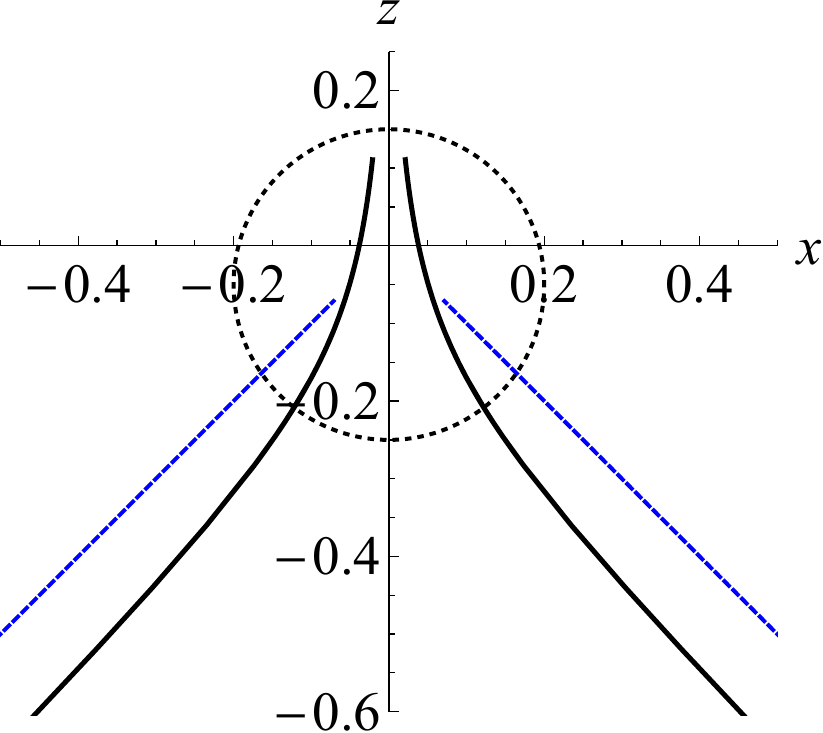}
\\[-1.8pt]
\end{minipage}
\end{center}
\vskip 2mm
\hskip 20mm (a) \hskip 40mm (b) \hskip 40mm (c) 
\vskip 2mm
\caption{(a) Early deformation of $C_1$ at $t=0.2$ (solid), based on the small-time approximation (\ref{initial deformation}); the initial form at $t=0$ is shown dotted. (b) Projection on $yz$-plane, showing increase of curvature at the tip; the initial projection on this plane is an ellipse. (c) Projection of both $C_1$ and $C_2$ on $xz$-plane, near the the region of strong interaction within the dotted circle of radius $\sim 0.2$. $(\kappa_{0} s_{0}=0.05, \,\kappa_{0} \,\delta_{0}=0.0025,\, \alpha = \pi/4)$.}
\label{Fig_deformed_vortex}
\end{figure}
\begin{figure}
\begin{center}
\begin{minipage}{0.99\textwidth}
\includegraphics[width=0.30\textwidth, trim=0mm 0mm 0mm 0mm]{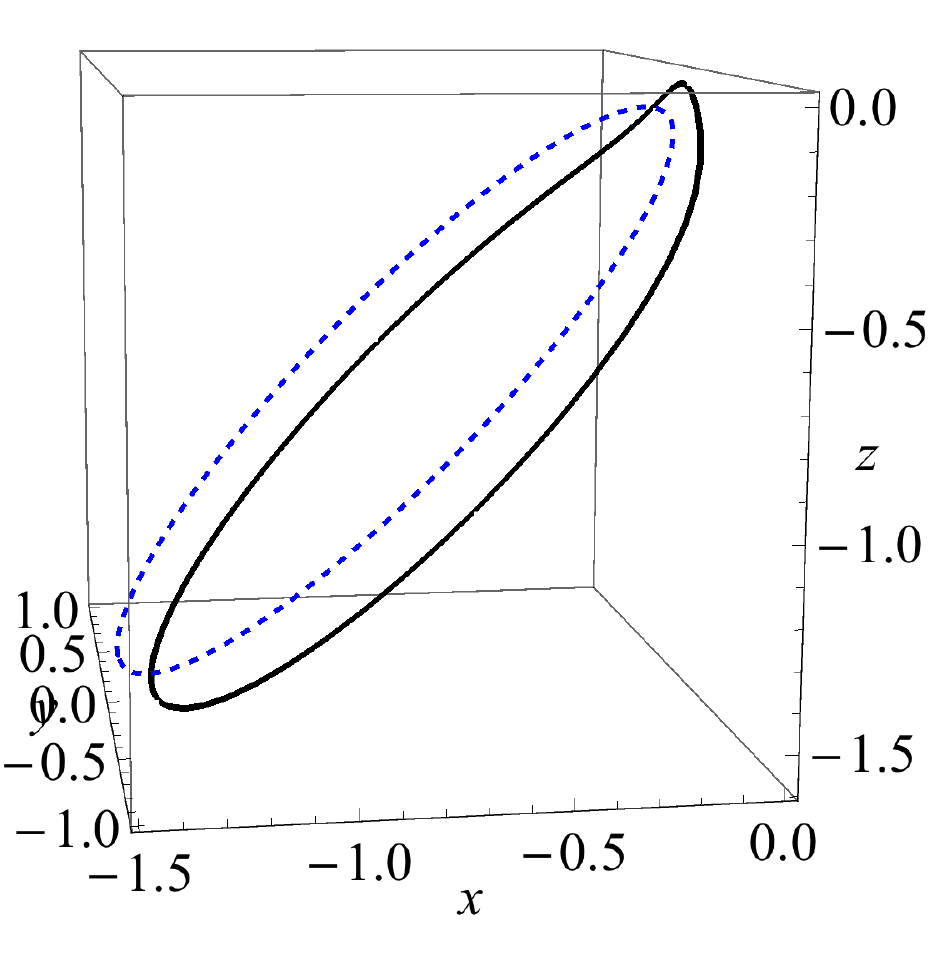}
\hspace*{12pt}
\includegraphics[width=0.30\textwidth,  trim=0mm 0mm 0mm 0mm]{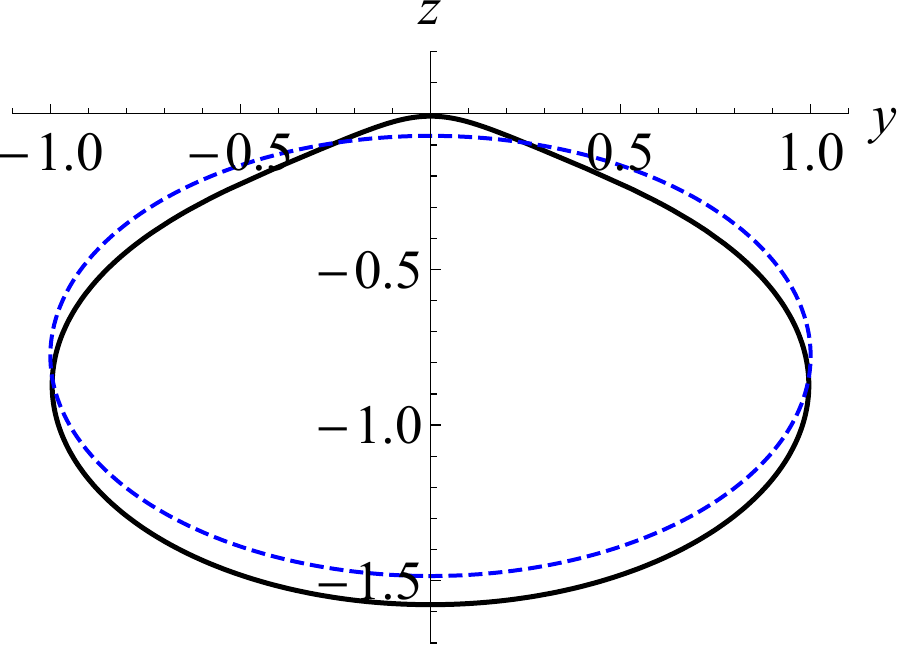}
\hspace*{12pt}
\includegraphics[width=0.30\textwidth,  trim=0mm 0mm 0mm 0mm]{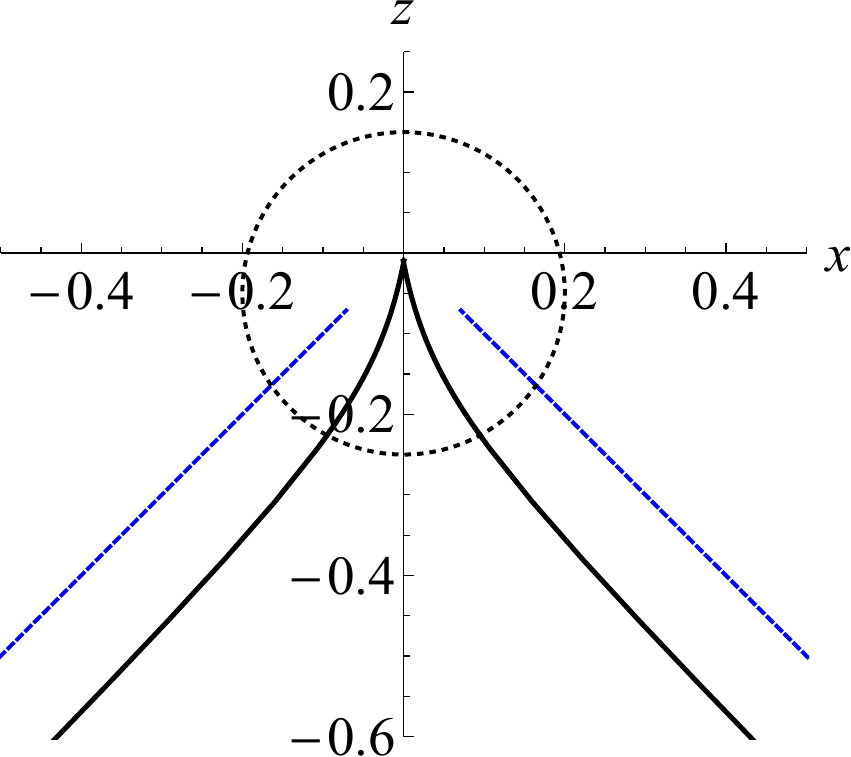}
\\[-1.8pt]
\end{minipage}
\end{center}
\vskip 2mm
\hskip 20mm (a) \hskip 40mm (b) \hskip 40mm (c) 
\vskip 2mm
\caption{Same as figure \ref{Fig_deformed_vortex}, still with $\kappa_{0} s_0=0.05$, but now with $\kappa_{0}\delta_{0}=3.303\times10^{-6}$; $t=0$ (dotted), and $t=0.115$ (solid) when, according to the linear approximation  (\ref{initial deformation}),
$C_1$ and $C_2$  `collide' at O; in this special case, the tips move on the diagonal lines $x=\pm z$.}
\label{Fig_deform_revised}
\end{figure}
The components of the initial induced velocity ${\bf v}_2({\bf x}_1)$=$({\varv}_{2x}$$(\theta_1),\, \varv_{2y}$$(\theta_1)$,\, $\varv_{2z}$$(\theta_1))$  on $C_1$ are found by taking 
${\bf x}={\bf x}_1 (\theta_1)$ in (\ref{transformed_velocity}); these  components  are shown in figure \ref {Fig_v_C1}(a), and the components of ${\bf v}({\bf x}_1)$= $(\varv_{x}$$(\theta_1), \,\varv_{y}$$(\theta_1)$,\,  $\varv_{z}$$(\theta_1))$  including the self-induced ingredients in figure \ref {Fig_v_C1}(b). Here, by way of illustration, we have chosen parameter values 
\begin{equation}
\kappa_{0} s_{0}=0.05, \quad\kappa_{0} \,\delta_{0}=\kappa_{0} s_{0}/20=0.0025, \quad \alpha = \pi/4,
\end{equation}
for which $\varv_{x}$ and $\varv_{z}$ are both positive at the tip 
$\theta_1=\pi$, as required.  Note also that the gradient  
$\tn{d}\varv_{y}$$(\theta_1)/\tn{d}\theta_1$ is positive at $\theta_1=\pi$, so the vortex  is being stretched at the tip. The resulting early displacement of the  vortex, based on the linearised  small-$t$  approximation
\begin{equation}
{\bf x}_{1}(\theta_{1},\,t) \approx {\bf x}_{1}(\theta_1,\,0) +t\,{\bf v}({\bf x}_1(\theta_1,\,0)) ,
\label{initial deformation}
\end{equation}
is shown in figure \ref{Fig_deformed_vortex}.  

Figure \ref{Fig_deform_revised}, by contrast, shows the same early deformation,  with $\kappa_{0} s_{0}=0.05, \kappa_{0} \delta_{0}= 3.303\times10^{-6}$.  For this much smaller value of $\delta_0$, the motion of the tip is directly towards the origin O; this is most evident in figure \ref{Fig_deform_revised}(c). The solid curves are for $t=0.115$, when the vortices  collide at the origin $(0,\,0,\,0)$; this is however pushing the  linear approximation (\ref{initial deformation}) too far, as it is accurate only for much smaller $t$. 
\subsection{The counter-intuitive increase of curvature at the tip}
\begin{figure}
\begin{center}
\includegraphics[width=4cm]{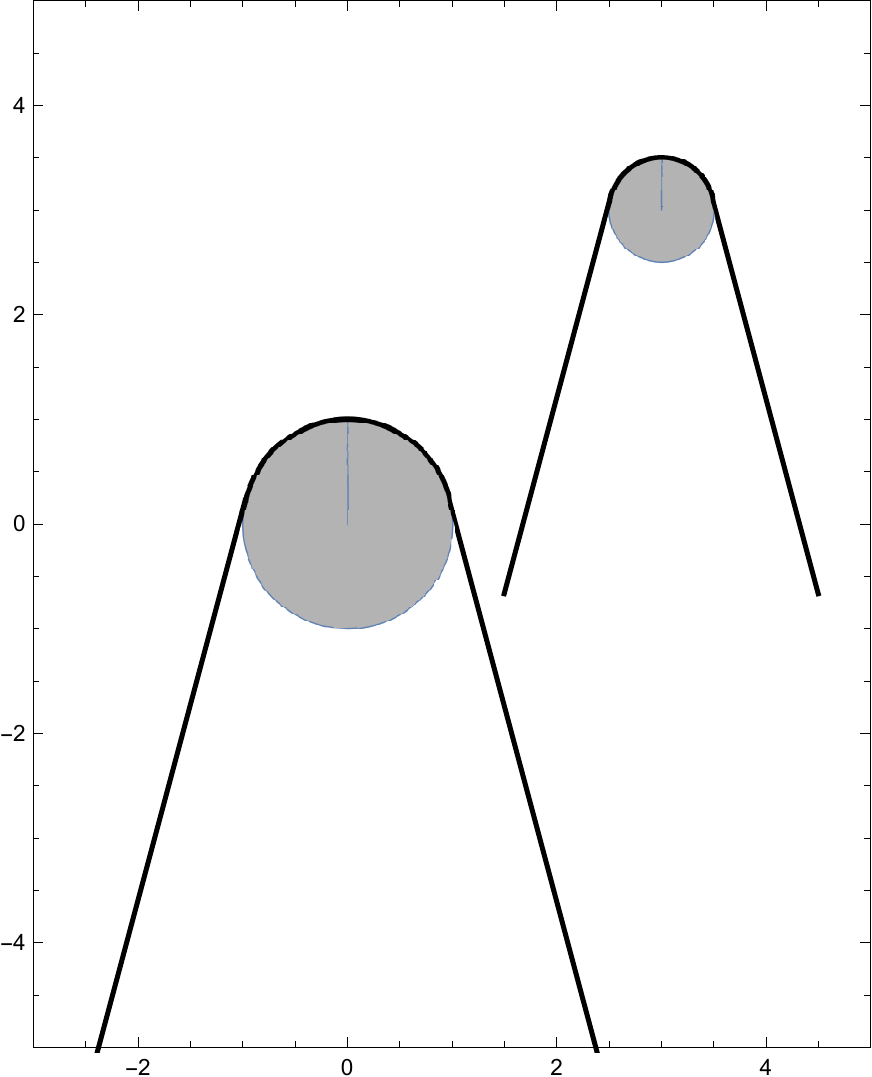}
\end{center}
\caption{Thought experiment in which an elastic cord  looped under tension round a horizontal rod is stretched by upward movement of the rod: if the cross-section of the rod is caused by some mechanism to decrease continuously, then the radius of curvature of the cord at the tip decreases despite the positive axial strain in its neighbourhood.}
\label{Fig_rod_stretch}
\end{figure}

It is evident from these figures that the stretching in the $y$-direction is coupled with an \emph {increase} of  curvature at the tip. This is counter-intuitive because one might expect the stretching to \emph{decrease} rather than increase the curvature.  It is perhaps helpful to consider an analogous situation as illustrated in figure \ref{Fig_rod_stretch}: imagine an elastic cord looped round a horizontal rod of circular cross-section, secured under tension  at some lower level, and stretched by upward movement of the rod;  imagine further that the radius of  cross-section of the rod is caused to decrease continuously by some mechanism (it could, for example, be a rapidly melting rod of ice).  Then the stretch of the cord increases at the tip, but its curvature also increases, remaining as it does under increasing tension in close contact with the rod. We shall find in \S \ref{increase_of_curvature} that this simple analogy does represent quite well the nature of the vortex stretching process at the tipping points. 
\begin{figure}
\begin{center}
\begin{minipage}{0.99\textwidth}
\includegraphics[width=0.45\textwidth, trim=0mm 0mm 0mm 0mm]{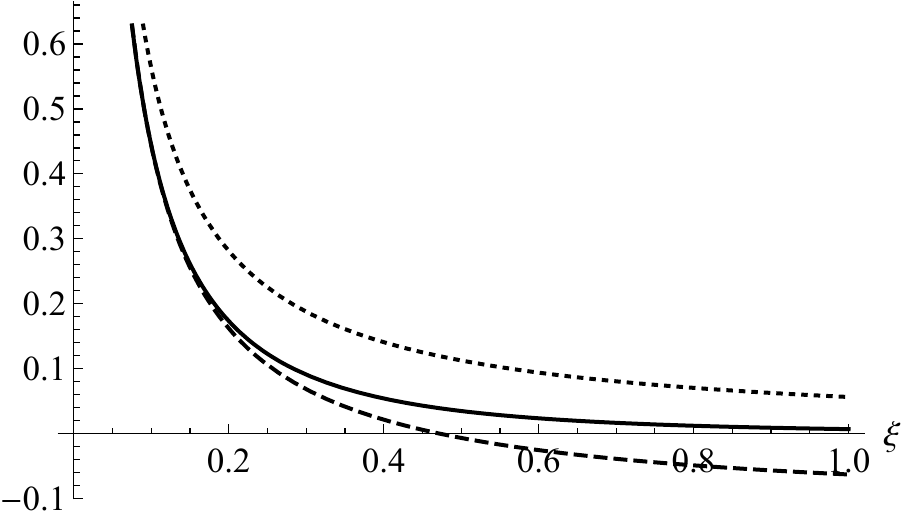}
\hspace*{20pt}
\includegraphics[width=0.45\textwidth,  trim=0mm 0mm 0mm 0mm]{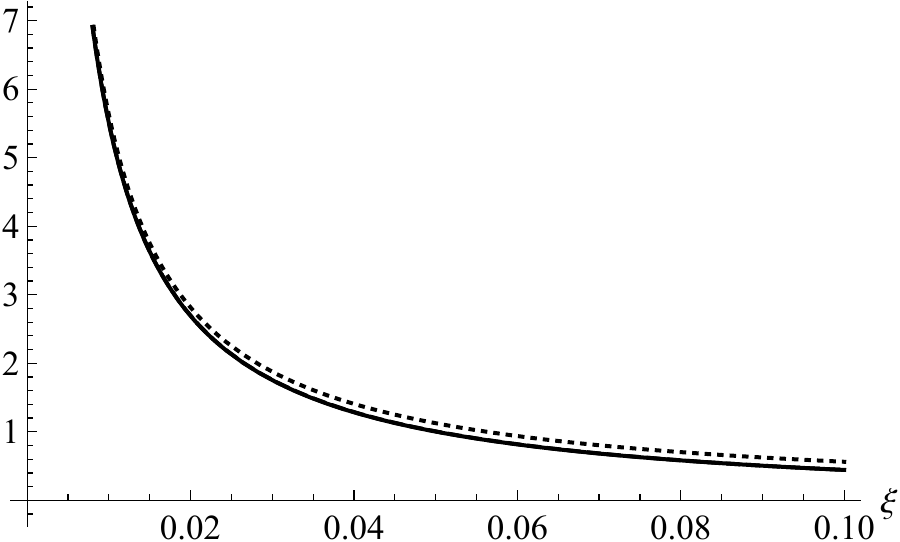}
\end{minipage}
\end{center}
\vskip 2mm
\hskip 30mm (a) \hskip 65mm (b) 
\vskip 2mm
\caption{Rate of stretching $\lambda_2(\xi)/\Gamma\kappa_{0}^2$ at the tipping point T$_{1}$ (solid); two-term small-$\xi$ asymptotic (dashed); one-term asymptotic (dotted); (a) in the range $0<\xi<1$  (b) the same, for the range $0\!<\!\xi\!<\!0.1$; in this range, the exact solution and the two-term small-$\xi$ asymptotic are indistinguishable.}
\label{Fig_lambda2}
\end{figure}
\subsection{Rate of stretching at the tipping point}
The rate of stretching of $C_1$ at the tipping point T$_{1}\,(-s,\,0,\,-s\cot\alpha)$\, is 
$\lambda_2\equiv \partial \varv_{y}/\partial y$,
and this may be evaluated from the exact solution.  The result, when simplified,  is 
\begin{equation}\label{lambda2}
\lambda_2=\frac{\Gamma\kappa_{0}^2}{4\pi}\frac{\left[(1+2 \xi ^{2}+2 \xi \sin\alpha)E(k)-2 \xi ^{2}K(k)\right]\cos\alpha}
{  \xi (1+2 \xi \sin\alpha)^{2}\sqrt{1+ \xi ^{2}+2 \xi \sin\alpha}}\,,
\end{equation}
with $k$ still given by (\ref{Tipping_point_k}). For small $ \xi $, this function has the asymptotic behaviour 
\begin{equation}\label{lambda2asym}
\lambda_2\sim\frac{\Gamma\kappa_{0}^2 \cos\alpha}{16\,\pi}\left[ \frac{4}{  \xi }-12\sin\alpha 
+ \xi \left\{6 \log\left(\frac{ \xi }{4}\right)\! +5(4\!-\!3\cos2\alpha)\right\}+\tn{O}\left( \xi ^2\right)\right]\,.
\end{equation}
Figure \ref{Fig_lambda2}(a) shows  $\lambda_{2}( \xi )/\Gamma\kappa_{0}^2$  as given by (\ref{lambda2}) for $0\!<\!\xi\!<\!1$, together with the small-$\xi$ two-term asymptotic result (\ref{lambda2asym}). Figure \ref{Fig_lambda2}(b) shows the same for 
$0\!<\!\xi\!<\!0.1$; over this range, the exact result and the two-term asymptotic result are virtually indistinguishable; the one-term asymptotic result $\lambda_2/\Gamma\kappa_{0}^2\sim  \cos\alpha/4\pi \,\xi$ is shown dotted.  It is perhaps significant that the $\xi$-dependence of $\lambda_2$ is so similar to that of $\varv_{2z}$, as shown in figure \ref{Fig_tip_comp_anal}(b); it is as if the relatively rapid upward movement of T$_{1}$ is the direct cause of a corresponding stretching of the vortex.

\subsection{Rate of strain in the $xz$-plane}\label{Sec_rate_of_strain}
Here, we return to the frame of reference O$'XYZ$.  We are concerned with the rate of strain that acts upon the vortex ${\mathcal V}_{1}$ at its tipping point T$_{1}$; the self-induced velocity of ${\mathcal V}_{1}$ makes no contribution to this,  so we need only calculate the rate-of-strain tensor $e_{ij}({\bf X})$ at T$_{1}$ induced by ${\mathcal V}_{2}$. This may be found explicitly from the exact solution, and evaluated at T$_{1}$ (see \ref{Tipping_Point_X}). At this point,  
  $e_{ij}\left({\tilde{\bf X}}\right)$ takes the form
\[\label{rate_of_strain_tensor}
 e_{ij}\left({\tilde{\bf X}}\right)= \left(\!\! \begin{array} {ccc}
e_{XX}\! &\! 0&\! e_{XZ}\\ 
0\! &  \! \lambda_2\! &\! 0\\
e_{ZX}\! &\! 0 \! &\! e_{ZZ}
\end{array}\!\! \right)\,, 
\] 
where $\lambda_{2}$ is as already determined by (\ref{lambda2asym}),  and
\begin{equation}\label{components_of_e_ij}
e_{XX}=\frac{\partial\varv_{2X}}{\partial X}\,,\quad e_{ZZ}=\frac{\partial\varv_{2Z}}{\partial Z}\,,\quad e_{XZ}=e_{ZX}=\thalf\left(\frac{\partial\varv_{2X}}{\partial Z}
+\frac{\partial\varv_{2Z}}{\partial X}\right)\,,
\end{equation}
evaluated at ${\tilde{\bf X}}$.

We shall need only the small-$\xi$ asymptotic expressions for the components $e_{XX}, e_{ZZ}$ 
and $e_{XZ}$, to the same order of approximation as for $\lambda_{2}$ in (\ref{lambda2asym}). After some simplification, these asymptotic results are
\begin{eqnarray}\label{e_ij_at_tip}
\frac{e_{XX}}{\Gamma\kappa_{0}^2}&=&-\frac{\cos\alpha \sin\alpha}{4\pi \xi^2}-\frac{\cos\alpha\cos 2\alpha}{8\pi \xi}
+\frac{(6+3\cos 2\alpha)\cos\alpha \sin\alpha}{16\pi}+\tn{O}(\xi)\,,\nonumber\\
\frac{e_{ZZ}}{\Gamma\kappa_{0}^2}&=&+\frac{\cos\alpha \sin\alpha}{4\pi \xi^2}+\frac{\cos\alpha(\cos 2\alpha-2)}{8\pi \xi}
+\frac{(15\sin\alpha-3\sin{3\alpha})\cos\alpha }{32\pi}+\tn{O}(\xi)\,,\nonumber\\
\frac{e_{XZ}}{\Gamma\kappa_{0}^2}&=&-\frac{\cos 2\alpha}{8\pi \xi^2}+\frac{\sin {3\alpha}-\sin\alpha}{16\pi \xi}
+\frac{12\log{(\xi/4)}+19+3\cos{4\alpha}}{64\pi}+\tn{O}(\xi)\,,
\end{eqnarray}
Thus, retaining only terms up to O$(\xi^{-1})$,  and still in the frame O$XYZ$, we have
\bea
 \frac{e\!_{ij}\left({\bf X}\right)}{\Gamma\kappa_{0}^2}
 = \!\frac{1}{8\pi \xi^2}\left(\!\! \begin{array} {ccc}
\!-\cos\alpha\,(2 \sin\alpha\!+\!\xi\cos 2\alpha)\! &\! \!0&\!\! -\cos 2\alpha\!+\!\thalf \xi(\sin {3\alpha}\!-\!\sin\alpha)\\ 
\!0\! &  \!\! 2 \xi\cos\alpha\!\! &\! 0\\
\!-\cos 2\alpha\!+\!\thalf \xi (\sin {3\alpha}\!-\!\sin\alpha)\! &\!\! 0 \!\! &\! \cos\alpha\,(2 \sin\alpha\!+\! \xi
(\cos 2\alpha\!-\!2))
\end{array}\!\!\! \right), 
\label{e_ij_leading_order}
\eea 
and we note that this satisfies $e_{ii}=0$, as required by incompressibility -- a useful check on the analysis. 
 
\begin{figure}
\begin{center}
\includegraphics[width=6cm]{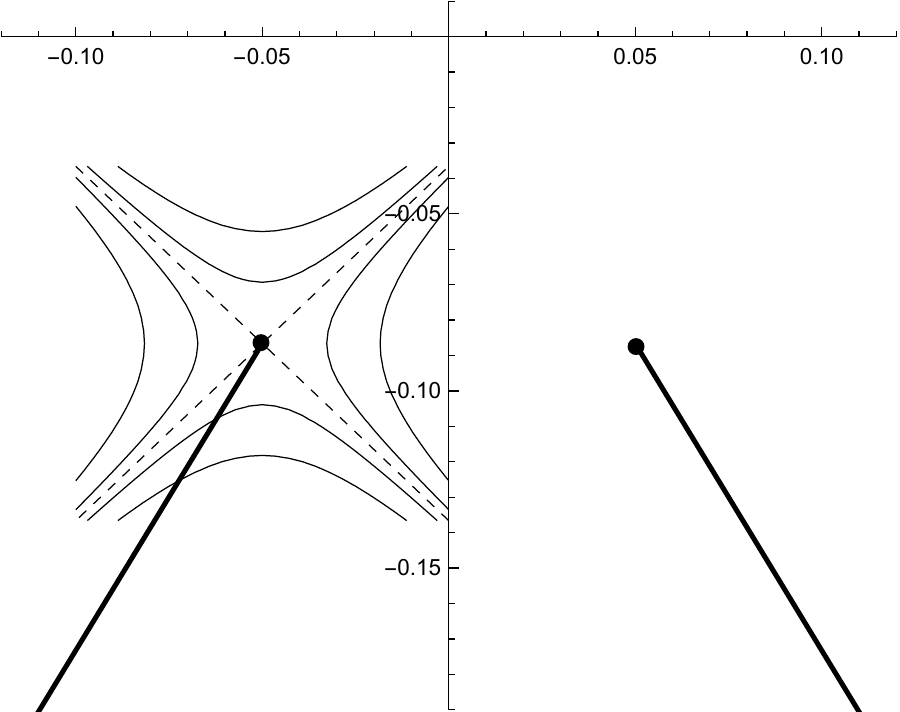}
\end{center}
\vskip 2mm
\caption{Sketch of projection in the $xz$-plane, showing the local strain field induced by $C_2$ near the tipping point of $C_1$. The stretching eigenvalue $\lambda_3>0$ (corresponding to the eigenvector $(1,0, 1)/\surd{2}$) is slightly less than $|\lambda_1|$,
where  $\lambda_1 <0$ is the contracting eigenvalue, the difference being compensated by the stretching of the vortex along its axial ($y$-)direction.}\label{Fig_local_structure}
\end{figure}
The eigenvalues of the tensor $e_{ij}$ (which are of course frame-independent) can now be calculated; correct to order $\xi^{-1}$ (for $\xi\ll 1$), they are
\begin{equation}\label{eigenvalues}
\frac{\lambda_1}{\Gamma\kappa_{0}^2}=\frac{-1}{8\pi \xi^2}+\frac{\sin\alpha -\cos\alpha}{8\pi \xi}, \quad \frac{\lambda_2}{\Gamma\kappa_{0}^2}=\frac{\cos\alpha}{4\pi \xi},\quad \frac{\lambda_3}{\Gamma\kappa_{0}^2}=\frac{1}{8\pi \xi^2}-\frac{\sin\alpha +\cos\alpha}{8\pi \xi},
\end{equation}
and note again that $\lambda_1+\lambda_2+\lambda_3 =0$.
The corresponding normalised eigenvectors ${\bf e}_{1},\, {\bf e}_{2}$ and ${\bf e}_{3}$ have components in O$'XYZ$
\begin{equation}\label{eigenvectors_OXYZ}
{\bf e}_{1}=  \frac{(1+\sin 2\alpha,\, 0,\,\cos 2\alpha)}{2^{1/2}(1+\sin 2\alpha)^{1/2}}, \quad{\bf e}_{2} = (0,\,1,\,0), 
\quad {\bf e}_{3}= \frac {(-1 +  \sin 2\alpha,\, 0,\,\cos 2\alpha)}{2^{1/2}(1+\sin 2\alpha)^{1/2}},
\end{equation}
or in the O$xyz$ frame
\begin{equation}\label{eigenvectors_Oxyz}
{\bf e}_{1}=  2^{-1/2}(-1,\,0,\,1), \quad {\bf e}_{2} =(0,\,1,\,0), 
\quad {\bf e}_{3}= 2^{-1/2}(1,\,0,\,1).
\end{equation}
Thus, whatever the value of $\alpha$, the direction of positive stretching in the $xz$-plane $(\lambda_{3}>0)$ is parallel to the diagonal $x=z$, and the contracting direction $(\lambda_{1}<0)$  is parallel to the diagonal $x=-z$. 

Relative to the tipping point $\tilde{{\bf x}}=(\tilde x, 0, \tilde z)$,  the induced flow near it is
\be \label{flow_near_T_1}
{\bf v}_2 \sim \left[-\thalf\lambda_{2}(x-{\tilde x})-(\lambda_{3}+\thalf\lambda_{1})(z-\tilde z),\,\lambda_{2}y, \,
(\lambda_{3}+\thalf\lambda_{1})(x-{\tilde x})-\thalf\lambda_{2}(z-\tilde z) \right] +\tn{O}(|{\bf x}-\tilde{{\bf x}}|^2)\,,
\ee
and under this irrotational uniform strain, points initially on the diagonal $x-\tilde x=z-\tilde z$ remain on it for all $t>0$. The situation is as sketched in figure \ref{Fig_local_structure}, or equivalently by the (near) hyperbolic curves of figure \ref{Fig_sketch_2}.  Clearly there is a persistent tendency for the flow (\ref{flow_near_T_1}) to move $\alpha$  to the asymptotic orientation $\pi/4$, on a time-scale $t_{o}=\lambda_{3}^{-1}\sim 8\pi s^{2}/\Gamma$.  The flow then serves to maintain $\alpha$ at the value $\pi/4$. [This effect does not show up in figures  \ref{Fig_deformed_vortex} and  \ref{Fig_deform_revised}, which take no account of the strain field near the tip of $C_{1}$; it would  show up if the expansion (\ref{initial deformation}) were continued to O$(t^2)$, because this would involve the term 
${\bf v}\!\cdot \!\nabla{\bf v}$ in the Navier-Stokes equation, bringing the local uniform strain flow into play.] We shall retain $\alpha$ explicitly in the formul\ae \, that follow, but adopt the value $\alpha=\pi/4$ in the figures and numerical calculations.

It is interesting to note here that when R$_{\Gamma}\gg1$ the leading-order deformation of each isovorticity contour  is to an ellipse whose principal axes are rotated through an angle $\pi/4$  relative to the principal axes of strain in the plane of cross-section, this rotation being in the same sense as the velocity within the vortex (an effect  first noted by \citealt{Robinson1984}). This means that in the situation under consideration here, the isovorticity contours near the tipping points may be expected to deform to ellipses with major axes parallel to the $z$-axis.  This effect has been observed in the recent DNS investigation of \cite{Brenner2016}, albeit in the context of the Euler equations. For the Navier-Stokes equations,  the effect is prominent at modest values of R$_{\Gamma}\sim 10^{2}$, 
but disappears as R$_{\Gamma}\rightarrow\infty$, as shown in figure \ref{Fig_isovorticity}.

\begin{figure}
\begin{center}
\begin{minipage}{0.99\textwidth}
\includegraphics[width=0.45\textwidth, trim=0mm 0mm 0mm 0mm]{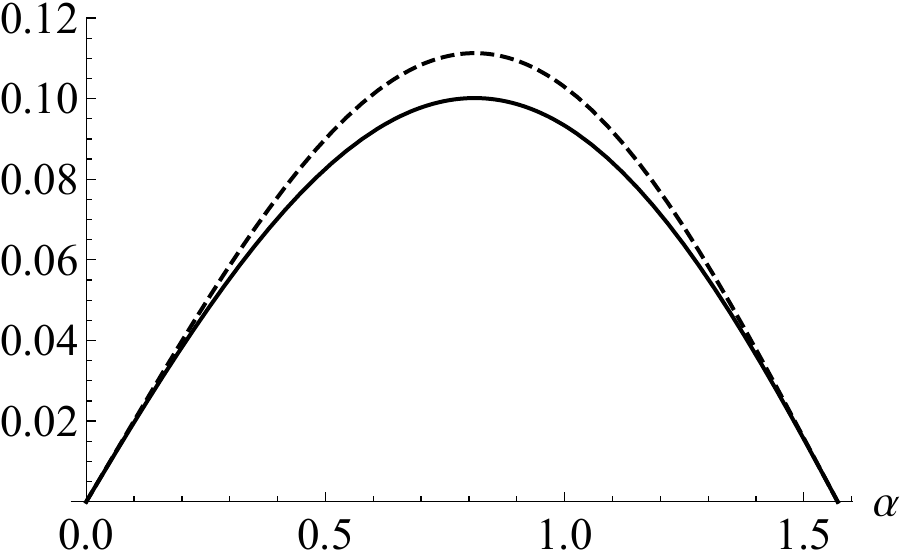}
\hspace*{20pt}
\includegraphics[width=0.45\textwidth,  trim=0mm 0mm 0mm 0mm]{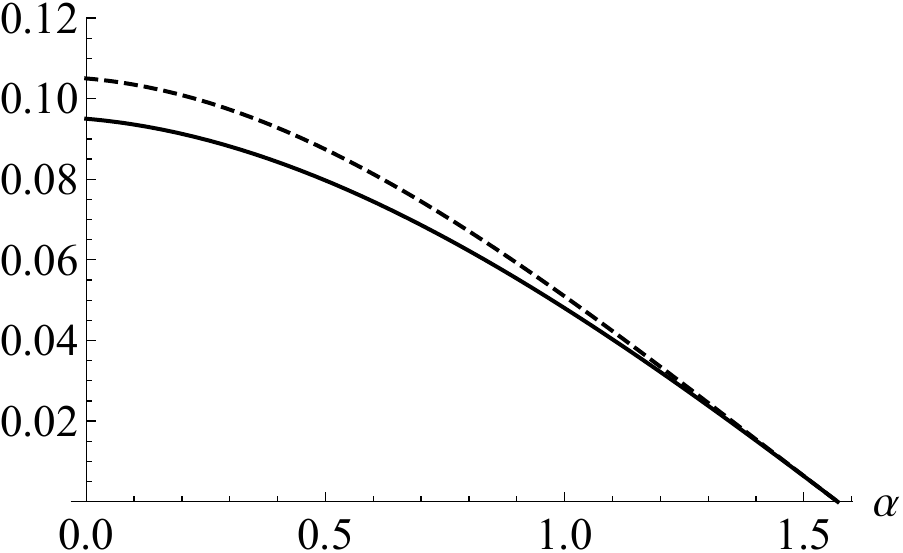}
\end{minipage}
\end{center}
\vskip 2mm
\hskip 30mm (a) \hskip 65mm (b) 
\vskip 2mm
\caption{Dependence of $\lambda_2/|\lambda_1|$ (solid) and $\lambda_2/\lambda_3$ (dashed) on the angle $\alpha$ $(0<\alpha < \pi/2)$; (a) with $d$ held constant at $0.1;$  (b) with $\xi$ held constant at $0.05$.}
\label{Fig_lambda_ratio}
\end{figure}

Using (\ref{lambda2asym}) and (\ref{eigenvalues}),  the ratios $\lambda_2/|\lambda_1|$  and $\lambda_2/\lambda_3$ are given, correct to order $\xi^2$  or $(d/R)^2$, by
\begin{equation}\label{ratios1}
\frac{\lambda_2}{|\lambda_1|}\sim 2\xi\cos\alpha[1\!-\!\xi(2 \sin\alpha\!+\! \cos\alpha)]
=\frac{d\sin2\alpha }{R}\left[1\!-\!\frac{d\sin\alpha (2\sin\alpha\!+\! \cos\alpha)}{R}\right],
\end{equation}
\begin{equation}\label{ratios2}
\frac{\lambda_2}{\lambda_3}\sim 2\xi\cos\alpha[1\!-\!\xi(2 \sin\alpha\!-\! \cos\alpha)]
=\frac{d\sin2\alpha }{R} \left[1\!-\!\frac{d\sin\alpha(2\sin\alpha\!-\! \cos\alpha)}{R}\right].
\end{equation}
These ratios, equal at leading order, depend on $\alpha$ as shown in figure \ref{Fig_lambda_ratio}(a) (with $d$ held constant as $\alpha$ varies),  and in figure \ref{Fig_lambda_ratio}(b) (with $\xi=d\sin\alpha/R$ held constant).  As indicated above, the positive eigenvalue $\lambda_3$ tends to deform the vortex core to elliptical form; the crucial question now is whether the core can nevertheless remain compact by the mechanism described in \S \ref{non-axisymmetric strain} if the Reynolds number R$_\Gamma$ is sufficiently large.

Note that, from (\ref{eigenvalues}), with $\alpha=\pi/4$,
\be\label{lambda_ratio}
\lambda_{3}/\lambda_{2}\sim\left(1/\!\surd{2}\right)\xi^{-1}\,,\quad (\xi \ll 1).
\ee
\subsection{Rate of increase of tip curvature}\label{increase_of_curvature}
We have already noted the initial increase of the tip curvature evident in figure \ref{Fig_deformed_vortex}(b); we now seek to determine its rate of increase.   If the point ${\bf x}_{0}(\theta_{1})$ on $C_1$ at time $t=0$ moves to ${\bf x}={\bf x}(\theta_1, t)$ at time $t$,  then the curvature at T$_{1}$ is given by
\be\label{curvature1}
\kappa (t)=\frac{|{\bf x'}(\theta_1, t)\times{\bf x''}(\theta_{1}, t)|}{|{\bf x'}(\theta_{1}, t)|^{3}}=
\frac{\left[({\bf x'}(\theta_{1}, t)\times{\bf x''}(\theta_{1}, t))\cdot({\bf x'}(\theta_{1}, t)\times{\bf x''}(\theta_{1}, t))\right]^{1/2}}{\left[{\bf x'}(\theta_{1}, t)\cdot{\bf x'}(\theta_{1}, t)\right]^{3/2}}\,,
\ee
where the dash denotes partial differentiation with respect to $\theta_{1}$. Substituting (\ref{initial deformation}), and expanding in powers of $t$, we obtain
\be\label{curvature2}
\kappa (t)=\left[\kappa_{0}+t\left(\frac{({\bf x'}_{0}\times{\bf x''}_{0})\cdot({\bf v'}_{0}\times{\bf x''}_{0}
+{\bf x'}_{0}\times {\bf v''}_{0})}{|{\bf x'}_{0}\times{\bf x''}_{0}|^{2}|{\bf x'}_{0}|^3}-
\frac{3({\bf x'}_{0}\cdot{\bf v'}_{0})|{\bf x'}_{0}\times{\bf x''}_{0}|}
{|{\bf x'}_{0}|^{5}}\right)\right] +\tn{O}(t^2)\,,
\ee
from which it follows that
\be\label{curvature3}
\left.\frac{\tn{d}\kappa}{\tn{d}t}\right|_{0}=\frac{({\bf x'}_{0}\times{\bf x''}_{0})\cdot({\bf v'}_{0}\times{\bf x''}_{0}
+{\bf x'}_{0}\times {\bf v''}_{0})}{|{\bf x'}_{0}\times{\bf x''}_{0}|^{2}|{\bf x'}_{0}|^3}-
\frac{3({\bf x'}_{0}\cdot{\bf v'}_{0})|{\bf x'}_{0}\times{\bf x''}_{0}|}
{|{\bf x'}_{0}|^{5}}\,.
\ee
In this expression, ${\bf x}'_{0}$ and ${\bf x}''_{0}$ are evaluated at T$_{1}$ at $t=0$, and are therefore simply
\be\label{initial_x}
{\bf x}'_{0}=R\left(0,\,1,\,0\right),\quad
{\bf x}''_{0}=-R\left(\sin\alpha,\,0,\,\cos\alpha\right),
\ee
and, by virtue of the symmetry about $\theta_{1}=\pi$, as evident in figure \ref{Fig_v_C1},
\be\label{initial_v}
{\bf v}_{0}=\left(\varv_{2x},\,0,\,\varv_{2z}\right)\,,\quad
{\bf v}'_{0}=\left(0,\,\varv'_{2y},\,0\right)\,,\quad
{\bf v}''_{0}=\left(\varv''_{2x},\,0,\,\varv''_{2z}\right)\,.
\ee
Substituting in (\ref{curvature3}) with $\kappa_{0}=R^{-1}$, we obtain
\be\label{curvature4}
\left.\frac{\tn{d}\kappa}{\tn{d}t}\right|_{0}=-\kappa_{0}^{2}\left(\varv''_{2x}\sin\alpha+2\varv'_{2y}+
\varv''_{2z}\cos\alpha\right)\,.
\ee

 With the notation ${\bf v_{2}}=(\varv_{2x}(\theta_1, s,\alpha),\,\varv_{2y}(\theta_1, s,\alpha),\,\varv_{2z}(\theta_1, s,\alpha))$  on $C_{1}$, and noting that $\partial \varv_{2y}/\partial \theta_{1}=\kappa_{0}^{-1}\partial \varv_{2y}/\partial y$ at T$_{1}$, we already have (from (\ref{Tipping_point_s_small})  and (\ref{lambda2asym})) the asymptotic results at the tipping point $\theta_1=\pi$, as $ \xi\rightarrow 0$
\be
{\varv}_{2x}(\pi,\xi,\alpha)\sim - \frac{\Gamma\kappa_{0}}{4\pi}\left(\log\frac{4}{\xi} \!-\!1\right) \cos\alpha\,,\quad
{\varv}_{2z}(\pi,\xi,\alpha)\sim\frac{\Gamma\kappa_{0}}{4\pi \xi}\,,\quad\varv'_{2y}(\pi,\xi,\alpha)\sim \frac{\Gamma\kappa_{0}}{4\pi \xi}\cos\alpha\,.
\ee
It remains to evaluate $\varv''_{2x}$ and 
$\varv''_{2z}$ to the same order.  The result, after hefty analysis, is
\be\label{v2x_double_prime}
\varv''_{2x}\sim \frac{\Gamma\kappa_{0}\cos\alpha}{8\pi  \xi} +\tn{O}(1),\quad
\varv''_{2z}\sim -\frac{\Gamma\kappa_{0}\sin\alpha}{4\pi  \xi^{2}}
+\frac{\Gamma\kappa_{0}(1\!-\!3\cos 2\alpha)}{8\pi \xi}+\tn{O}(1)\,.
 \ee
As expected, $\varv''_{2x}>0$ and $\varv''_{2z}<0$, as already evident in the plots of figure \ref{Fig_v_C1}. 

\begin{figure}
\begin{center}
\includegraphics[width=0.4\textwidth,  trim=0mm 0mm 0mm 0mm]{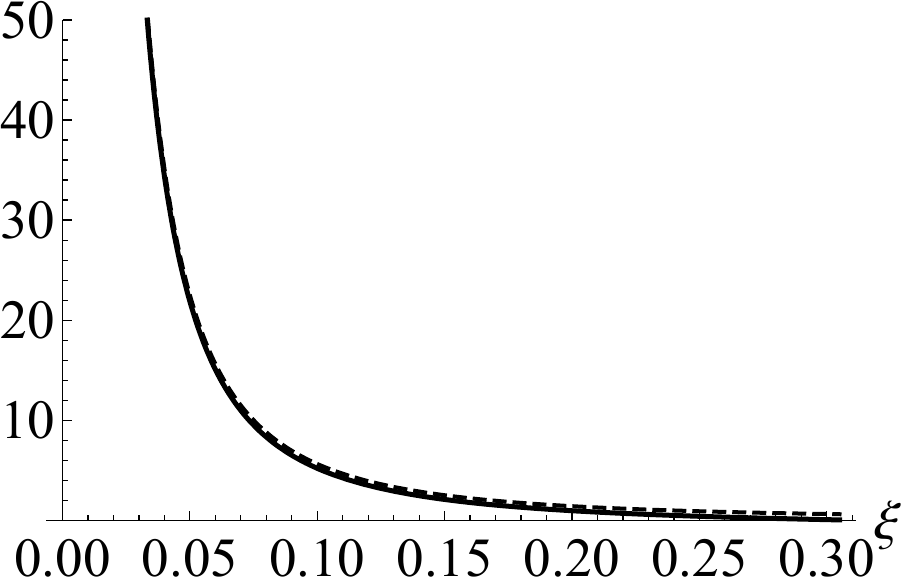}
\end{center}
\vskip 2mm
\caption{Variation of $\tn{d}\kappa/\tn{d}t$ as given by (\ref{curvature4}) in the range $0.02\lesssim \xi \lesssim 0.3$; the black curve includes  all three terms, while the dashed curve includes only the dominant O$(\xi^{-2})$ term of (\ref{v2x_double_prime}); the curves are virtually indistinguishable for $\xi \lesssim 0.1$.}
\label{Fig_curvature_dot}
\end{figure}
Returning now to (\ref{curvature4}), it is evident that the term $-2\varv'_{2y}$  tends to \emph{decrease} $\kappa$ as intuition might suggest; less obviously, the term 
$-\varv''_{2x}\sin\alpha$ also tends to decrease $\kappa$.  However the term 
$-\varv''_{2z}\cos\alpha$ is positive and an order of magnitude larger than the other two terms of 
(\ref{curvature4}) at small $\xi$, and this dominant contribution provides the increase of curvature evident in figure \ref{Fig_deformed_vortex}, in conformity with the interpretation suggested by figure \ref{Fig_rod_stretch}. 

Figure \ref{Fig_curvature_dot} shows $\tn{d}\kappa/\tn{d}t|_{t=0}$ as given by
(\ref{curvature4}); the solid curve includes the contributions from all three terms, while the dashed curve includes only the dominant O$(\xi^{-2})$ contribution; the two curves are virtually indistinguishable for $\xi\lesssim 0.1$  indicating that the terms $-\varv''_{2x}\sin\alpha$ and $-2\varv'_{2y}$ really do make negligible contributions in this range to the rate-of-change of curvature.  We conclude then that 
\be\label{curvature5}
\left.\frac{\tn{d}\kappa}{\tn{d}t}\right|_{0}\sim-\kappa_{0}^{2}\,\varv''_{2z}\cos\alpha\sim\frac{\Gamma\kappa_{0}^{3}\sin\alpha\cos\alpha}{4\pi\, \xi^{2}}\,,
\ee
to good approximation in the range $\xi\lesssim 0.1$.

\section{Rescaling for times $t>0$}\label{Sec_rescaling}
So far, we have considered only the situation at time $t=0$; this  however has revealed the fact that, when $\xi=\kappa_{0} s\lesssim 0.1$, the velocity field and initial deformation in the neighbourhood of the tipping points are determined solely by $s, \,\alpha$  and the curvature $\kappa_{0}$ of 
$C_1$ at T$_{1}$ (and $C_2$ at T$_{2}$).  We have also found good reason to focus on the situation when $\alpha=\pi/4$. For times $t\!>\!0$, the curves $C_{1}$ and $C_{2}$ are of course no longer circular, but we may reasonably assume that  the behaviour in the crucial neighbourhood of  the tipping points will continue to be controlled by the current values of $s(t),\, \alpha(t)$ and curvature $\kappa(t)$;  this behaviour is then just as if $C_{1}$ and $C_{2}$ were replaced by the circles of curvature at T$_{1}$ and T$_{2}$ respectively.  This `circle-of-curvature assumption', 
 which we now adopt, is the key assumption of the present analysis,

Let us first summarise the main results obtained so far.  From  (\ref{self-induced_velocity2}) and (\ref{Tipping_point_s_small}), the velocity components at the tip T$_{1}$ at time $t=0$ are given at leading order by
\be\label{velocity_components}
\tilde{\varv}_{x}= \tilde{\varv}_{1x}\!+\!\tilde{\varv}_{2x}\sim\frac{\kappa_{0}\Gamma}{4\pi}\left[\log{\left(\frac{1}{\kappa_{0}\delta}\right)}+\beta-\left(\log{\frac{4}{\xi}}\!-\!1\right)\right] \cos\alpha = 
\frac{\kappa_{0}\Gamma}{4\pi}\left[\log\left(\frac{s}{\delta}\right)+\beta_{1}\right]\cos\alpha\,,
\ee
\vskip -2mm
\noindent and
\be
\label{velocity_components_z}
\tilde{\varv}_{z}= \tilde{\varv}_{1z}+\tilde{\varv}_{2z}\sim 
\frac{\kappa_{0}\Gamma}{4\pi }\left[ \frac{1}{\kappa_{0}s}
-\left\{\log \left(\frac{4}{\kappa_{0}^{2}s\delta}\right)+\beta\right\}\sin\alpha\right]\,,
\ee
where $\beta_{1}=\beta-\log{4}+1$; for a Gaussian core, $\beta=0.828$ and so $\beta_{1}=0.442$, and for a uniform vorticity core, $\beta=1.136$, $\beta_{1}=0.750$.
The coordinates $\tilde{x}, \,\tilde{z}$ of T$_{1}$  then change initially according to the equations
\be
\left.\frac{\tn{d}\tilde x}{\tn{d}t}\right|_{0}= -\left.\frac{\tn{d}s}{\tn{d}t}\right|_{0} =\tilde{\varv}_{x},\quad
\left.\frac{\tn{d}\tilde z}{\tn{d}t}\right|_{0} =\tilde{\varv}_{z}\,.
\ee
We have also obtained asymptotic results for the initial rate of stretching $\lambda\, 
(\equiv\!\lambda_2)$ and rate of increase of curvature $\tn{d}\kappa/\tn{d}t$ at T$_{1}$:
\be\label{lambda_kappa_dot}
\lambda\sim\frac{\Gamma\kappa_{0}^2\cos\alpha}{4\pi \xi},\quad \frac{\tn{d}\kappa}{\tn{d}t}\sim\frac{\Gamma\kappa_{0}^3\sin\alpha\cos\alpha}{4\pi\, \xi^{2}}\,,\quad\tn{at}\,\,t=0.
\ee
while the $x$ and $z$ components of the locally uniform straining flow (\ref {flow_near_T_1})  maintain the orientation angle at $\alpha=\pi/4$.

For $t>0$, under the above circle-of-curvature assumption,  these results continue to hold with $s_{0}=s(0)$ and $\kappa_{0}$ simply replaced by $s(t)$ and $\kappa(t)$, for so long as  $\xi(t)\equiv\kappa(t) s(t)\ll1$; in effect, it is now the circles of curvature $C_{1}(t)$ at T$_{1}$ and $C_{2}(t)$ at T$_{2}$, each of radius  $\kappa(t)^{-1}$, that fulfil the role of
 the initial circles $C_1$ and $C_2$ of radius $\kappa(0)^{-1}$.  From (\ref{velocity_components}) and (\ref{lambda_kappa_dot}), we then have at leading order
\be\label{dynamical_system_1}
\frac{\tn{d} s}{\tn{d}{ t}}=-\frac{\kappa\Gamma}{4\pi}\left[\log\left(\frac{s}{\delta}\right)+\beta_{1}\right]\cos\alpha\,,\quad \frac{\tn{d} \kappa}{\tn{d}{ t}}=\frac{\Gamma\kappa\sin\alpha\cos\alpha}{4\pi {s}^2}\,,
\ee
and 
\be\label{lambda_circle}
\lambda(t)=\Gamma\kappa\cos\alpha/4\pi s\,,
\ee
these equations now holding for all $t\ge 0$, at least for so long as the condition $\xi(t)\ll 1$ is satisfied.  From (\ref{velocity_components_z}),  we also have
\be\label{eqn_z_decoupled}
\frac{\tn{d} z}{\tn{d} t}=\frac{\Gamma}{4\pi}\left[\frac{1}{s}
-\kappa\left\{\log \left(\frac{4}{\kappa^{2}s\delta}\right)+\beta\right\}\sin\alpha\right]\,.
\ee

We must also consider the important question as to how the vortex core in the neighourhood of T$_{1}$ responds to the rate of stretching  $\lambda(t)$ to which it is subjected.  Here, the model  described in \S \ref{Sec_time-dependent strain} is relevant, because, under the condition  $s(t)\kappa(t)\ll1$, the vortex ${\mathcal V}_{1}$ can be considered to be nearly rectilinear, even although it is the non-zero curvature of its twin vortex ${\mathcal V}_{2}$  that gives $\lambda(t)\ne 0$.  Of course, we have here the additional complication that the strain tensor  is non-axisymmetric, tending to distort the vortex core from circular form, but, as discussed in \S \ref{non-axisymmetric strain}, this effect is nullified at sufficiently large R$_\Gamma$.  
With $\lambda(t)$ now given by (\ref{lambda_circle}), eqn.~(\ref{gamma_delta}) gives
\be\label{gamma_delta2}
\frac{\tn{d} \,\delta^2}{\tn{d}{ t}}=\nu -\frac{\Gamma\kappa\cos\alpha}{4\pi {s}} \,\delta^{2},
\ee
Together with (\ref{dynamical_system_1}), we have thus arrived at a third-order dynamical system describing the possible collapse to a singularity.  If we non-dimensionalise with respect to length-scale $R=\kappa_{0}^{-1}$ and with dimensionless time $\tau=\Gamma\kappa_{0}^2\,t$, this system is
\be \label{system1}
\frac{\tn{d} s}{\tn{d}{ \tau}}=-\frac{\kappa\cos\alpha}{4\pi}\left[\log\left(\frac{s}{\delta}\right)+\beta_{1}\right] \,,\quad \frac{\tn{d} \kappa}{\tn{d}{\tau}}=\frac{\kappa\cos\alpha\sin\alpha}{4\pi {s}^2}\,,\quad \frac{\tn{d} \,\delta^2}{\tn{d}{\tau}}=\epsilon -\frac{\kappa\cos\alpha}{4\pi {s}} \,\delta^{2}\,,
\ee
where $\epsilon=\nu/\Gamma=R^{-1}_{\Gamma}$.  This `tip dynamical system' controls the evolution of  the tip variables $s(\tau),\kappa(\tau)$ and $\delta(\tau)$.  Equation (\ref{eqn_z_decoupled}) in dimensionless form
\be\label{eqn_z_decoupled_dim}
\frac{\tn{d} z}{\tn{d} \tau}=\frac{1}{4\pi}\left[\frac{1}{s}
-\kappa\left\{\log \left(\frac{4}{\kappa^{2}s\delta}\right)+\beta\right\}\sin\alpha\right]\,,
\ee
is decoupled from the system (\ref{system1}), and may be treated separately.

\section{An alternative initial vorticity field}
\begin{figure}
\begin{center}
\begin{minipage}{0.99\textwidth}
\includegraphics[width=0.40\textwidth, trim=0mm 0mm 0mm 0mm]{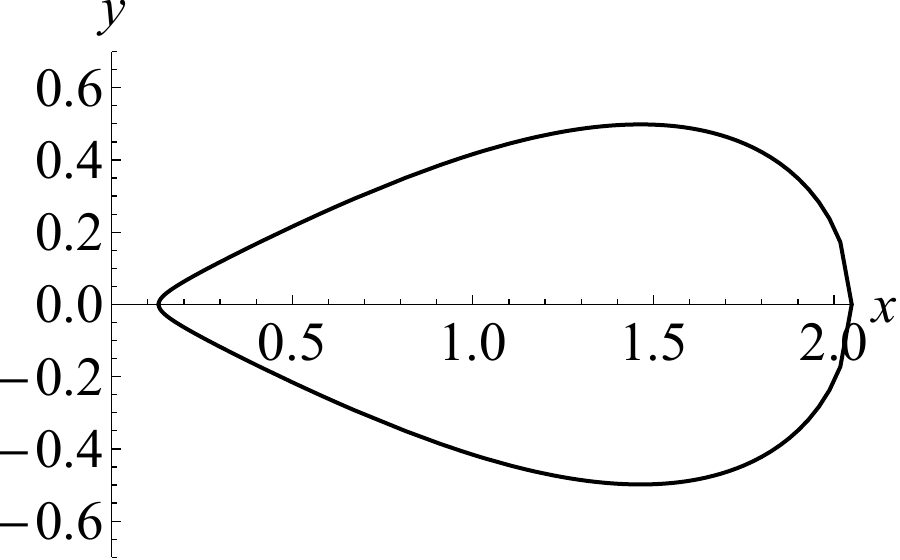}
\hspace*{30pt}
\includegraphics[width=0.45\textwidth,  trim=0mm 0mm 0mm 0mm]{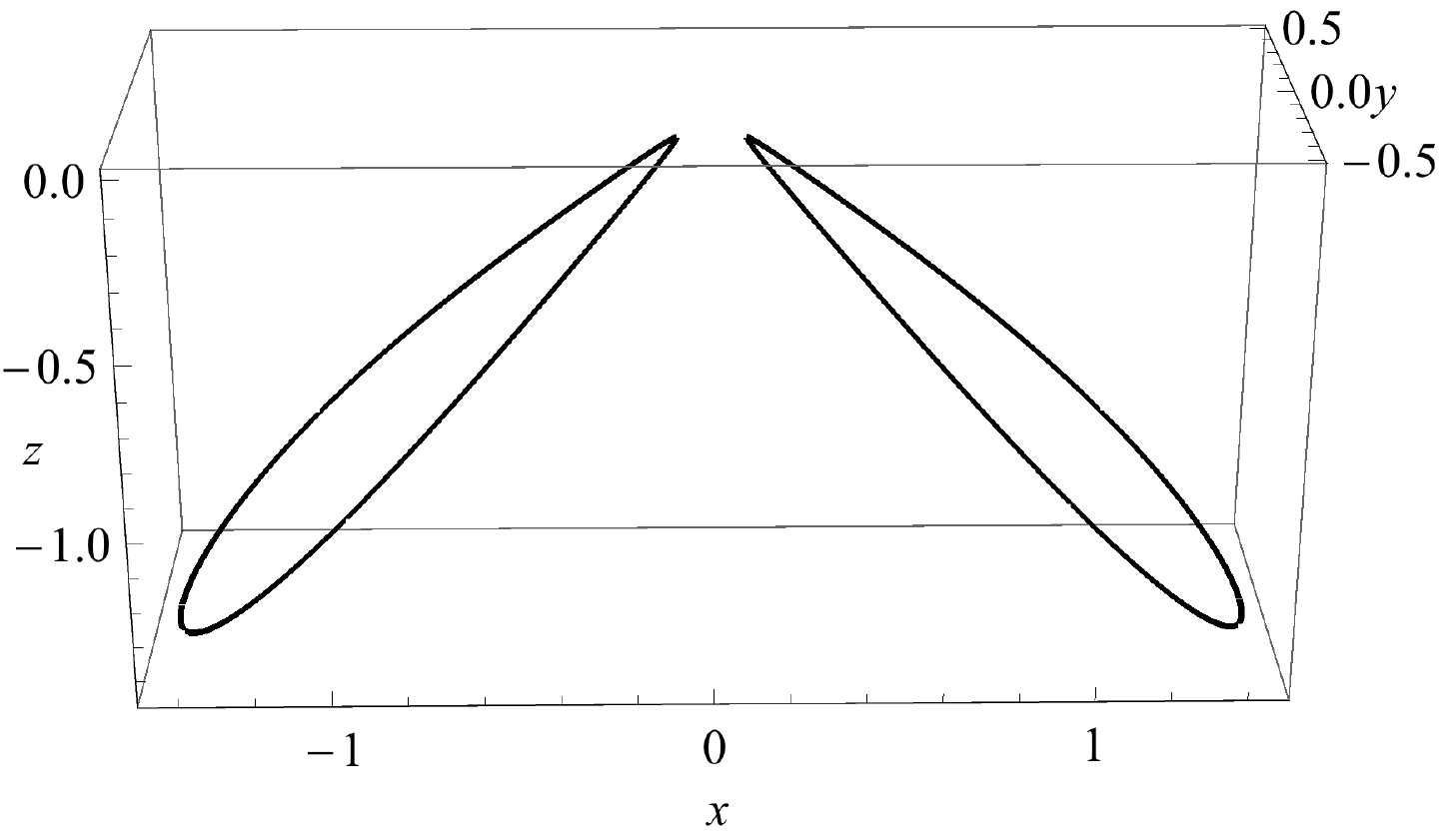}
\end{minipage}
\end{center}
\vskip 2mm
\hskip 30mm (a) \hskip 65mm (b) 
\vskip 2mm
\caption{(a) Ovoid (\ref{ovoid_eqn}) with $m=0.5, d=0.0707, \tn{and}\, m\kappa=25$; (b) corresponding curves $C'_{1}$ and $C'_{2}$ on planes inclined at angles $\pm \pi/4$. }
\label{Fig_Ovoids}
\end{figure}

 We note here that the assumed initial configuration of two circular vortices is not the only possibility.  We have  investigated also the situation when the initial state consists of two `ovoidal' vortices, as shown in figure \ref{Fig_Ovoids}.
The ovoid  in figure \ref{Fig_Ovoids}(a) is described by the equation
  \be\label{ovoid_eqn}
y^{2}= m^{2}(x-d)^{2}\left[1-m^{2}(x-d)^{2}\right]-(m\kappa)^{-2},\quad m\kappa>2\,,
 \ee
which intersects the $x$-axis at  points
\be \label{ovoid_zeros}
x=x_{1, 2}=d+m^{-1}\sqrt{\thalf\left(1\mp \sqrt{1 -4(m \kappa)^{-2}}\right)}\,.
\ee
Figure \ref{Fig_Ovoids}(b) shows two such ovoids symmetrically placed on planes inclined at angles $\alpha=\pm\pi/4$.  Near the tipping points, these ovoids are locally hyperbolic (as  investigated by \citealt{Kimura2018}).  The advantage here is that the initial separation $2s(0)$ and the initial tip curvature $\kappa(0)$ (non-dimensionalised by the maximum span of the ovoid in the 
$y$-direction) may be independently prescribed, i.e.~$\kappa(0)$ is not restricted to be unity.

However it is still only the circle of curvature at either tipping point that is relevant when 
$s(0)\!\ll\! 1$; so the equations  (\ref{system1}) and (\ref{eqn_z_decoupled_dim}) are still applicable, but now with this greater freedom in the initial conditions.

\section{Scaling properties of the tip dynamical system}\label{Sec_Leray_scaling}
The logarithmic term in (\ref{system1}), 
\be\label{A_def}
\Lambda=\log\left(s/\delta\right)+\beta_{1},
\ee
varies quite slowly, and some useful information can be gained by regarding it 
 as at least approximately constant and equal to its initial value, i.e.~ $\Lambda\approx\Lambda_{0}=\log\left(s_{0}/\delta_{0}\right)+\beta_{1}$. If then, anticipating a singularity at some time $\tau=\tau_{c}$\,, we further suppose that 
$s\sim (\tau_c-\tau)^{p}$ and $\kappa\sim (\tau_c-\tau)^{q}$, then the first two equations of
(\ref{system1})  require that $p-1 =q$ and $q-1=q-2p$, so that $p=1/2,\, q=-1/2$. Thus we have
\be\label{scalings_Leray}
s^{2}(\tau)=s_{0}^{2}(1-\tau/\tau_{c})\quad \tn{and}\quad\kappa^{-2}(\tau)=\kappa_{0}^{-2}(1-\tau/\tau_{c})\,,
\ee
and we recognise that, at least as far as the variables $s(\tau)$ and $\kappa(\tau)$ are concerned, this is none other than Leray scaling with  $\tn{d}s^{2}/\tn{d}\tau=-s_{0}^{2}/\tau_{c}$ and $\tn{d}\kappa^{-2}/\tn{d}\tau=-\kappa_{0}^{-2}/\tau_{c}$.   We shall describe this as `partial Leray scaling', because  the cross-sectional scale $\delta(\tau)$ does not conform to this scaling (see below).

We shall for the moment adopt the initial condition $\kappa(0)\equiv\kappa_{0}=1$. We may  then immediately estimate the singularity time $\tau_{c}$ from the fact that, from the second equation of 
(\ref{system1}), still assuming $\Lambda=$ const.,
\be\label{sing_time}
\tau_{c}\approx\frac{2\pi s_{0}^2}{\sin\alpha\cos\alpha}\,.
\ee
Substituting (\ref{scalings_Leray})  in the first of  (\ref{system1}), we find that this partial Leray scaling is achieved only if the initial conditions are such that
\be\label{in_cond_Leray}
s_{0}=\frac{\sin\alpha}{\Lambda}=\frac{\sin\alpha}{\log\left(s_{0}/\delta_{0}\right)+\beta_{1}}\,.
\ee
This determines the required value of $\delta_{0}$ to achieve this scaling:
\be\label{in_cond_Leray2}
\frac{\delta_{0}}{s_{0}}=\exp\left[\beta_{1}-\frac{\sin\alpha}{s_{0}}\right]\,.
\ee
With $\alpha=\pi/4$ and $s_{0}\ll 1$, this is an exceedingly small number. For example, if  $s_{0}=0.01$, then $\delta_{0}/s_{0}\approx 3.03879\times 10^{-31}$; if $s_{0}=0.1$ (about the largest value for which  the use of `low-$s$ asymptotics' is still just acceptable), it takes the more modest value $\delta_{0}/s_{0}\approx 1.32139 \times10^{-3}$. 

Turning  to the third equation of (\ref{system1}), this now takes the form
\be\label{delta_Leray_2}
\frac{\tn{d} \,\delta^2}{\tn{d}{\tau}}=\epsilon -\frac{\mu}{\tau_{c}-\tau} \,\delta^{2}\,,
\ee
where, with $\kappa_{0}=1$,
\be\label{mu_value}
 \mu=\frac{\tau_{c}}{s_0}\frac{\cos\alpha}{4\pi}=\frac{s_{0}}{2\sin{\alpha}} \,.
\ee
This is precisely as in \S \ref{Sec_time-dependent strain}, where we found a significant change of behaviour when  $\mu$ increases through the critical value  $\mu=1$; there, Leray scaling for $\delta$ was possible only for $\mu>1$ and for a particular choice of $\delta_{0}^2$.  But here, with $\alpha=\pi/4$ and $s_{0}\ll 1$, we are definitely in the range $\mu<1$, so Leray scaling for $\delta$ is not possible, and indeed the indications are that $\delta^2$ must plunge towards zero as $\tau\rightarrow\tau_{c}$ like $(1-\tau/\tau_{c})^{\mu}$ (cf. (\ref{delta_squared(t)})).  When $\alpha=\pi/4$, $\mu\approx 0.7071 s_{0}$, and this plunge is sharp and sudden.


We have yet to determine an appropriate value of $\epsilon$.  If we adopt the reasonable condition that 
$\tn{d} \,\delta^2/\tn{d}{\tau}$ be negative at $\tau=0$, this requires that
\be\label{epsilon_limit}
\epsilon< \epsilon_{c}=\frac{\cos\alpha}{4\pi}\left(\frac{\delta_{0}}{s_0}\right)^{2}s_{0}\,,
\ee
again an exceedingly small number.  With $\alpha=\pi/4$ and $s_{0}=0.01$ as above, $\epsilon_{c}=5.196\times10^{-65}$, the corresponding Reynolds number being $R_{\Gamma}\approx 2\times10^{64}$.  [This may seem unduly large, but  it is of course small compared with the value $R_{\Gamma}=\infty$ adopted in investigations that focus exclusively on Euler evolution!]

\begin{figure}
\begin{center}
\begin{minipage}{0.99\textwidth}
\includegraphics[width=0.45\textwidth, trim=0mm 0mm 0mm 0mm]{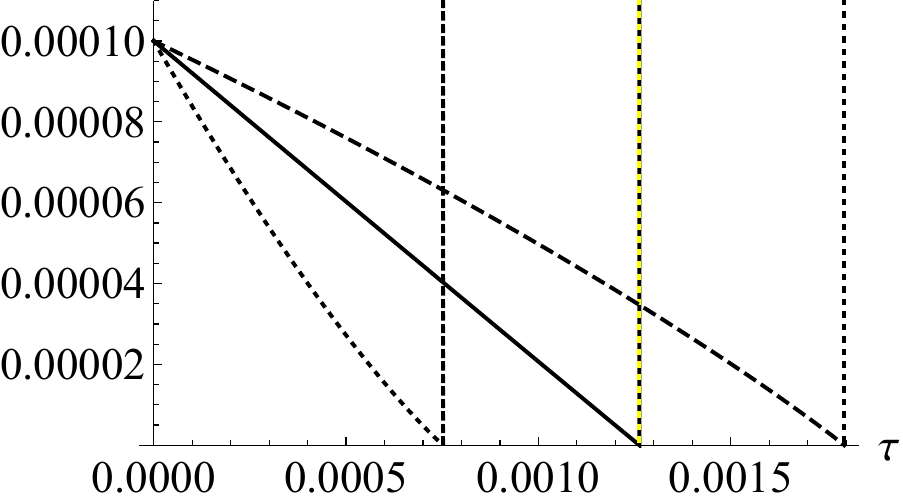}
\hspace*{20pt}
\includegraphics[width=0.45\textwidth,  trim=0mm 0mm 0mm 0mm]{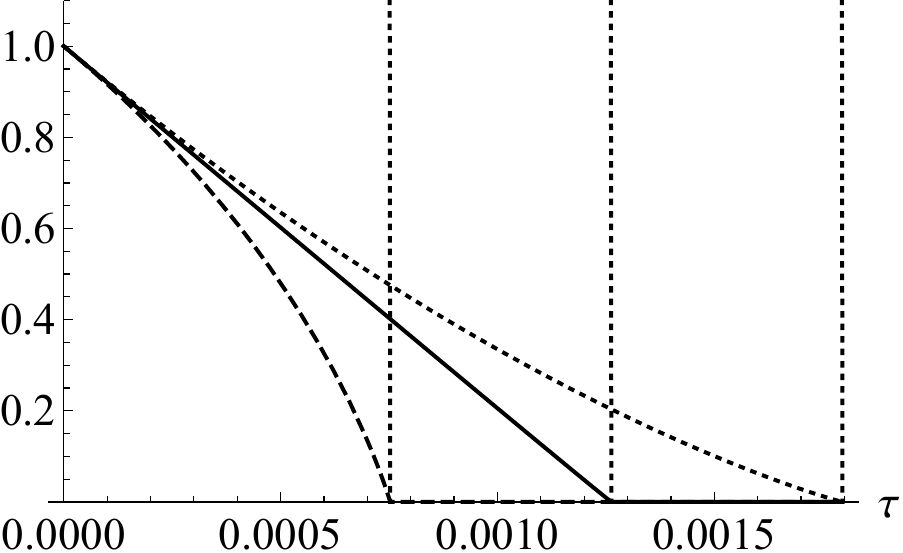}
\end{minipage}
\end{center}
\vskip 3mm
\hskip 30mm (a) \hskip 65mm (b) 
\vskip 1mm
\caption{(a) Curves of $s^2(\tau)$ for three initial values of $\delta_{0}$, with $s_{0}=0.01$ and $\epsilon\ll \epsilon_{c}$ in each case:  $\delta_{0}=\delta_{0c}=3.03879 \times 10^{-33}$ (solid),\,$10^{13}\delta_{0c}$ (dotted), and $10^{-33}\delta_{0c}$ (dashed); (b) corresponding curves of $\kappa^{-2}(\tau)$.}
\label{Fig_sd_Leray}
\end{figure}
\vskip 10mm
\begin{figure}
\begin{center}
\includegraphics[width=0.45\textwidth, trim=0mm 0mm 0mm 0mm]{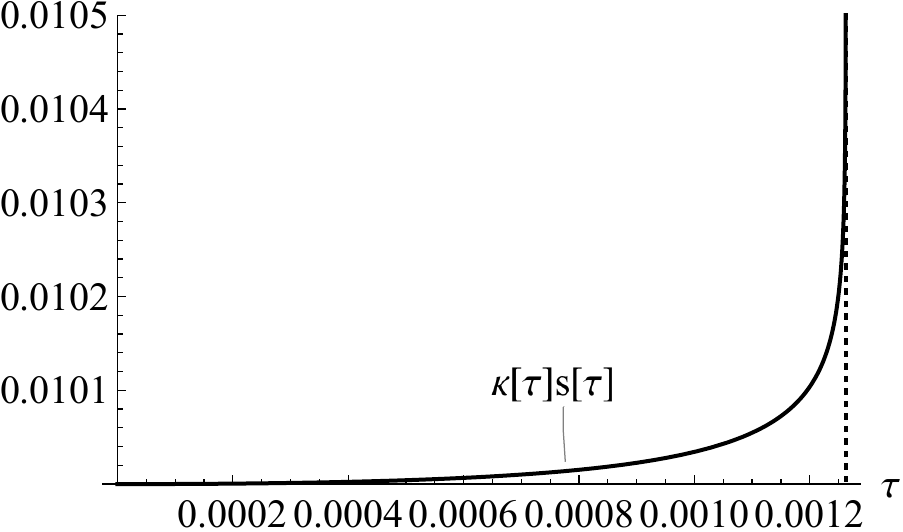}
\includegraphics[width=0.45\textwidth,  trim=0mm 0mm 0mm 0mm]{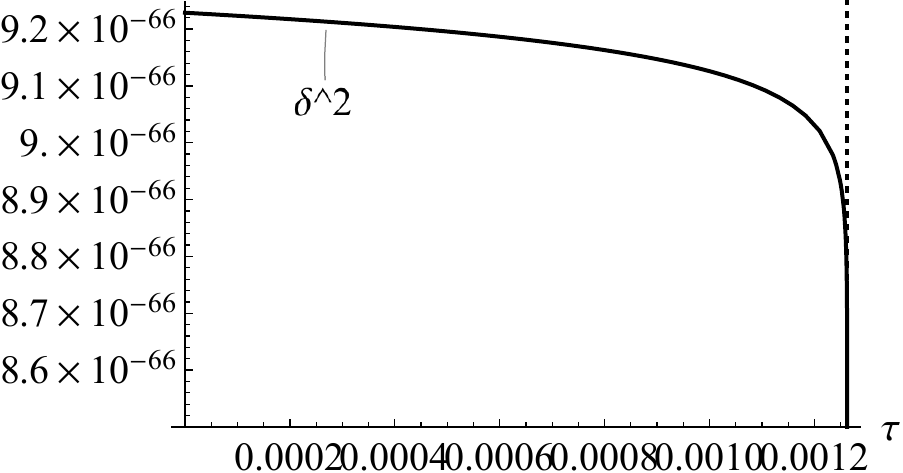}
\end{center}
\hskip 30mm (a) \hskip 65mm (b) 
\vskip 2mm
\caption{With $s_{0}=0.01$ as in figure \ref{Fig_sd_Leray}, this figure shows  (a) the product $\kappa(\tau)s(\tau)$ which would be constant under exact Leray scaling; the increase near  $\tau=\tau_{c}$ is due to the weak variation of the logarithmic factor $\Lambda$; it was not possible, within the computational resolution available, to determine the extent of this increase;  (b) the decrease of $\delta^{2}(\tau)$, showing the beginning of the plunge towards zero; here only an $8\%$ reduction is visible; again, within the computational limits, it was not possible to determine the true limiting behaviour.}
\label{Fig_sk_d_Leray}
\end{figure}
\section{Numerical treatment of the tip dynamical system}
First, we verify numerically the correctness of the partial Leray scaling found in \S \ref{Sec_Leray_scaling}.
Figure \ref{Fig_sd_Leray}(a) shows the function $s^{2}(\tau)$  found from numerical solution of (\ref{system1}); here, we have chosen $s_{0}=0.01$ so that the asymptotic results for $s_{0}\ll 1$ are reliable,
and in each case, we have chosen $\epsilon \ll \epsilon_{c}$, where $\epsilon_{c}$ is given by (\ref{epsilon_limit}). Actually, provided this condition is satisfied, the results are very insensitive to variation of $\epsilon$ down to the limit $\epsilon = 0$.  The black curve corresponds to the critical theoretical value $\delta_{0}=\delta_{0c}=3.03879 \times 10^{-31}s_{0}$ for Leray scaling, and it is indeed a straight line. It intersects the axis $s=0$ at $\tau=\tau_{c}=0.00126257$, which may be compared with the theoretical value $\tau_{c}=4\pi s_{0}^{2}=0.00125664$; the small difference is due to the weak variation of the parameter $\Lambda$.  The  dashed and dotted curves correspond to very different values of $\delta_{0}$ far removed from $\delta_{0c}$:  $\delta_{0}= 10^{13}\delta_{0c}$ and $10^{-33}\delta_{0c}$ respectively.  Here, the curvature, convex and concave respectively, indicates departure from partial Leray scaling. Figure \ref{Fig_sd_Leray}(b) shows corresponding curves for $\kappa^{-2}(\tau)$, again a straight line for $\delta_{0}=\delta_{0c}$.  Here, the departure from partial Leray scaling in the dashed and dotted curves appears more marked.

Even for $\delta_{0}=\delta_{0c}$, the partial Leray scaling is not perfect, as can be seen from the curve of  $\kappa(\tau)s(\tau)$ shown in figure \ref{Fig_sk_d_Leray}(a); this should be constant on the basis of exact Leray scaling, but it shows a significant increase near $\tau=\tau_{c}$; this is associated with the fact that $\Lambda$ is not exactly constant --- the logarithm varies significantly with the dramatic increase of $s/\delta$  as $\tau\rightarrow\tau_{c}$. This increase occurs when $\delta$ plunges towards zero;  the beginning of this plunge is shown in figure \ref{Fig_sk_d_Leray}(b) which shows an $\sim 8 \%$ decrease in $\delta^{2}$; it is difficult to visualise more than this, because, when  $s_{0}=0.01$, the plunge is so steep that it cannot be resolved numerically. For this reason it is necessary to examine more closely the behaviour of solutions of the system (\ref{system1}) as $\tau\rightarrow \tau_{c}$, as done in the following section.

\section{Asymptotics as $\tau\rightarrow \tau_{c}$}\label{Sec:Asymptotics_near_tau_c}
In the limiting situation $\tau_{c}-\tau \ll \tau_{c}$, we may assume that the term $\epsilon$ in the third equation of the system (\ref{system1}) is negligible compared with the term $\kappa\,\delta^{2}\cos\alpha/4\pi \,s$, which is after all responsible for the plunge towards zero.  The system then takes the asymptotic form
\be\label{system_asymptotic}
\frac{\tn{d} s}{\tn{d}{\hat{\tau}}}=-\kappa\Lambda_{1} \,,\quad \frac{\tn{d} \kappa}{\tn{d}{\hat{\tau}}}=\frac{\kappa\sin\alpha}{s^{2}}\,,\quad \frac{\tn{d} \,\delta^2}{\tn{d}{\hat{\tau}}}= -\frac{\kappa}{s} \,\delta^{2}\,,
\ee
where ${\hat{\tau}}=\tau\cos\alpha/4\pi$ and $\Lambda_{1}$ is the value of $\Lambda$ during this limit period.  Remarkably, this system admits the similarity solution
\be\label{exact_Leray}
s\sim s_{1}(1-\hat\tau/\hat{\tau}_{c})^{1/2}\,,\quad \kappa\sim \kappa_{1}(1-\hat\tau/\hat{\tau}_{c})^{-1/2}\,,\quad \delta \sim \delta_{1}(1-\hat\tau/\hat{\tau}_{c})^{1/2}\,,
\ee
with $\Lambda_{1}=\log[s_{1}/\delta_{1}]+\beta_{1}$. In fact, the three equations of 
(\ref{system1}) then give respectively
\be
\frac{s_{1}}{2\hat{\tau}_{c}}=\kappa_{1}\Lambda_{1},\quad \frac{1}{2\hat{\tau}_{c}}=
\frac{\sin\alpha}{s_{1}^{2}},\quad -\frac{\delta_{1}^2}{\hat{\tau}_{c}}=-\frac{\kappa_{1}\delta_{1}^2}{s_{1}},
\ee
from which we deduce
\be
\hat{\tau}_{c}=\frac{s_{1}}{\kappa_1}=\frac{s_{1}^2}{2\sin\alpha},
\ee
and so, with $\alpha=\pi/4$,
\be
s_{1}\kappa_1=2\sin\alpha=\surd{2}.
\ee
It follows also that 
\be\label{limiting_gaussian}
\Lambda_{1}\equiv\log\left[\frac{s_{1}}{\delta_{1}}+\beta_{1}\right]=\frac{\sin\alpha}{\kappa_{1}s_1}=\frac{1}{2},\quad\tn{so}\quad \delta_{1}/s_{1}
=\tn{e}^{-\left[0.5\,-\,\beta_{1}\right]}= 0.943367.
\ee
This is therefore the limiting value of $\delta(\tau)/s(\tau)$ at $\tau=\tau_c$ irrespective of the initial conditions at $\tau=0$.

It is apparent therefore that there are two stages in the collapse process: first a stage during which $\kappa(\tau)$ and $s(\tau)$ are given by (\ref{scalings_Leray}),  and
\be
\delta^{2}(\tau)\approx\delta^{2}(0) (1-\tau/\tau_{c})^{\mu},\quad\tn{with}\quad\mu=s_{0}/2\sin\alpha.
\ee 
This stage ends when $\delta^{2}\sim s^{2}\approx s_{0}^2(1-\tau/\tau_{c})$.  Using (\ref{sing_time}), this gives
\be\label{stage_one_ends}
\tau_{c}-\tau \sim \frac{2\pi s_{0}^2}{\sin\alpha\cos\alpha}\exp{\left[\frac{2}{1-\mu}\left(\beta_{1}-\frac{\sin\alpha}{s_{0}}\right)\right]}=\tau_{1}, \,\,\,\tn{say}.
\ee
At this `changeover' stage, exact Leray scaling as described by (\ref{exact_Leray}) becomes established and persists until the singularity time $\tau=\tau_{c}$.  With $s_{0}=0.01$, (\ref{stage_one_ends}) gives $\tau_{1}=2.49300\times10^{-61}\tau_{c}$; it is easy to see why computation fails to resolve this changeover.

\section{Tipping point trajectories}\label{Tipping point trajectories}
\begin{figure}
\begin{center}
\begin{minipage}{0.99\textwidth}
\includegraphics[width=0.30\textwidth, trim=0mm 0mm 0mm 0mm]{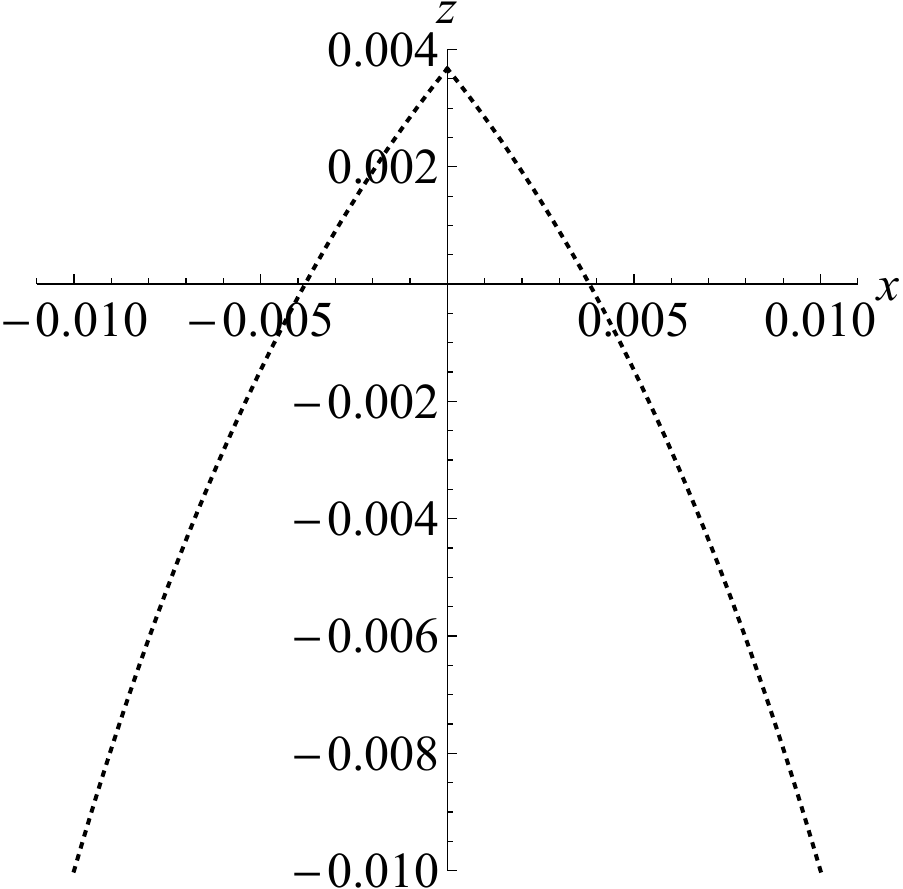}
\hspace*{12pt}
\includegraphics[width=0.30\textwidth,  trim=0mm 0mm 0mm 0mm]{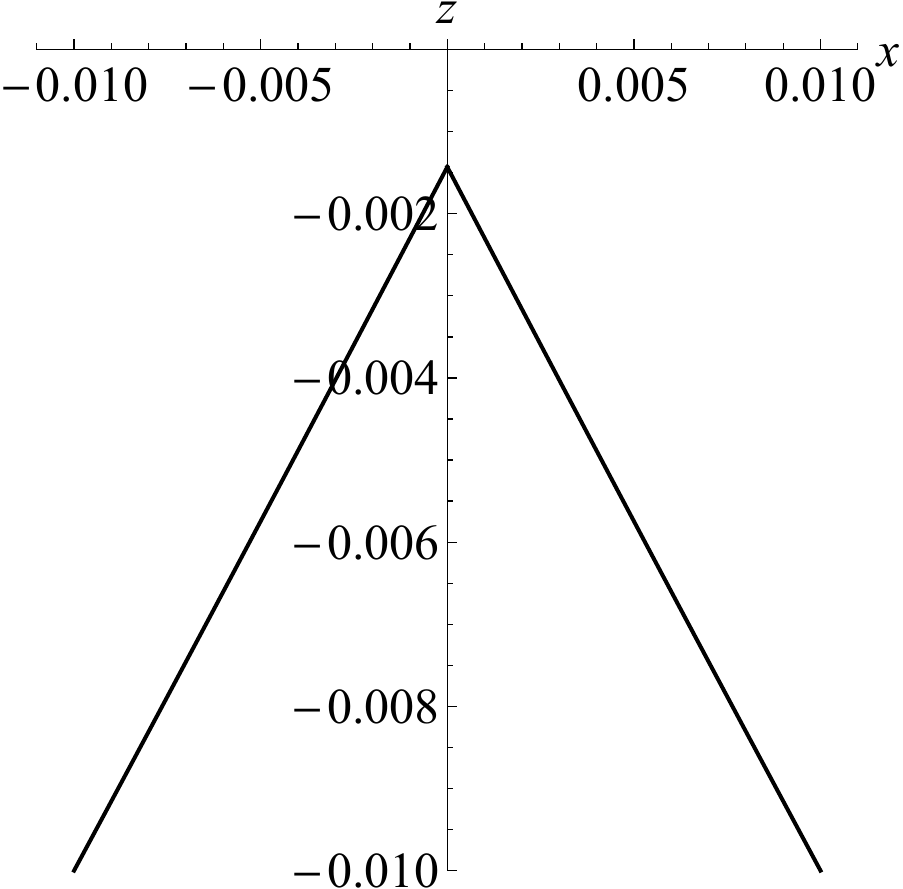}
\hspace*{12pt}
\includegraphics[width=0.30\textwidth,  trim=0mm 0mm 0mm 0mm]{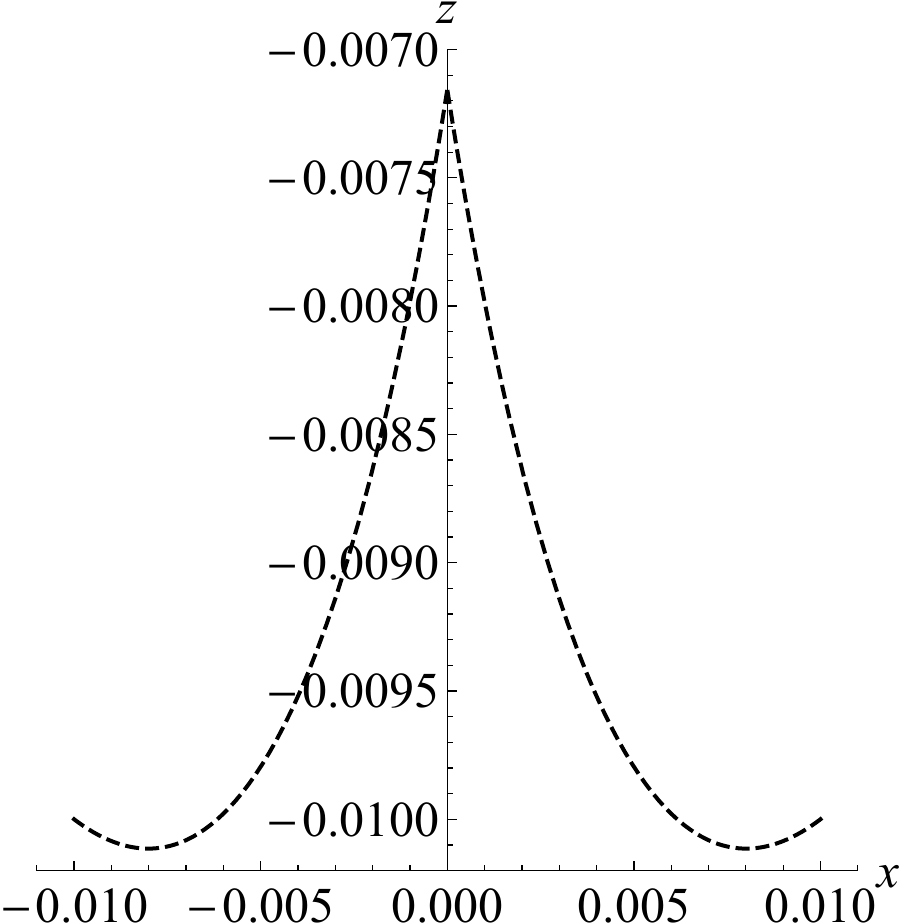}
\\[-1.8pt]
\end{minipage}
\end{center}
\hskip 20mm (a) \hskip 40mm (b) \hskip 40mm (c) 
\vskip 0mm
\caption{Tipping point trajectories: in each case $\alpha=\pi/4,\, s_{0}=0.01$, $\epsilon\ll \epsilon_{c}$,  and the trajectories start from $(\mp\, 0.01,\,-0,01)$; (a) $\delta_{0}=3.03879 \times10^{-20}$; (b) the critical case of Leray scaling $\delta_{0}=3.03879\times 10^{-33}$ when the trajectories are almost straight lines, with gradients $\sim\!\pm0.85$; (c) $\delta_{0}=3.03879 \times10^{-66}$; this is so small that the condition (\ref{delta_second_constraint}) is not satisfied, and the tipping points initially move downwards; from (\ref{eqn_z_decoupled_dim}), there is a  turning point in this case (where $\tn{d}z/\tn{d}\tau=0$)  at $\tau/\tau_{c}\approx 0.31$.}
\label{Fig_traj}
\end{figure}

When $s$ and $\kappa$  are  given by the partial-Leray-scaling results (\ref{scalings_Leray}) and (\ref{in_cond_Leray2}),  the motion of the tipping points $(\tilde{x},\tilde{z})$ are determined from (\ref{scalings_Leray}) and (\ref{eqn_z_decoupled_dim}) by \footnote{This speed variation is such as to conveniently overcome Zeno's paradox ``There is no motion, because that which is moved must arrive at the middle before it arrives at the end, and so \emph{ad infinitum}." (Aristotle \emph {Physics} 239b 11-13)}
\be
\frac{\tn{d}\tilde{x}}{\tn{d}\tau}=\pm\frac{s_{0}}{2\tau_{c}}\left(1-\frac{\tau}{\tau_{c}}\right)^{-1/2}\,,\quad
\frac{\tn{d} \tilde{z}}{\tn{d} \tau}=\frac{1}{4\pi}\left[\frac{1}{s}
-\kappa\left\{\log \left(\frac{4}{\kappa^{2}s\delta}\right)+\beta\right\}\sin\alpha\right]\,.
\ee
Hence, with $\alpha=\pi/4$, 
\be
\frac{\tn{d}\tilde{z}}{\tn{d}\tilde{x}}=\pm \frac{\tau_{c}}{2\pi s_{0}^{2}}\left(1-\Lambda_{2}\,\kappa_{0}s_{0}\,\sin\alpha\right)=\pm( 2-\surd{2}\,\kappa_{0}s_{0}\,\Lambda_{2})\,,
\ee
where, after simplification using (\ref{in_cond_Leray2}), 
\be
\Lambda_{2}=\log [4/\kappa^{2}s\,\delta]+\beta =\sin\alpha/s_{0}-2\log \kappa_{0}s_{0}+4\log{2}-1.
\ee
Hence the tipping points which start from $(\mp s_{0},\,s_{0})$ move on the straight lines 
\be
\tilde{z}+s_{0}=\pm\left( 2-\surd{2}\,\kappa_{0}s_{0}\,\Lambda_{2}\right)(\tilde{x}\pm s_{0}),
\ee 
which intersect at time $\tau=\tau_{c}$ at the `singularity point'
\be
(\tilde{x}_{s},\,\tilde{z}_{s})=\left(0,\, -s_{0}+s_{0}\left(2-\surd{2}\,\kappa_{0}s_{0}\, \Lambda_{2}\right)\right)\,.
\ee
For $s=0.01$, this evaluates to $\tilde{z}_{s}=-0.0015532$, and the straight-line trajectories meet at an angle of $131.85^{\scriptsize{\tn{{o}}}}$. (Note that this angle of intersection is not $\pi/2$, and there is no contradiction here because the tipping point trajectories are quite distinct from the projection at any time $\tau \le \tau_{c}$ of $C_{1}$ and $C_{2}$ on the plane $y=0$; this is related to the distinction between Lagrangian and Eulerian descriptions.)

When the initial conditions are not compatible with partial Leray scaling, the trajectories can be determined by solving (\ref{system1}) together with (\ref{eqn_z_decoupled_dim}), and constructing parametric plots of $\left(\pm\tilde{x}(\tau),\, \tilde{z}(\tau)\right)$.  The results corresponding to the three cases of figure \ref{Fig_sd_Leray} are shown in figure \ref{Fig_traj}. As expected from the above argument, the trajectories for case (b) of partial Leray scaling $(\delta_{0}=3.03879\times 10^{-33}=\delta_{0c})$  are indeed straight lines.  For case (a) 
($\delta_{0}=10^{13}\delta_{0c}$),  the curves are significantly curved, but still meet the axis at a finite angle.  For case (c) ($\delta_{0}=10^{-33}\delta_{0c}$), they are even more curved, and they initially slope downwards because the condition (\ref{delta_second_constraint}) is not satisfied; however there is a turning point when $\tn{d}z/\tn{d}\tau=0$ (here at $\tau/\tau_{c}\approx 0.31$, and they then move upwards, meeting the $z$-axis at a smaller nonzero angle when $\tau=\tau_c$.

\begin{figure}
\vspace*{12pt}
\begin{center}
\begin{minipage}{0.99\textwidth}
\includegraphics[height=0.17\textheight,width=0.47\textwidth]{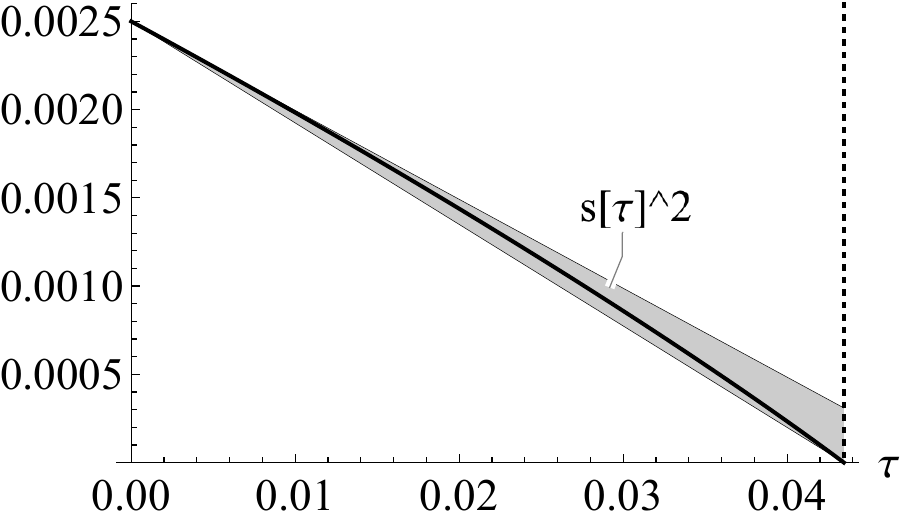}
\hspace*{10pt}
\includegraphics[height=0.17\textheight,width=0.47\textwidth]{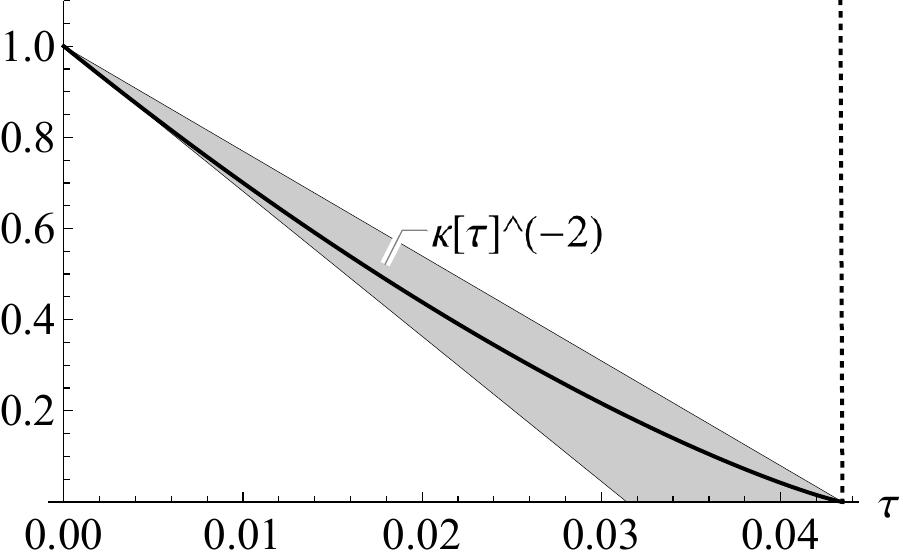}
\\
\vskip -1mm
\hskip 30mm (a) \hskip 65mm (b) 
\vskip 3mm
\includegraphics[height=0.17\textheight,width=0.47\textwidth]{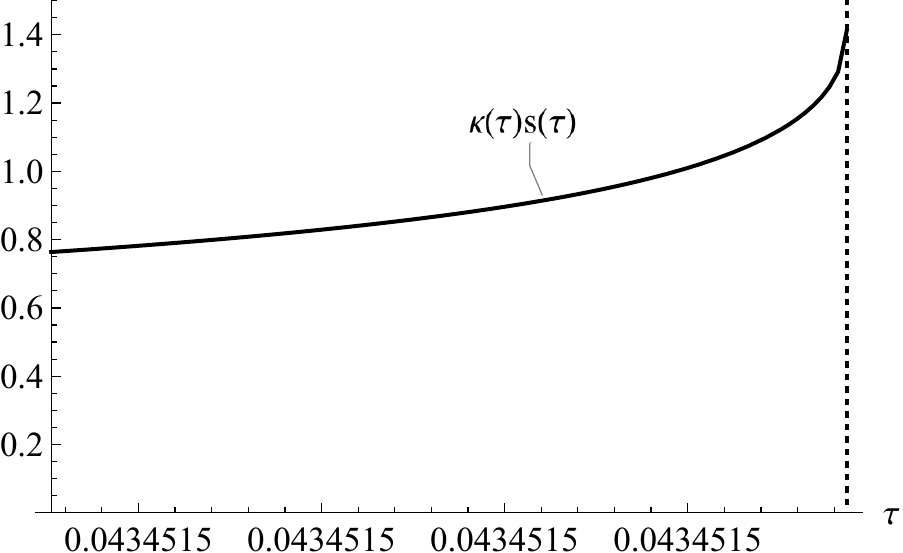}
\hspace*{2.80pt}
\includegraphics[height=0.17\textheight,width=0.4886\textwidth]{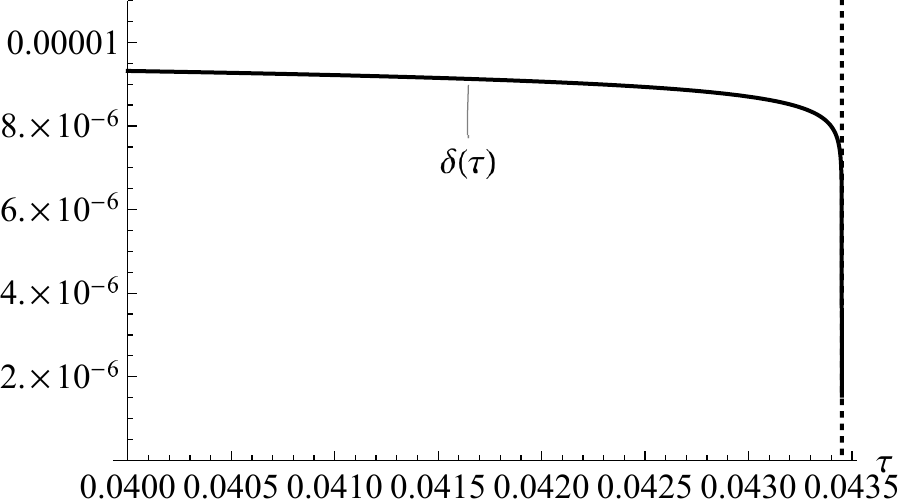}
\\
\vskip -1mm
\hskip 30mm (c) \hskip 65mm (d) 
\vskip 3mm
\includegraphics[height=0.17\textheight,width=0.47\textwidth]{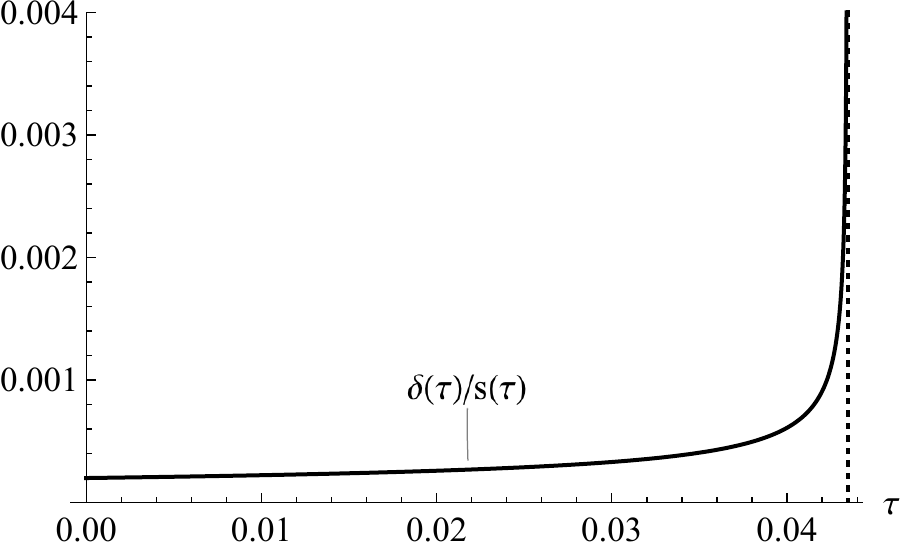}
\hspace*{-2.6pt}
\includegraphics[height=0.17\textheight,width=0.502\textwidth]{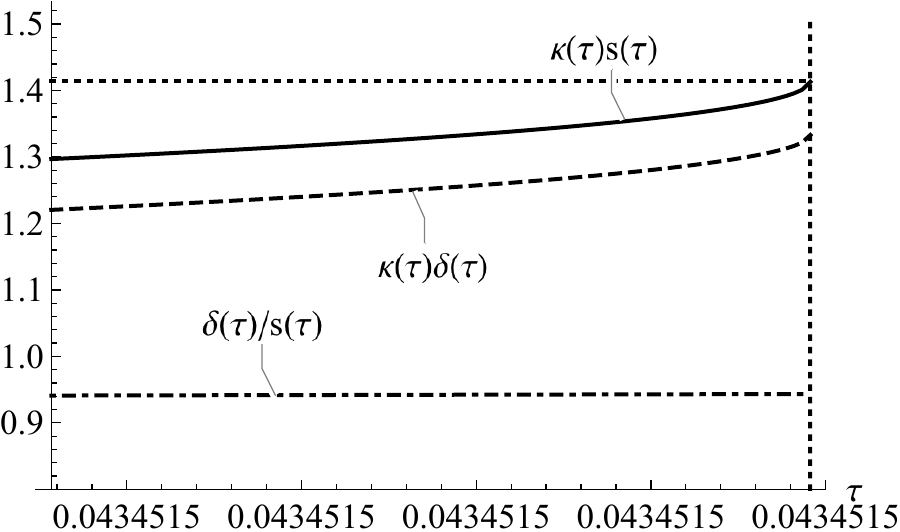}
\vskip 1mm
\hskip 30mm (e) \hskip 65mm (f) 
\vskip 2mm
\end{minipage}
\end{center}
\vspace*{4pt}
\caption{Numerical solution of the system (\ref{system1}) for the vortex separation parameter $s(\tau)$, core size $\delta(\tau)$, and curvature $\kappa(\tau)$ for $0\!<\!\tau\!<\!\tau_c\!\approx\! 0.0434515450$ (marked in all panels by the vertical dotted line); $s(0)\!=\!0.05,\, \delta(0)\!=\!10^{-5}$, $\kappa(0)\!=\!1, \epsilon=10^{-20}$; 
(a) $s(\tau)^2$ (solid curve) decreases to zero, with small negative curvature; (b) $\kappa(\tau)^{-2}$ decreases to zero, with significant positive curvature;
(c) $\kappa(\tau)s(\tau)\equiv\xi(\tau)$ increases near $\tau=\tau_{c}$ but is actually bounded (see panel (f)); 
(d) $\delta(\tau)$ initially decreases slowly but plunges towards zero for $\tau\gtrsim 0.04$; 
(e) $\delta(\tau)/s(\tau)$ increases rapidly near $\tau=\tau_{c}$, but remains bounded;  
(f) evolution of $\kappa(\tau) s(\tau)$ (solid),  $\kappa(\tau) \delta(\tau)$ (dashed) and 
$ \delta(\tau)/ s(\tau)$ (dash-dotted)  in the interval $(1-10^{-11})\, \tau_{c}\!<\!\tau\!<\!\tau_{c}$ extremely close to $\tau_{c}$; the level $\!\surd{2}$ is indicated by the horizontal dotted line. The limiting values  are $\kappa(\tau_{c}) s(\tau_{c})=1.414213,\, \kappa(\tau_{c}) \delta(\tau_{c})=1.334122, \, \delta(\tau_{c})/ s(\tau_{c})=0.943367$.}
\label{Fig_Evolution}
\end{figure}

\section{Evolution towards a singularity}\label{Sec_Evolution_sing}
With the choice $s(0)=0.05$, (\ref{stage_one_ends}) gives the more reasonable value 
$\tau_{1}=3.31652\times10^{-12}$; as pointed out earlier, this is within the range where it is still reasonable to use the results of the low-$s$ asymptotics leading to the system (\ref{system1}).  With this choice, the behaviour at the changeover time and as $\tau\rightarrow\tau_{c}$ can be resolved numerically.   Figure \ref{Fig_Evolution}  shows results of  numerical integration of the system (\ref{system1}) when $s(0)=0.05$. We also take $\delta(0)=10^{-5}$, deliberately much greater than the value $\sim 5.6\times 10^{-8}$ for which partial Leray scaling would apply. 

The first panel (a) shows a decrease to zero of $s(\tau)^{2}$, almost linear with concave curvature as for the dashed curve in figure (\ref{Fig_sd_Leray})(a); this falls to zero at $\tau=\tau_{c}\approx 0.043\,451\,545$, the singularity time marked by the vertical dotted line (the line that cannot be crossed).  It lies entirely in the  shaded area between two straight lines, the upper one being tangent to the curve at $\tau=0$. \, Panel (b) shows $\kappa(\tau)^{-2}$, which similarly decreases to zero,  the departure from linear behaviour being much more marked.  

Panel (c) shows the product $\kappa(\tau) s(\tau)$, which  increases slowly over much of the range but rapidly near $\tau=\tau_{c}$.\,  Panel (d) shows $\delta(\tau)$, which decreases slowly over much of the range, but plunges towards zero for $\tau\gtrsim 0.0433$. \,  Panel (e) shows $\delta(\tau)/s(\tau)$ which increases very slowly over much of the range, but rapidly for $\tau\gtrsim 0.04$. \, Panel (f) shows $\kappa(\tau) s(\tau)$, $\kappa(\tau) \delta(\tau)$ and $\delta(\tau)/s(\tau)$ in an extremely small left-neighbourhood $(1\!-\!10^{-11})\tau_{c}<\tau\le\tau_{c}$ of  $\tau_{c}$, where each of these quantities reaches a limiting value ($1.414\,213,\, 1.334\,122$ and $0.943\,367$ respectively) in perfect agreement with the values obtained from the local Leray similarity solution obtained above in \S\ref{Sec:Asymptotics_near_tau_c}. 

Note that, since $\xi(\tau)\equiv\kappa(\tau)s(\tau)$ increases with increasing $\tau$, the ratio $\lambda_3/\lambda_2$ as given by (\ref{lambda_ratio}) decreases, so the condition (\ref{strain-field_parameter}), $\epsilon_1\equiv c\epsilon \ll 1$,  is satisfied for all $\tau$ if satisfied at $\tau=0$.

\section{The singularity question}\label{Sec_Singularity}
Although it is clear from the foregoing analysis that we have a Biot-Savart singularity, this cannot  be claimed  as a Navier-Stokes singularity, because  the apparent limiting value of the ratio $\delta(\tau_{c})/s (\tau_{c})\approx 0.943\,367$, although less than unity, is inconveniently large for validity of the Biot-Savart description.  The ratio 
$\delta/s $ increases inexorably from its initial low level  ($0.0002$  in figure \ref{Fig_Evolution}(e)), and the Biot-Savart description is reasonable only for so long as 
$\delta/s \lesssim 0.1$, say.  For the initial conditions $ s(0)=0.05, \delta(0)=10^{-5}, \kappa(0)=1$, as in figure \ref{Fig_Evolution}, the value $\delta/s=0.1$  is attained at $\tau=\tau_{0.1}\approx .999\,999 \tau_{c}$ (actually $\delta(\tau_{0.1})/s (\tau_{0.1})\approx 0.099$), and presumably some detectable reconnection of the vortices begins soon after this stage is reached, as anticipated in the `tent model' of \cite{Kimura2018}.  At this stage, some of the circulation $\pm\Gamma$  in the vortices is stripped away in reconnected vorticity flux; it will then be necessary to consider in detail the nature of the reconnection process.

\section{An Euler singularity}\label{Sec_Euler_limit}
The above formalism applies equally to evolution under the Euler equations for which we 
 simply set $\nu=0$ (equivalently $\epsilon=0$).
Here, we may for definiteness suppose that the vorticity $\omega$ is uniform over the cross-section of the vortex, so that $\pi\,\delta^2 \,\omega=\Gamma=$ const.  For this vorticity distribution, the parameters $\beta$ and $\beta_{1}$ take the modified values 
\be
\beta=1.136,\quad \beta_{1}=0.750.
\ee
The asymptotic analysis of \S \ref{Sec:Asymptotics_near_tau_c} still applies, but now
\be
\delta_{1}/s_{1}=\tn{e}^{-[0.5-\beta_{1}]}\approx 1.28403.
\ee
\begin{figure}
\begin{center}
\includegraphics[width=0.45\textwidth, trim=0mm 0mm 0mm 0mm]{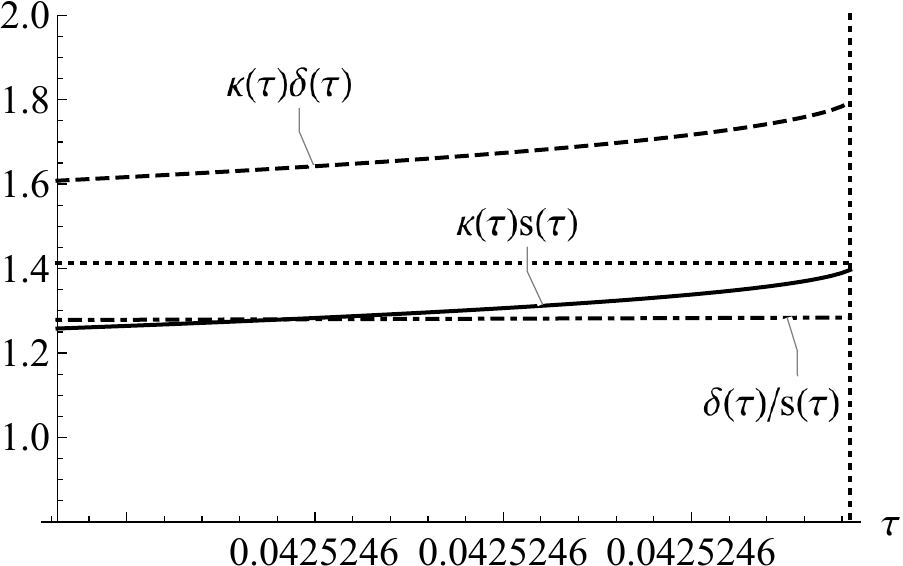}
\hskip 10mm
\includegraphics[width=0.3\textwidth, trim=0mm 0mm 0mm 0mm]{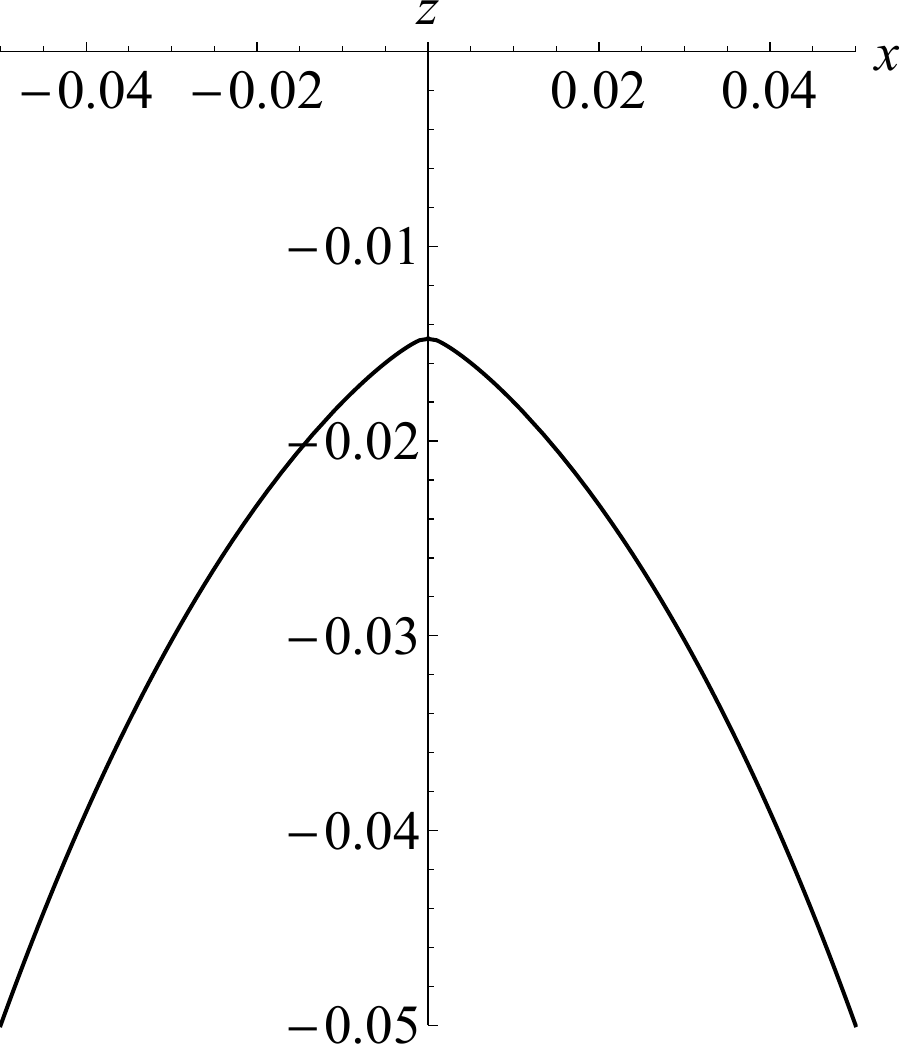}\\
\vskip 2mm
\hskip 5mm (a) \hskip 60 mm (b)
\end{center}
\caption{Euler evolution: (a) asymptotic behaviour of  $\kappa(\tau)s(\tau),\,\,\kappa(\tau)\delta(\tau)$ and $\delta(\tau)/s(\tau)$ as $\tau\rightarrow \tau_{c}\approx 0.042\,524\,599\,087$, for $(1-10^{-11})\tau_{c}<\tau<\tau_{c}$, computed from (\ref{system1}) with $\beta=1.136,\, \beta _{1}=0.750$, and with initial conditions $s(0)=0.05,\,\delta(0)=10^{-5}, \kappa(0)=1 $, and with $\alpha=\pi/4$ and $\epsilon = 0$; the abscissa is not resolved; (b) corresponding trajectories of the tipping points.}
\label{Fig_Euler_Asymp}
\end{figure}
Computation confirms this asymptotic value (figure \ref{Fig_Euler_Asymp}) with reasonable accuracy; in fact, the computed limiting values are $\kappa(\tau_{c})s(\tau_{c})=1.40608, \kappa(\tau_{c})\delta(\tau_{c})=1.80543, \delta(\tau_{c})s(\tau_{c})=1.28401$.
\begin{figure}
\begin{center}
\includegraphics[width=0.50\textwidth, trim=0mm 0mm 0mm 0mm]{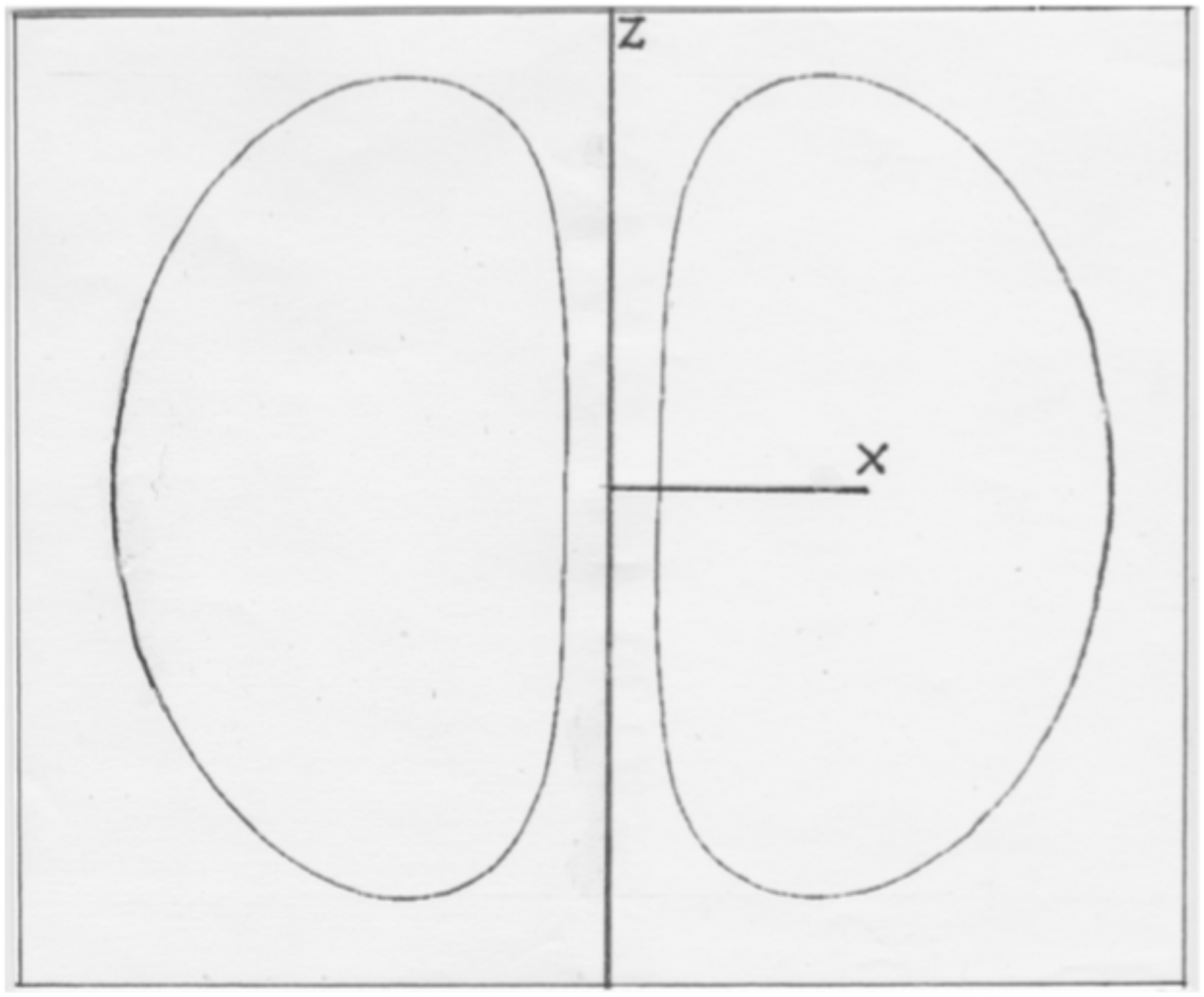}
\end{center}
\caption{Vortex pair  at parameter value $\delta/s =1.22$; $\delta$ is defined so that the enclosed area is $\pi\delta^2$, and $2s$ is the distance between the two centres of vorticity.  [Adapted from \citealt{Pierrehumbert1980}.]}
\label{Fig_vortex_pair}
\end{figure}

The strain field at the tipping point T$_{1}$ is still given by (\ref{e_ij_leading_order}), with  eigenvalues (\ref{eigenvalues}). This strain field initially distorts the vortex core at T$_{1}$  to elliptic form, with major axis in the z-direction.  As the two vortices approach, they do not remain elliptic, because the strain field in the neighbourhood of each vortex is no longer uniform.  The analogous two-dimensional steady problem was solved by  \cite{Pierrehumbert1980} (with a small correction by \citealt{Saffman1982}), who showed how the shape changes as the parameter $\delta/s$ increases.  The limiting value for steady flow is  $\delta/s =2.16$, corresponding to the situation when the vortices are in contact on the plane $x=0$. Our critical value  
$\delta(\tau_{c})/s  (\tau_{c}) \sim 1.28$  gives a vortex pair whose boundaries are  close to that corresponding to the value $\delta/s  = 1.22$ shown in figure \ref{Fig_vortex_pair}; this is one of the values for which the shape was computed by \cite{Pierrehumbert1980}; for our problem, this value is attained when $\tau\approx (1-1.5\times 10^{-10}) \,\tau_{c}$.  Our vortices are of course curved out of the plane $y=0$; but we conjecture that they will evolve until $\delta/s \approx 1.28$ and  will then continue to shrink at this fixed ratio to a singularity at the finite time $\tau=\tau_{c}$. This is on the reasonable assumption that the equations (\ref{system1}) do continue to hold during this final stage, reasonable because the essential mechanisms of reduction of $s(\tau)$, increase of  
$\kappa(\tau)$ and decrease of $\delta(\tau)$, all associated with the curvature of the vortex pair in the $y$-direction, continue to operate;  of course it is the volume form of the Biot-Savart law (rather than its line-integral form) that should be used in order to confirm this.  Nevertheless, our claim is that this establishes a clear route to a possible finite-time Euler singularity.

We note that \cite{Habibah2018} have recently developed an asymptotic theory for vortex-pair propagation in powers of the parameter $\hat\epsilon=\delta/s $, and have shown that the speed of propagation of the pair is given by the beautifully simple formula
$U=\Gamma/4\pi s +\hat{\epsilon}^{2}Q_{2}/2 s ^{3}$, where $Q_{2}$ is the strength of the quadrupole at O$(\hat{\epsilon}^{2})$ associated with the elliptical deformation of each vortex;  in other words, the speed is slightly increased as a result of this deformation. They show streamlines and vorticity contours for $\hat{\epsilon}=0.3$, which, even for this relatively low value, exhibit a slight flattening of each vortex on the side near to the central plane.

\section{Discussion and conclusions}\label{Conclusions}
The analysis of this paper has built on two key ideas developed in \S 2:  first, that a Burgers-type  stretched vortex can exhibit a finite-time singularity despite the smoothing action of viscosity if the imposed rate of stretching is proportional to $(t_{c}-t)^{-1}$, where $t_{c}$ is an artificially imposed singularity time;  and second, that the Gaussian core of such a vortex can resist deformation and remain compact provided the vortex Reynolds number $R_{\Gamma}=\Gamma/\nu$ is large enough.

On this basis, we have analysed the evolution of two initially circular vortices of radius $R$ propagating towards each other on planes tilted at angles $\pm \alpha$ to the symmetry plane $x=0$, at  $R_{\Gamma}\equiv \epsilon ^{-1}\gg 1$.  We assumed that 
$0<\delta\ll s \ll R\equiv\kappa_{0}^{-1}$, where $2s$ is the separation of the `tipping points' of the vortices and $\delta$  the initial scale of the Gaussian vortex cores; under these circumstances, we obtained an exact expression for the velocity field induced by either vortex, and found its asymptotic behaviour (for small $\xi\equiv \kappa_{0} s$) near the tipping points.  We then argued that, for $t>0$, the local evolution is governed by the separation $s(\tau)$,\,tip curvature $\kappa(\tau)$ and core radius $\delta(\tau)$ as these develop in dimensionless time $\tau=(\Gamma/R^2)t$, and we showed that the angle $\alpha$ tends to $\pi/4$ under the action of the induced strain field near each tipping point. In this way, we derived the dynamical system (\ref{system1}).  The solution showed an approach to a Biot-Savart singularity at a finite time $\tau_c$ as shown in figure \ref{Fig_Evolution}, with approximate `partial Leray scaling' of the variables $s(\tau)$ and $\kappa(\tau)$, but with a breakdown of the assumptions $\delta\ll s\ll \kappa^{-1}$ in the very final stage of evolution. Conditions for partial Leray scaling  were determined, and it was shown that similar scaling for $\delta(\tau)$ is not possible for so long as $\kappa(\tau)s(\tau)\ll 1$.  However,  exact full Leray scaling is achieved in the asymptotic limit $\tau\rightarrow\tau_c$, as described in \S10.  The trajectories of the tipping points were determined in \S11; they meet at a finite angle at the point of singularity.

In \S \ref{Sec_Euler_limit}, we considered the the Euler flow situation ($\epsilon=0$), drawing on  results of \cite{Pierrehumbert1980} concerning vortex-pair evolution. Although the vorticity is now spread uniformly over a finite vortex core, we argue that the dynamical system (\ref{system1}) remains valid (possibly with change of numerical coefficients) with $2s$ now interpreted as the distance between the two centres of vorticity, and $\pi \delta^2$ as the area of each `vortex patch'.  The same mechanisms that decrease $s$ and $\delta$ and increase $\kappa$ (due to the curvature of the vortices in the $y$-direction) are still present, and this justifies our conjecture that this system will indeed collapse to an Euler singularity.    

Finally, in \S 15, we have calculated explicitly the rate of viscous reconnection on the symmetry plane $x=0$, and have shown that, although the tip vorticity can increase by an arbitrarily large factor within a finite time, reconnection ultimately drives the `surviving circulation' to zero, so rapidly that the maximum vorticity remains bounded.

A number of points in the analysis deserve particular attention, and point the way to possible future investigations:
\vskip 1mm
\noindent (i) Our key assumption has been that progress towards a singularity is controlled by the three variables $s(\tau), \kappa (\tau)$ and $\delta(\tau)$. This `circle-of-curvature assumption' is supported by the fact that the resulting dynamical system (\ref{system1}) yields scaling for $s(\tau)$ close to the Leray scaling $(t_{c}-t)^{1/2}$ that has been found in a number of previous Biot-Savart computations.  We describe this as `partial Leray scaling' because it does not apply to $\delta(\tau)$.  For particular tuning of the initial value $\delta(0)$ of $\delta$, we have found that this partial Leray scaling is exact.  
\vskip 1mm
\noindent (ii) Nevertheless, the question arises as to whether torsion of the vortex tubes (due to distortion out of planes parallel to $x=\pm z$) may also have an influence on the evolution near the tipping points.  For our configuration, this torsion is antisymmetric in the variable $y$ (and zero at the tipping points), so is unlikely to affect this evolution. Torsion is coupled with internal twist of the vortex tubes, and, as described in \S\ref{Preferential_twist_decay}, is attenuated by the stretching of the tubes. We believe that this torsion is the source of Kelvin waves that propagate away from the reconnection region, as observed in some Biot-Savart computations;  the symmetries  (\ref{symmetry_x}) and (\ref{symmetry_y}) that we have imposed do not however permit the Kelvin-wave instability.  A stability analysis of the flow in the tipping-point region would perhaps be amenable to numerical treatment.
\vskip 1mm
\noindent (iii) We have argued that there is always a tendency to restore the angle $\alpha$ to the value $\pi/4$, due to the rate-of strain at the tipping point.  This is hard to reconcile with Biot-Savart computations like that shown in figure \ref{Fig_rings_3D}, and the initial deformation shown in figure \ref{Fig_deformed_vortex}, which show a tendency for $\alpha$ to decrease from $\pi/4$. The footnote on p.~19 provides a partial explanation for this apparent disagreement.  It could be that the equilibrium angle is achieved as a compromise between the rapid extension in the $z$-direction and the very local restoring rate-of-strain field.  A smaller value of $\alpha$ could easily be assumed in the analysis, without  changing the qualitative nature of the results;  further analysis could shed light on this. 
\vskip 1mm
\noindent (iv) We have argued that core flattening is very limited if the Reynolds number is very high -- much higher than has been achieved in direct numerical simulations. Core flattening may however become more marked when $\delta/s$ increases to order unity, irrespective of Reynolds number;  use of  the volume form of the Biot-Savart law (rather than the line-integral form) might shed light on this also.  We may note that core-flattening cannot occur for quantised vortices in liquid helium; moreover, in this context $\delta$ is constant, and $s$ must decrease until $s\sim\delta$, when quantum reconnection  occurs, as described by \cite{Bewley2008}.
\vskip 1mm
\noindent (v) As $\delta/s$ increases to O$(1)$, the parameter $\xi\equiv \kappa s$ does so also, as evident in figure \ref{Fig_Evolution}(c) and (f), and \S\ref{Sec:Asymptotics_near_tau_c}.  The small-$\xi$ asymptotic expressions that we have used to obtain equations (\ref{system1}) must therefore fail, and should perhaps be replaced by the exact expressions for velocity, rate of stretching, and rate of increase of curvature in terms of elliptic integrals, as obtained in \S\S4 and 5 (e.g. (\ref{Tipping_point_vel2x}) for $\varv_{2x}$). This makes the system (\ref{system1}) much more complicated, but is unlikely to change the qualitative character of the solutions. Again this deserves further investigation.
\vskip 1mm
\noindent (vi) As we have pointed out, and as recognised in previous investigations, the Biot-Savart approach breaks down as and when the parameter $\delta/s$ increases to order unity. At this stage, the two vortices begin to overlap, and the onset of reconnection is inevitable.  The incident flux of vorticity $\Gamma$ then splits into surviving flux $\Gamma\!_{s}(\tau)\equiv \gamma(\tau)/\Gamma$ and reconnected flux $\Gamma\!_{r}(\tau)$.  In principle it is still possible that a Navier-Stokes singularity could occur if 
$\omega_{m}(\tau)\equiv\gamma(\tau)/\delta(\tau)^2$ were to increase without limit as $\tau\uparrow \tau_{c}$.  This reconnection process is the subject of a follow-up paper in preparation.
\vskip 1mm
Since the configuration that we have studied is generally agreed to be the most favourable for the development of a Navier-Stokes singularity, it seems well worth pursuing the implications of the simplified dynamical-system approach that we have developed in this paper.  Through this, we have succeeded in explaining how the Biot-Savart evolution can be governed by partial Leray scaling if the initial conditions are well tuned, and we have provided strong evidence that an Euler finite-time singularity can occur with the geometry considered.  This must remain conjectural at present; it will require further investigation that will again need to use the volume-integral form of the Biot-Savart law rather than its line-integral limit.

\vskip 2mm
\noindent
YK acknowledges support from JSPS KAKENHI Grant Numbers JP18H04443,\,JP24247014,\,  JP16H01175.
We acknowledge  use  of \emph {Mathematica} for the algebraic manipulations and plotting routines used throughout this paper.  We are grateful to three referees, whose helpful comments have led  to significant improvements in the presentation.
\bibliographystyle{jfm}
\providecommand{\NOOPSORT}[1]{}

\end{document}